\documentclass{article}

\usepackage{arxiv}

\usepackage[utf8]{inputenc} 
\usepackage[T1]{fontenc}    
\usepackage[hidelinks]{hyperref}       
\usepackage{url}            
\usepackage{booktabs}       
\usepackage{amsfonts}       
\usepackage{nicefrac}       
\usepackage{microtype}      
\usepackage{graphicx}
\usepackage{natbib}
\usepackage{doi}
\usepackage{authblk}
\usepackage{amsthm} 
\usepackage[mathscr]{euscript}

\newtheorem{theorem}{Theorem}
\newtheorem{proposition}{Proposition}
\newtheorem{lemma}{Lemma}

\newtheorem{definition}{Definition}
\newtheorem{assumption}{Assumption}

\usepackage{physics}
\usepackage[hidelinks]{hyperref} 
\usepackage[capitalise]{cleveref}
\crefname{assumption}{Assumption}{assumptions}
\Crefname{assumption}{Assumption}{assumptions}
\usepackage{tikz}
\usepackage{bbm}
\newcommand{\indicator}{\mathbbm{1}}

\renewcommand{\ip}[2]{#1\cdot #2}
\newcommand{\ipalt}[2]{\left\langle #1,#2 \right\rangle}
\DeclareMathOperator{\sinc}{sinc}

\newcommand{\NN}{\mathbb{N}}
\newcommand{\ZZ}{\mathbb{Z}}
\newcommand{\RR}{\mathbb{R}}

\newcommand{\EE}[1]{\mathbb{E} \left[ #1 \right]}
\newcommand{\CC}{\mathbb{C}}
\newcommand{\QQ}{\mathbb{Q}}
\newcommand{\RRd}{{\RR^d}}

\newcommand{\Var}{{\rm var}}
\newcommand{\Cov}{{\rm cov}}
\newcommand{\de}{\mathrm{d}}

\newcommand{\cov}[1]{\, {\rm cov}\left( #1 \right) }

\renewcommand{\var}[1]{\, {\rm var}\left( #1 \right) }
\newcommand{\cum}[1]{\, {\mathscr C}\left[ #1 \right] }
\newcommand{\set}[1]{\left\{#1\right\}}
\newcommand{\pred}{\bullet}

\newcommand{\freq}{{k}}
\newcommand{\lag}{u}
\newcommand{\loc}{s}

\newcommand{\randmeasurebase}{{\xi}}
\newcommand{\randmeasure}[1]{\randmeasurebase_{#1}}
\newcommand{\sampledmeasurebase}{\zeta}
\newcommand{\sampledmeasure}[1]{\sampledmeasurebase_{#1}}
\newcommand{\randmeasurecenter}[1]{\randmeasurebase_{#1}^0}
\newcommand{\momentmeasure}[1]{M_{#1}}
\newcommand{\cumulantmeasure}[1]{C_{#1}}
\newcommand{\reducedmeasure}[2]{\Breve{#1}_{#2}}
\newcommand{\reducedmomentmeasure}[1]{\reducedmeasure{M}{#1}}
\newcommand{\reducedcumulantmeasure}[1]{\reducedmeasure{C}{#1}}
\newcommand{\reducedcumulantdens}[1]{\reducedmeasure{c}{#1}}

\newcommand{\sdf}[1]{f_{#1}}
\newcommand{\asdf}[1]{\tilde{f}_{#1}}
\newcommand{\coh}[1]{r_{#1}}
\newcommand{\acoh}[1]{\tilde{r}_{#1}}
\newcommand{\phase}[1]{\theta_{#1}}

\newcommand{\intensitybase}{\lambda}
\newcommand{\intensity}[1]{\intensitybase_{#1}}
\newcommand{\intensityest}[1]{\hat{\intensitybase}_{#1}}

\newcommand{\region}{\mathcal{R}}

\newcommand{\grid}{\mathcal{G}}
\newcommand{\gridoffset}{v}
\newcommand{\aliasfreq}{\psi}
\newcommand{\Aliasfreq}{\Psi}

\newcommand{\elemprod}[1]{{\textstyle\prod}\left(#1\right)}

\newcommand{\conj}[1]{\overline{#1}}

\newcommand{\dft}[2]{J_{#1;#2}}
\newcommand{\pgram}[2]{I_{#1;#2}}
\newcommand{\mtpgram}[1]{\hat{f}_{#1}}

\newcommand{\Th}{th }	

\newcommand{\borel}[1]{\mathcal{B}(#1)}
\newcommand{\ball}[1]{B_{#1}}

\newcommand{\suppsection}[2]{\cref{#2}}

\usepackage{xcolor}

\title{Spectral estimation for point processes and random fields}
\renewcommand{\shorttitle}{Spectral estimation for point processes and random fields}

\newbox{\orcid}\sbox{\orcid}{\includegraphics[scale=0.06]{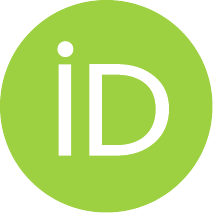}} 
\author[1]{\href{https://orcid.org/0000-0002-8808-4821}{\usebox{\orcid}}\hspace{1mm}{J. P. GRAINGER}\thanks{\texttt{jake.grainger@epfl.ch}}}
\author[2]{\href{https://orcid.org/0000-0002-1343-8058}{\usebox{\orcid}}\hspace{1mm}{T. A. RAJALA}\thanks{\texttt{tuomas.rajala@iki.fi}}}
\author[3]{\href{https://orcid.org/0000-0002-4830-8966}{\usebox{\orcid}}\hspace{1mm}{D. J. MURRELL}\thanks{\texttt{d.murrell@ucl.ac.uk}}}
\author[1]{\href{https://orcid.org/0000-0003-0061-227X}{\usebox{\orcid}}\hspace{1mm}{ S. C. OLHEDE}\thanks{\texttt{sofia.olhede@epfl.ch}}}
\affil[1]{Institute of Mathematics, \'Ecole Polytechnique F\'ed\'erale de Lausanne, Station 8, 1015 Lausanne, Switzerland }
\affil[2]{Natural Resources Institute Finland, 00790 Helsinki, Finland }
\affil[3]{Research Department of Genetics, Evolution and Environment, Centre for Biodiversity and Environment Research, University College London, UK }

\hypersetup{
pdftitle={Spectral estimation for point processes and random fields},
pdfsubject={stat.ME},
pdfauthor={J. P. GRAINGER, T. A. RAJALA, D. J. MURRELL,  S. C. OLHEDE},
pdfkeywords={Spatial point pattern, random field, spectral representation, spatial multitapering, coherence, partial coherence.},
}

\begin{document}
\maketitle

\begin{abstract}
Spatial variables can be observed in many different forms, such as regularly sampled random fields (lattice data), point processes, and randomly sampled spatial processes. 
Joint analysis of such collections of observations is clearly desirable, but complicated by the lack of an easily implementable analysis framework. 
We fill this gap by providing a multitaper analysis framework using coupled discrete and continuous data tapers, combined with the discrete Fourier transform for inference. 
Using this set of tools is important, as it forms the backbone for practical spectral analysis.
In higher dimensions it is important not to be constrained to Cartesian product domains, and so we develop the methodology for spectral analysis using irregular domain data tapers, and the tapered discrete Fourier transform.
We discuss its fast implementation, and the asymptotic as well as large finite domain properties. 
Estimators of partial association between different spatial processes are provided as are principled methods to determine their significance, and we demonstrate their practical utility on a large-scale ecological dataset.
\end{abstract}

\keywords{Spatial point pattern \and  random field \and  spectral representation \and  spatial multitapering \and  coherence \and  partial coherence.}

\section{Introduction}\label{sec:introduction}
Collections of spatial variables studied in geostatistics, ecology and other spatial sciences involve complex interactions between a variety of different components.
Often we need to jointly analyze data of different types, such as spatial point patterns, marked point patterns and realizations of random fields.
We therefore need a common framework to include all data types in our analysis.
Spectral analysis provides a convenient way to construct notions of correlation and partial correlation between these different types of processes.
In this paper, we develop methodology to estimate such quantities for any combination of point patterns, marked point patterns and realizations of random fields, when the processes may be recorded with differing sampling methods, and when the observational region is not necessarily rectangular, but common to all observed spatial variables.
Existing methodology, with the exception of that for univariate Gaussian random fields \citep{anden2020multitaper}, cannot handle arbitrary observational regions or different sampling mechanisms (including \citealt{rajala2023what}), and the existing spectral estimation methodology for marked point processes can be seen to be biased.
All of these issues are resolved by the novel methodology that we propose in this paper.

Probabilistically, a spectral representation for general multivariate random measures is available \citep{brillinger1972spectral,daley2003introduction}, providing the theoretical background for our work.
However, statistical estimation of the spectral density matrix function for such processes has not yet been developed.
Our introduction of multitapering is a necessary step to develop statistical methodology for the spectral analysis of multivariate spatial data.
We show in simulations that the large sample theory developed in this paper is applicable to data which is similar to data of practical interest, on which we also illustrate our methodology.
This enables us to use principled thresholds to determine significance, and gives confidence in the quality of the proposed methodology.

Although methodology for spectral estimation exists for spatial point processes \citep{bartlett1964spectral,diggle1987nonparametric,mugglestone1996practical,mugglestone1996exploratory,rajala2023what} and random fields \citep{bandyopadhyay2009asymptotic,matsuda2009fourier} separately, the extension to multivariate spatial data is more challenging.
\cite{kanaanCrossspectralPropertiesSpatial2008} proposed an estimator for the cross-spectral density function between a random field and an unmarked point pattern, that is limited to random fields continuously sampled in a rectangular region.
\cite{eckardt2019spatial} proposed a periodogram when the random field is recorded on an integer grid within a rectangular domain, but do not study its properties or discuss smoothing or tapering.
In reality, random fields can never be sampled continuously, and often we have multiple random fields recorded on different grids.
Handling this is not trivial, and getting it wrong can result in substantial bias in the estimated spectra.
We also consider marked point processes, where each point is associated with an additional random variable called a mark (e.g. size of a tree).
A periodogram estimator has been proposed in the marked setting \citep{renshaw2002two,eckardt2019partial}, however, this estimator is biased (see \suppsection{8.1}{app:marked:expectation} for details).
Hence, several key issues must be resolved to develop a practical, unified framework for spectral analysis of multivariate spatial data.

Whilst \cite{anden2020multitaper} consider spectral estimation for univariate Gaussian random fields on non-rectangular domains, performing such analysis in the case of multivariate random measures is more challenging.
In particular, one first needs to construct continuous families of taper functions, in order to analyse the point processes, and then also build discrete families of taper sequences which are appropriate for each grid used to record the random fields, but that are related to the continuous tapers (otherwise estimates of cross statistics will be biased).
We start from the discrete tapers of \cite{simons2011spatiospectral} and use these to generate continuous tapers, utilising the low wavenumber concentration already required to retain the same desirable properties.
We then construct discrete taper sequences from these continuous tapers, again exploiting the low wavenumber concentration.
As a result, we can combine all of these different kinds of data together, with no need for aggregation or interpolation, and requiring no additional tuning parameters beyond the single bandwidth parameter already used for multitaper estimation in the case of time series and random fields \citep{walden2000unified}.
This results in a methodology which neatly handles all of these additional complexities presented by the richer class of spatial processes.

Understanding dependence between spatial processes is challenging, making it valuable to have diverse tools for analysing observations. 
However, existing exploratory methods may fail to clearly reveal underlying processes. 
A common approach is to model point processes given some covariates and then examines pairwise residual dependence, but this ignores interactions with other point processes in the system. 
Partial coherence (see \cref{sec:background:coherence}) offers a measure of dependence that accounts for all other processes, not just covariates.

\begin{figure}[h]

\includegraphics{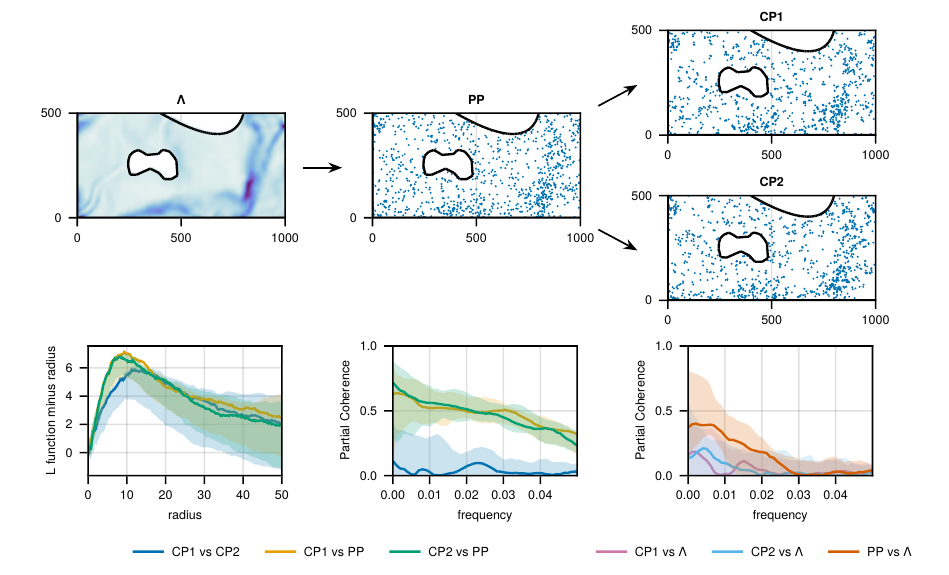}
    \caption{
        Simulated patterns (top), the inhomogeneous cross-L function (bottom left), and the partial coherence between the point processes (bottom middle) and between the covariate and the processes (bottom right).
    }
    \label{fig:fully_inhom_example}
\end{figure}

Consider an inhomogeneous Poisson process (PP) and two cluster processes (CP1, CP2) that share the Poisson process as a parent but are otherwise independent. 
Suppose the Poisson process intensity ($\Lambda$) is observed as a covariate (see the Appendix for details). 
\Cref{fig:fully_inhom_example} illustrates a sample process on an irregular domain (see \cref{sec:experiment:application}), along with the inhomogeneous cross-L function \citep{baddeley2000non}, and the partial coherence between processes. 
These statistics are standardized to zero when no interaction remains after accounting for other processes. 
Pointwise extrema over 100 replications are shown as bands; the lines correspond to the realization in the top row. 
The inhomogeneous cross-L function incorrectly suggests interactions among all processes, while partial coherence (which we estimate in this paper) correctly identifies the independence of the cluster processes once the Poisson process is accounted for. 
Partial coherence also captures the association between the covariate ($\Lambda$) and the Poisson process (PP), but not with the cluster processes (CP1, CP2), as expected. 
This toy example highlights how partial coherence complements existing spatial statistics tools, and why its accurate estimation is essential.

In this paper, we provide a single unified multitaper framework for estimation of the spectra and cross-spectra of random fields, point processes and marked point processes.
The methodology is discussed in detail, and we provide theoretical justifications for the proposed estimators.
We verify the properties of the proposed estimator through simulation studies, the first of their kind for spectral analysis of multivariate spatial data, and apply the methodology to forest ecology data from Barro Colorado Island \citep{condit2019complete}.
Several outstanding issues, long eluding the successful development of spectral analysis in this setting are resolved, including the extension to non-rectangular regions, the bias in the marked case, and handling different kinds of grid sampling.
We then discuss how to use this framework to compute estimates of coherence and partial coherence between the different processes, with principled significance thresholds.
We further demonstrate the promise of spectral analysis to summarize dependence in complex systems, and provide methodology to perform inference.

For readers from a point process background, \cite{percival1993spectral} provide an introduction to spectral analysis and, in particular, the multitaper method.
For readers from a time series background, \cite{illian2008statistical} provide an introduction to point processes.
In either case, Chapter 8 of \cite{daley2003introduction} provides a detailed treatment of the spectra of random measures, which is the most relevant theoretical background for this paper.

\section{Background}\label{sec:background}
\subsection{Basic notation}
We write $\NN=\set{1,2,\ldots}$ and for $n\in\NN$, $[n]=\set{1,\ldots,n}$.
Given a set $A$, and a binary operator on $A$, say $*$, then for $a\in A$ we write $A * a = \set{b * a \mid b\in A}$  for the right coset and $a * A$ for the left coset.
For two vectors $a$ and $b$, we will write $a\circ b$ for the elementwise (Hadamard) product, $a\oslash b$ for elementwise division, and $\ip{a}{b}$ for the dot product.
Furthermore, we will write $\elemprod{a}=\prod_{j=1}^d a_j$ for the product of elements of $a=(a_1,\ldots,a_d)^\top$.

\subsection{Random measures and spectral density functions}\label{sec:background:randommeasures}
A \emph{point process} is a random set of locations in some space, say $\RRd$, such that there are finitely many points within a given bounded Borel set \citep{moller2003statistical}.
A point process is said to be \emph{simple} if points do not occur in the same location almost surely.
A \emph{marked point process} is a point process with additional information at each of the points, called a mark, which we shall take to be real-valued and non-negative. 
For a marked point process, the point process without the marks is often referred to as the \emph{ground process}.
The unmarked case is recovered when the marks are independent and where the conditional distribution of the mark given a point is present at a location, the \emph{mark kernel}, is just a point mass at 1. 
Let $X$ be the set of random point locations, and $W(x)$ be the mark at  $x$, the point location.
The \emph{mark-sum measure}, or \emph{count measure} in the unmarked case (setting $W(x)=1$), is
\begin{align*}
    \randmeasurebase(A) &= \sum_{x \in X \cap A} W(x), \qquad A\in\borel{\RRd},
\end{align*}
where $\borel{\RRd}$ denotes the Borel sets of $\RRd$ \citep{daley2003introduction}.

A \emph{random field} is a random function, say $Y$, on $\RRd$ \citep{adler2010geometry}.
Assuming the random field is almost surely continuous and non-negative, we can define a random measure from $Y$ by
\begin{align*}
    \randmeasurebase(A) &= \int_A Y(u) \de u, \qquad A \in \borel{\RRd}.
\end{align*}
The two preceding equations show that random measures provide a unified framework in which to study these different spatial processes \citep{daley2003introduction}.

Assume that we have $P$ such processes, and augment our previous notation by writing $\randmeasure{p}$ for the $p$\Th random measure.
We assume that these processes are homogeneous \citep[Chapter 12]{daley2007introductionV2} and that their second-order moment measures exist and are finite (the equivalent of finite variance for time series).
From \cite{daley2003introduction}, for $1\leq p, q \leq P$ the first and second moment measures are given for $A,B \in \borel{\RRd}$ by
\begin{align*}
    \momentmeasure{p}(A) &= E\left\{\randmeasure{p}(A)\right\}, \qquad
    \momentmeasure{p,q}(A\times B) = E\left\{\randmeasure{p}(A) \randmeasure{q}(B)\right\},
\end{align*}
with the appropriate extension of $\momentmeasure{p,q}$ to $\borel{\RR^{2d}}$.
Under stationarity these moment measures have reduced forms so that for any $A\in\borel{\RRd}$ and any $g$ a bounded measurable function of bounded support
\begin{align*}
    \momentmeasure{p}(A) = \intensity{p} \ell(A),\qquad
    \int_{\RR^{2d}}  g(x, y) \momentmeasure{p,q}(\de x \times \de y) &= \int_\RRd \int_\RRd g(s + u, s) \ell(\de s) \reducedmomentmeasure{p,q}(\de u),
\end{align*}
where $\lambda_p$ is called the mean density and $\reducedmomentmeasure{p,q}$ is the reduced second-order moment measure and $\ell$ is the Lebesgue measure \citep[Chapter 8]{daley2003introduction}.\footnote{We write the lag in the first index to match the notation used in the time series literature, in contrast to \cite{daley2003introduction}.}

If $\randmeasure{p}$ is a simple point process, then $\intensity{p}$ is the intensity, which describes the average number of points per unit area.
If $\randmeasure{p}$ is a random field $\intensity{p}$ is the mean at any given location.
If $\randmeasure{p}$ is a marked point process with a simple ground process, then $\intensity{p}$ is the product of the mean mark and the intensity of the ground process \citep[Equation 5.1.19]{illian2008statistical}.

The reduced covariance (signed) measure between the $p$\Th and $q$\Th process is
\begin{align*}
    \reducedcumulantmeasure{p,q}(A) &= \reducedmomentmeasure{p,q}(A) - \intensity{p}\intensity{q}\ell(A), \qquad A \in \borel{\RRd}.
\end{align*}
Whilst the reduced covariance measures do not necessarily have densities, we write the reduced covariance density, $\reducedcumulantdens{p,q}$, a generalized density (which may have point masses), satisfying
\begin{align*}
    \reducedcumulantmeasure{p,q}(A) &= \int_A \reducedcumulantdens{p,q}(u)\de u, \qquad A \in \borel{\RRd}.
\end{align*}
If $p=q$ and the process in question is a simple point process with a reduced factorial moment measure that admits a density, which we call $\rho_{p,p}$, then, $\reducedcumulantdens{p,p}(\cdot) = \rho_{p,p}(\cdot)-\intensity{p}^2 + \intensity{p} \delta(\cdot)$ where $\delta(\cdot)$ is the Dirac delta function.
This is referred to as the complete covariance function by \cite{bartlett1963spectral}, who first introduced the spectra of point processes.
If the processes in question are both random fields, then $\reducedcumulantdens{p,q}(\cdot) = \cov{Y_{p}(\cdot), Y_{q}(0)}$ the usual autocovariance function.

The notion of spectra exists in a more general form than the one given here \citep{daley2003introduction}, but we are interested in processes for which the spectral density function exits.
To ensure this existence, we require that reduced covariance measure $\reducedcumulantmeasure{p,q}$ is totally finite.
In the random field case, this corresponds to the standard assumption that $\reducedcumulantdens{p,q}$ is integrable, e.g.~\cite{brillinger1974timeseries}.
In the point process case, this corresponds to assuming that $\rho_{p,q}(\cdot)-\intensity{p}\intensity{q}$ is integrable, as in \cite{rajala2023what} for example.

The (cross) spectral density function between the $p$\Th and $q$\Th processes is defined as
\begin{align*}
    \sdf{p,q}(\freq) &= \int_\RRd  e^{-2\pi i \ip{\freq}{u}} \reducedcumulantmeasure{p,q}(\de u), \qquad \freq \in \RRd.
\end{align*}
We call the matrix-valued function $f(\cdot)=[f_{p,q}(\cdot)]_{1\leq p,q\leq P}$ the spectral density matrix function.
At a given wavenumber, $f(\freq)$ plays the role of a wavenumber domain covariance matrix \citep{brillinger1972spectral,daley2003introduction}.

This definition generalizes the usual notion of spectral density matrix function from time series and random fields, as well as including the point process case as introduced by \cite{bartlett1963spectral}.
However, in the marked case this differs slightly from the definition introduced by \cite{renshaw2002two}.
In particular, \cite{renshaw2002two} defines the spectral density to be proportional to the Fourier transform of the reduced factorial moment density of the mark-sum measure, whereas we define it to be the Fourier transform of the reduced covariance measure.
Importantly, our definition corresponds to a special case of the definition for random measures, so we inherit all the properties of the spectral density matrix function listed by \cite{daley2003introduction}, such as positive semi-definiteness.
In addition, setting the mark kernel to a point mass at one recovers the unmarked case, in that the spectral densities as we define them are the same.
For a more thorough discussion, see \suppsection{8.2}{app:marked:spectra}.

In order to handle sampling of the random fields, we will need an additional assumption on the decay of their covariance function and spectral density function.
\begin{assumption}[Covariance decay]\label{assumption:spectra:exists}
    The reduced covariance measure $\reducedcumulantmeasure{p,q}$ is totally finite.
    In the case where both processes are random fields, the covariance density (function) $\reducedcumulantdens{p,q}$ is continuous and there exists $C,\delta > 0$ such that for all $x\in\RRd$, $|\reducedcumulantdens{p,q}(x)| + |\sdf{p,q}(x)| \leq {C}{(1+\norm{x}_2)^{-d-\delta}}$.
\end{assumption}
This latter condition ensures that we have an aliasing relation between the spectral density function of the continuous process and its sampled counterpart.
For example, random fields with Mat\'ern (cross-)covariance functions satisfy \cref{assumption:spectra:exists}.

So far, we only considered almost-surely non-negative random fields; however, this condition can be relaxed by viewing everything as random-signed measures.
Not all of the theory necessarily follows, see Section 8.4 of \cite{daley2003introduction} for details.
However, if these random signed measures do have reduced covariance measures satisfying \cref{assumption:spectra:exists}, then we can also perform estimation with our framework (e.g. for many Gaussian processes).

\subsection{Coherence and partial coherence}\label{sec:background:coherence}
Typically, we standardise the cross spectral density functions to complex \emph{coherence}, which is the wavenumber-domain correlation between the two processes.
Because coherence is complex-valued, we take the magnitude, which is called magnitude coherence, and argument, which is called phase \citep{carter1987coherence}. 
In particular, for $\freq\in\RRd$ define the coherence and phase as
\begin{equation*}
    \coh{p,q}(\freq) = \frac{\abs{\sdf{p,q}(\freq)}}{\left\{\sdf{p,p}(\freq)\sdf{q,q}(\freq)\right\}^{1/2}},\qquad
    \phase{p,q}(\freq) = \arg \sdf{p,q}(\freq).
\end{equation*}

Coherence has both benefits and limitations in this setting.
One benefit is that the coherence between a random field and a point process is natural to define.
In contrast, when considering spatial-domain cross-statistics between a random field and a point process, the random field needs to be made into a point process (or vice versa) and then standard spatial cross statistics for point processes (or random fields) can be computed \citep[Section 6.11.2]{illian2008statistical}.
This requires choices around how we convert one process into the other, which is not necessary with the wavenumber-domain approach.
However, there are some clear limitations.
In particular, the coherence assumes that the processes in question are pairwise homogeneous, and does not account for potential confounding from other observed processes.

One way to address both limitations is to consider partial coherence, which is the wavenumber-domain equivalent of partial correlation.
This is equivalent to computing the coherence of residual processes, after first removing the linear effect of the other processes \citep{eichler2003partial}.
In other words, this approach can account for inhomogeneity in the mean of the processes, as is also the case with the intensity reweighted stationary approaches introduced by \cite{baddeley2000non}, and the parametric approach proposed by \cite{waagepetersen2016analysis}.

More formally, define $\mathcal{V}_{p,q} = [P]\setminus\set{p,q}$ and let $\sdf{p,q\pred \mathcal{V}_{p,q}}$ be the cross-spectral density function of two residual processes formed by producing the best linear prediction of $\randmeasure{p}$ and $\randmeasure{q}$ from the other processes $\set{\randmeasure{r} \mid r \in \mathcal{V}_{p,q}}$.
As with partial correlation, the magnitude partial coherence and partial phase for $\freq\in\RRd$ are
\begin{align*} 
    \coh{p,q \pred \mathcal{V}_{p,q}}(\freq) = \frac{\abs{\sdf{p,q\pred \mathcal{V}_{p,q}}(\freq)}}{\left\{\sdf{p,p\pred \mathcal{V}_{p,q}}(\freq)\sdf{q,q\pred \mathcal{V}_{p,q}}(\freq)\right\}^{1/2}},\qquad
    \phase{p,q\pred \mathcal{V}_{p,q}}(\freq) = \arg \sdf{p,q\pred \mathcal{V}_{p,q}}(\freq),
\end{align*}
respectively \citep{dahlhaus2000graphical,eckardt2019partial}.
Importantly the magnitude partial coherence (and phase) can be computed efficiently by inverting the spectral matrix and making appropriate transformations \citep{dahlhaus2000graphical}.
This approach is then able to account for conditional inhomogeneity in the processes, which is not possible with coherence alone.
As demonstrated in \cref{fig:fully_inhom_example}, it is easy to account for the other observed point processes as well as the random fields.

\section{Estimation}
\subsection{Multitapering}
Assume we observe the random measures $\randmeasure{1},\ldots,\randmeasure{P}$ on some bounded region $\region\subset\RRd$.
Given a single realisation, we aim to construct estimators of the spectral density matrix function.
Of course, when the processes are random fields we cannot observe them everywhere in $\region$, but must instead sample them discretely in space.
In \cref{sec:estimation:irregular_sampling}, we discuss irregular sampling approaches, however, here we will assume that the random fields have been recorded on a regular grid.
In particular, if the $p$\Th process is a random field, assume it is observed on the intersection between $\region$ and the grid $\grid_p = \ZZ^d\circ \Delta_p + \gridoffset_p$ where $\Delta_p\in \QQ^d_{>0}$ denotes the sampling interval and $\gridoffset_p\in\QQ^d$ the offset of the grid.

To estimate the spectral density function, we first construct multiple tapered Fourier transforms, whose covariance matrix is asymptotically $f(\freq)$, and then compute their sample covariance matrix.
All of the process in which we are interested can be thought of generally as random measures \citep{daley2003introduction}.
Similarly, all of the tapered Fourier transforms can be seen as special cases of the tapered Fourier transform of a general random measure.

Consider a function $h:\RRd\rightarrow \RR$, bounded in magnitude and zero outside $\region$, which we will call a taper.
Then the tapered Fourier transform of $\randmeasure{p}$ is defined as
\begin{align*}
    J_p(\freq; h) &= \int_{\RRd} h(\loc) e^{-2\pi i \ip{\freq}{\loc}} \randmeasurecenter{p}(\de\loc), \qquad \freq\in\RRd
\end{align*}
where $\randmeasurecenter{p}$ denotes the centred random measure, i.e. $\randmeasurecenter{p}(A) = \randmeasure{p}(A) - \intensity{p} \ell(A)$.
Because $h$ is only non-zero inside $\region$, it is possible to compute the tapered Fourier transform from our observations.

If the $p$\Th process is a (marked) point process, with locations $X_p$ and marks $W_p$, then
\begin{align*}
    J_p(\freq; h)
    &= \sum_{x\in X_p} h(x)W_p(x)e^{-2\pi i \ip{\freq}{\loc}} - \intensity{p} H(\freq), \qquad \freq\in\RRd,
\end{align*}
where for $\freq\in\RRd$, $H(\freq)=\int_{\RRd} h(\loc)e^{-2\pi i \ip{\freq}{\loc}} \de\loc$ is the Fourier transform of $h$. 
When the process is not marked ($W_p(x)=1$ a.s.) this is the tapered Fourier transform proposed in \cite{rajala2023what}. 
When the process is marked, our approach represents a generalisation to the marked setting.

If the $p$\Th process is a random field, with the field denoted by $Y_p$, then
\begin{align*}
    J_p(\freq; h) &= \int_\RRd h(\loc)\{Y_p(\loc)-\intensity{p}\}e^{-2\pi i \ip{\freq}{\loc}} \de\loc, \qquad \freq\in\RRd.
\end{align*}
However, computing this would require us to record the random field everywhere in $\region$.
For some function $g$, let the grid sampled function $g^{(\grid_p)}$ be 
$
    g^{(\grid_p)}(\loc) = g(\loc) \elemprod{\Delta_p} \sum_{z\in\grid_p} \delta(\loc-z),
$
for $\loc\in\RRd$, where $\delta(\cdot)$ is the Dirac delta function. 
Then if we instead consider
\begin{align*}
    J_p(\freq; h^{(\grid_p)}) = \elemprod{\Delta_p} \sum_{\loc\in\grid_p} h(\loc)\{Y_p(\loc)-\intensity{p}\}e^{-2\pi i \ip{\freq}{\loc}}, \qquad \freq\in\RRd
\end{align*}
we see that this is essentially the usual tapered Fourier transform for random fields \citep[for example]{anden2020multitaper}.
It is important that the tapers $h$ and $h^{(\grid_p)}$ behave similarly, see \cref{assumption:tapers:concentration}, and this is easiest to achieve when the tapers used for the random fields are subsampled from the continuous base taper (with appropriate rescaling).

The tapered periodogram between processes $p$ and $q$ is
\begin{equation*}
    I_{p,q}(\freq; h_p, h_q) = J_{p}(\freq; h_p)\conj{J_q(\freq; h_q)}, \qquad \freq \in \RRd,
\end{equation*}
for the appropriate choices of $h_p$ and $h_q$.
The periodogram is the sample variance of one observation (with known mean zero).
The rational for this choice being that the tapered Fourier transform has a variance related to $f$.
However, the clear issue is that the sample variance of one observation is a very poor estimate of the population variance.
One technique to resolve this problem is multitapering, first proposed for the time series setting by \cite{thomson1982spectrum}, which constructs multiple different tapered Fourier transforms, and then computes the sample variance of this collection.
In particular, consider a family of tapers for each process $\set{h_{p;m}}_{m\in[M]}$, and define the shorthands
\begin{equation*}
    \dft{p}{m}(\freq) = J_{p}(\freq; h_{p;m}), \qquad \pgram{p,q}{m}(\freq) = I_{p,q}(\freq; h_{p;m}, h_{q;m}), \qquad \freq\in\RRd,
\end{equation*}
where $h_{p;m}$ is the appropriate taper for the given sampling regime, i.e. if the $p$\Th process is a (marked) point process $h_{p;m} = h_m$ and if the $p$\Th process is a random field sampled on the grid $\grid_p$ then $h_{p;m} = h_m^{(\grid_p)}$.
For a given $m$, the taper is always the same for two point processes, but for random fields will depend on the sampling grid.
Then the multitaper estimator is 
\begin{equation*}
    \mtpgram{p,q}(\freq) = \frac{1}{M} \sum_{m=1}^M \pgram{p,q}{m}(\freq), \qquad \freq\in\RRd.
\end{equation*}
The estimator performs well if the family of vectors of tapered Fourier transforms $J_m(\freq)=[\dft{p}{m}(\freq)]_{p\in[P]}$ are iid across $m$,
which is true asymptotically (see \cref{theorem:normality_dft}).

\subsection{A general characterisation for growing observational regions}\label{sec:estimation:general}

Whilst we can make stronger statements about specific region shapes and asymptotic regimes, we start with the generic setting in which we have a sequence of increasing regions, and ask what properties we would need tapers to satisfy in order to construct consistent estimators.
Such properties will place some implicit constraints on the kinds of sequences of regions for which we can use multitapering. In \suppsection{2.2}{sec:estimation:template}, we give explicit forms of growing domain and taper choice that satisfy all of the assumptions we introduce here.
In general these properties are useful, firstly to abstract some of the proofs, and secondly because they give a good indication of the desirable taper properties in practice, when given a single region.

Consider a sequence of regions $\{\region_n\}_{n\in \NN}$ such that for $n\in\NN$, $\region_{n} \subset \region_{n+1}$, and $\ell\left(\bigcup_{n=1}^{\infty}\region_{n}\right) = \infty$.
We aim to construct a sequence of families of tapers $\{h_{m,n}\}_{m\in[M_n]}$, where the number of tapers $M_n$ may depend on $n$.
For consistency of the multitaper estimate (\cref{res:consistency}), we will need $M_n$ to grow with $n$, but for the asymptotic normality of the tapered Fourier transform (\cref{theorem:normality_dft}), we will need $M_n$ to be fixed. 
This will be made clear in the relevant results.
\begin{assumption}[Finite sample taper properties]\label{assumption:tapers:finitesample}
    The family of tapers $\{h_{m,n}\}_{m\in[M_n]}$ are such that for all $n\in\NN$, for all $m\in[M_n]$, $h_{m,n}: \RRd \rightarrow \RR$ is bounded, continuous, supported on a subset of $\region_n$ with $\norm{h_{m,n}}_2^2=1$ and $\norm{H_{m,n}}_1<\infty$, where $H_{m,n}$ is the Fourier transform of $h_{m,n}$.
\end{assumption}
Boundedness and continuity ensure the tapered Fourier transforms are well defined.
The $L^2$ norm assumption ensures that the tapers are normalized.
The finite $L^1$ norm allows us to rearrange the order of integration in the proof of some results and to invert various Fourier transforms, and is satisfied by all taper families constructed in this paper.\footnote{Using no taper (a scaled indicator function), we would not satisfy this assumption. However, due to the substantial leakage bias \citep{percival1993spectral}, we never use no taper.}

There are two forms of bias present in spectral estimation, which we will refer to as leakage bias and aliasing bias.
The leakage bias results from the boundary effects, and is importantly a property of the taper which we control.
The aliasing bias comes from the sampling of the random fields, and is a property of both the sampling grid and the underlying random field.
Intuitively, aliasing results from only being able to sample a function on some regular grid.
In this case, we are thinking about this function being the autocovariance of a random field (or between two random fields).
The impact of such sampling in space is clear: we cannot observe the covariance at all possible lags, but just on a subset of them.
In wavenumber, grid sampling results in aliasing, which is a more complicated phenomena, where the Fourier transform we observe is the Fourier transform we care about plus some erroneous values from higher wavenumbers.

The bias effects of tapers are most easily understood in terms of a smoothing in the wavenumber domain.
To understand this, we will need to consider the Fourier transform of the taper, and the aliased spectral density function.
In particular, write $H_{p;m,n}$ for the Fourier transform of the taper so that if the $p$\Th process is a point process
\begin{align*}
    H_{p;m,n}(\freq) &= H_{m,n}(\freq) = \int_\RRd h_{m,n}(\loc) e^{-2\pi i \ip{\freq}{\loc}} \de\loc,
\end{align*}
and for a random field is
\begin{align*}
    H_{p;m,n}(\freq) &= H_{m,n}^{(\grid_p)}(\freq) = \elemprod{\Delta_p} \sum_{\loc\in\grid_p} h_{m,n}(\loc) e^{-2\pi i \ip{\freq}{\loc}},
\end{align*}
for any $\freq \in \RRd$.
In the latter case, $|H_{p;m,n}|$ is periodic on $K_{p}=[-1/2,1/2]^d \oslash \Delta_p$, which we will call the Nyquist box.
For convenience, if the process is sampled continuously we set $K_p=\RRd$.
Aliasing for processes on a single grid \citep{percival1993spectral} is simpler compared to multiple processes sampled in different ways.
Therefore we need a more general notion of aliased spectra.

\begin{definition}[The aliased spectral density]\label{def:aliased_sdf}
    Introduce a set of aliasing wavenumbers $\Aliasfreq_p$ and a phase adjustment function $w_p:\RRd\rightarrow \CC$, so that if the $p\Th$ process is sampled on a grid, then $\Aliasfreq_p = \ZZ^d \oslash \Delta_p$ and $w_p(x)=e^{-2\pi i \ip{\gridoffset_p}{x}}$ and otherwise $\Aliasfreq_p=\set{0}$ and $w_p(x)=1$.
    Let the aliased spectral density function be
    \begin{align*}
        \asdf{p,q}(\freq) &= \sum_{\aliasfreq\in\Aliasfreq_{p}\cap\Aliasfreq_{q}} \sdf{p,q}(\freq + \aliasfreq) w_{p}(\aliasfreq) \conj{w_{q}(\aliasfreq)}, \qquad \freq\in\RRd.
    \end{align*}
\end{definition}
Aliasing is not always present, but it is notationally convenient to define the aliased spectral density function in all cases.
Since zero is always in the set $\Aliasfreq_p\cap\Aliasfreq_q$, we have $\sdf{p,q}(\freq)=\asdf{p,q}(\freq)$ if there is no aliasing, in particular, if one of the processes is sampled continuously in space.
When both processes $p$ and $q$ are sampled on a grid, $|\asdf{p,q}|$ is periodic on the set $K_{p,q} = [-1/2,1/2]^d\oslash\Delta_{p,q}$ where $\Delta_{p,q}$ is elementwise the largest number such that both $\Delta_p\oslash \Delta_{p,q}\in\NN^d$ and $\Delta_q\oslash \Delta_{p,q}\in\NN^d$.
For example, if $\Delta_{p} = (4, 1/3)^\top$ and $\Delta_{q} = (6, 1/2)^\top$, then $\Delta_{p,q} = (2, 1/6)^\top$.
When one or both of the processes are sampled continuously in space, then we write $K_{p,q} = \RRd$.
The product $H_{p;m,n}(\freq)\conj{H_{q;m,n}(\freq)}$, which is important for leakage bias, is also periodic in magnitude on $K_{p,q}$.

\begin{figure}[h]

\includegraphics{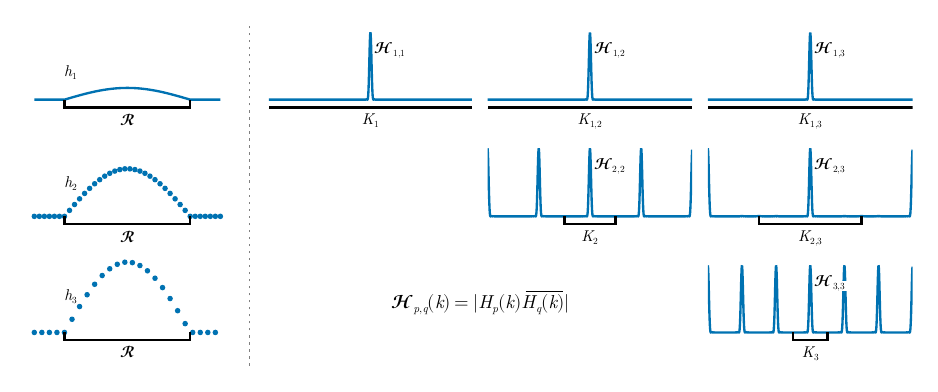}
    \caption{
        An illustration of the aliasing effects in one dimension, assuming the first process was recorded continuously, and the second and third recorded on grids with sampling intervals of 4 and 6 respectively.
        The taper in question is a minimum bias taper \citep{riedel1995minimum}.
    }
    \label{fig:aliasing_illustration}
\end{figure}

The left half of \cref{fig:aliasing_illustration} shows an example taper and its discrete counterparts on two different grids.
The right half of \cref{fig:aliasing_illustration} shows the products of the different Fourier transforms of the taper and its discrete counterparts.
In particular, we see from the diagonal terms that the discrete version repeat on different intervals.
In addition, from the off-diagonals we see that pairs including the continuous taper do not repeat, and the cross term between the two discrete tapers is periodic with a longer period than either of the originals, specifically $K_{p,q} \supseteq K_p$ for all $p,q\in[P]$.

Unless specified otherwise, we shall assume for simplicity that the intensity/mean $\intensity{p}$ is known, which we refer to as the ``oracle case''.
Estimating the intensity has a negligible effect, except near zero wavenumber.
However, the finite sample equations become more complicated, and so we consider the oracle case here, and the non-oracle case is discussed in \suppsection{6}{app:nonoracle}.
We can now obtain a general expression for the effect of the tapers and grid sampling on the bias of the periodogram.
This generalizes the standard result for multivariate time series \citep{walden2000unified}, random fields \citep{guillaumin2022debiased} and univariate point processes \citep{rajala2023what}.
\begin{proposition}[Expectation of the periodogram]\label{res:bias:finite}
    Given the processes satisfy \cref{assumption:spectra:exists} and the tapers satisfy \cref{assumption:tapers:finitesample}, the expectation of the periodogram is
    \begin{align*}
        E\left\{\pgram{p,q}{m,n}(\freq)\right\} &= \int_{K_{p,q}} H_{p;m,n}(\freq') \conj{H_{q;m,n}(\freq')} \asdf{p,q}(\freq-\freq') \de\freq',
    \end{align*}
    for all $n$, for all $m\in[M_n]$, $\freq\in\RRd$.

    \begin{proof}
        See \suppsection{3.1}{proof:res:bias:finite}.
    \end{proof}
\end{proposition}
The only way to reduce aliasing bias is to sample the process at a higher resolution, or to sample it randomly and treat it as a marked point process (see \suppsection{9}{app:irregular_sampling}).
Since we typically do not control the sampling mechanism, we cannot control the aliasing bias, and should just be aware that it exists.
However, since the expectation is a convolution between the aliased spectral density function and a property of the taper, we can control leakage bias by constraining the behaviour of the taper.

The underlying idea is to construct the tapers such that their Fourier transforms concentrate within a bandwidth around zero, and that this concentration increases appropriately with growing domain \citep{walden2000unified}.
Combined with continuity of the spectral density function, this will yields asymptotic unbiasedness (up to an aliasing effect).

\begin{assumption}[Taper concentration]\label{assumption:tapers:concentration}
    Let $b_n\rightarrow0$ as $n\rightarrow \infty$ be the taper bandwidth, and $\ball{b_n}$ be a ball centred at zero of radius $b_n$.
    Then we assume that for all $p,q\in [P]$ as $n\rightarrow\infty$,
    \begin{align*}
        \max_{m\in[M_n]} \abs{1-\int_{\ball{b_n}} H_{p;m,n}(\freq) \conj{H_{q;m,n}(\freq)} \de\freq} \rightarrow 0, \quad
        \max_{m\in[M_n]} \int_{K_{p}\setminus\ball{b_n}} \abs{H_{p;m,n}(\freq)}^2 \de\freq &\rightarrow 0.
    \end{align*}
\end{assumption}
In the case when every process is sampled continuously, the latter condition is not necessary, as $\norm{H_{m,n}}_2=\norm{h_{m,n}}_2=1$. 
However, for the processes sampled on a grid it is necessary.
If the former condition holds, then for the grid sampled cases the latter condition is equivalent to $\max_{m\in[M_n]}|1-\elemprod{\Delta_p} \sum_{u\in\grid_p} h_{m,n}(u)^2|\rightarrow 0$ as $n\rightarrow\infty$.

\begin{theorem}[Asymptotic bias of the periodogram]\label{res:bias:asymp}
    Let the processes satisfy \cref{assumption:spectra:exists}, and the tapers satisfy \cref{assumption:tapers:finitesample,assumption:tapers:concentration} for some fixed number of tapers $M$.
    For all $\freq\in\RRd$, $p,q\in[P]$, for all fixed $m\in [M]$,
    $$E\left\{\pgram{p,q}{m,n}(\freq)\right\} \rightarrow \asdf{p,q}(\freq),$$
    as $n\rightarrow\infty$.
    \begin{proof}
        See \suppsection{3.2}{proof:res:bias:asymp}.
    \end{proof}
\end{theorem}

Theorem \ref{res:bias:asymp} shows that the only asymptotic bias of the periodogram is due to aliasing.
Importantly, in the case where at least one of the processes is not sampled on a grid, the periodogram is asymptotically unbiased. 
This includes marked point processes, meaning that we have resolved the bias present in the current state-of-the-art estimator \citep{renshaw2002two}.
The two main results we will establish are consistency of the multitaper periodogram under a growing number of tapers, and asymptotic normality of the tapered discrete Fourier transform under a finite number of tapers.
In both cases, we need the tapers to be asymptotically orthogonal.
\begin{assumption}[Asymptotic orthogonality]\label{assumption:tapers:orthogonality}
    For all $p,q\in[P]$, as $n\rightarrow \infty$,
    \begin{align*}
        \max_{m\in[M_n]} \max_{m'\in[M_n]\setminus \set{m}}\abs{\int_{\ball{b_n}} H_{p;m,n}(\freq) \conj{H_{q;m',n}(\freq)} \de\freq} &\rightarrow 0.
    \end{align*}
\end{assumption}
In general, we will construct tapers that are close to being orthogonal, but not quite, in particular when constructing tapers for unusual region shapes.
In practice, \cref{assumption:tapers:concentration,assumption:tapers:orthogonality} can be checked numerically, as will be discussed in \cref{sec:estimation:interpolate}.
As well as yielding variance reduction for consistency, this assumption also ensures that the different tapered Fourier transforms are asymptotically uncorrelated when we study asymptotic normality.
We also assume that the $L^1$ norms of the Fourier transforms of the tapers shrink to zero as the observational region grows.
\begin{assumption}[Shrinking taper Fourier transform]\label{assumption:tapers:consistency}
    For all $p\in[P]$, there is $C_p<\infty$ such that
    \begin{align*}
        \max\limits_{m\in [M_n]} \int_{K_p} \abs{H_{p;m,n}(\freq)} \de \freq &\rightarrow 0, \qquad \max_{m\in[M_n]} \sum_{\aliasfreq \in\Aliasfreq_{p}} \left(\int_{K_p+\aliasfreq} \abs{H_{m,n}(\freq)}^{2} \de\freq\right)^{1/2} \rightarrow C_p
    \end{align*}
    as $n\rightarrow \infty$.
\end{assumption}
The first part of \cref{assumption:tapers:consistency} is commonly used when dealing with spectral estimation on unusual domains or more abstract settings \citep{brillinger1982asymptotic}.
The second part is a technical condition to handle the grid sampled processes.
See \suppsection{2.2}{sec:estimation:template} for examples of constructions where both conditions are satisfied.
In particular, both conditions are implied if we assume that the Fourier transform of the taper is bounded by a decreasing function with appropriate tail decay.

Finally we need to make an assumption about the higher-order cumulants of the processes.
At a high level, we just need to ensure that fourth order terms that appear in the variance of the periodogram (due to the periodogram being a product of two tapered Fourier transforms) are negligible.
It will be convenient for the theory to embed all of the grids on one fine grid.
In particular, let $\grid = \Delta\circ \ZZ^d$ to be a grid which contains all of the other grids we consider (e.g. by taking $\Delta$ to be one over the product of the denominators of all $\Delta_p$ and $\gridoffset_p$ in each dimension).
Weighted sums on the original grid can always be constructed by multiplying the desired weights by $\indicator_{\grid_p}$.
Now it is easier to work with the sampled random measures, which will include the potential effect of the grid sampling. 
Define the sampled random measures $\sampledmeasure{p}$ for $A\in\borel{\RRd}$ by $\sampledmeasure{p}(A)=\randmeasure{p}(A)$ if the $p$\Th process is sampled continuously, and $\sampledmeasure{p}(A) = \sum_{u \in \grid\cap A} Y_p(u)$ if the $p$\Th process is sampled on a grid.

Let $\cum{X_1,\ldots,X_r}$ denote the $r$th-order cumulant of some random variables $X_1,\ldots,X_r$.
Define the $r$\Th-order cumulant measure of the sampled process as 
$$\cumulantmeasure{p_1,\ldots,p_r}^{(\sampledmeasurebase)}(A_1\times \cdots\times A_r) = \cum{\sampledmeasure{p_1}(A_1), \ldots, \sampledmeasure{p_r}(A_r)},$$ for $A_1,\ldots,A_r\in \borel{\RRd}$, appropriately extended to $\borel{\RR^{rd}}$ \citep{daley2007introductionV2}.
Now we need a minor extension of Proposition 12.6.III in \cite{daley2007introductionV2} to include the grid sampling.

\begin{lemma}[Reduced cumulant measure]\label{prop:reduced_cumulant_measure}
    For $p_1,\ldots,p_r\in[P]$, assume that the $r$\Th joint moment measures of $\randmeasure{p_1},\ldots,\randmeasure{p_r}$ exist and are finite.
    Say that either all of the processes are sampled continuously, or they are enumerated so that at least $\randmeasure{p}$ is sampled on a grid.
    For any bounded measurable function of bounded support $g:\RR^{rd}\rightarrow \RR$ there exists a reduced cumulant measure $\reducedcumulantmeasure{p_1,\ldots,p_r}^{(\sampledmeasurebase)}$ such that
    \begin{align*}
        &\int_{\RR^{rd}} g(x_1,\ldots,x_r) \cumulantmeasure{p_1,\ldots,p_r}^{(\sampledmeasurebase)}(\de x_1\times\cdots\times \de x_r)  \\
        &\quad = \int_{\RRd} \int_{\RR^{(r-1)d}} g(x+\lag_1,\ldots,x+\lag_{r-1},x) \reducedcumulantmeasure{p_1,\ldots,p_r}^{(\sampledmeasurebase)}(\de u_1\times\cdots\times \de u_{r-1}) \ell_{p_r}(\de x),
    \end{align*}
    where $\ell_{p_r}$ is the Lebesgue measure on $\RR^d$ if $\randmeasure{p}$ is sampled continuously (meaning all the processes are sampled continuously), and $\ell_{p_r}$ is the counting measure on $\grid$ if $\randmeasure{p}$ is sampled on a grid.

    \begin{proof}
        See \suppsection{3.3}{proof:prop:reduced_cumulant_measure}.
    \end{proof}
\end{lemma}

More concretely, when all of the processes are sampled on a grid, in the fourth-order case
\begin{align*}
    \reducedcumulantmeasure{p,q,r,s}^{(\sampledmeasurebase)}(A_1 \times A_2 \times A_3) &= \sum_{\lag_1\in\grid\cap A_1} \sum_{\lag_2\in\grid\cap A_2} \sum_{\lag_3\in\grid\cap A_3} \cum{Y_p(\lag_1), Y_q(\lag_2), Y_r(\lag_3), Y_s(0)}.
\end{align*}
so that in particular, the summand is the usual cumulant function of a stationary multivariate time series or random field, see for example \cite{brillinger1974timeseries}.

\begin{assumption}[Fourth-order cumulants]\label{assumption:cumulant:totallyfinite}
    For $p,q\in[P]$, the first four joint moment measures of $\sampledmeasure{p}$, $\sampledmeasure{q}$ exist and are finite, and the reduced cumulant measure $\reducedcumulantmeasure{p,p,q,q}^{(\sampledmeasurebase)}$ is totally finite.
\end{assumption}

\cref{assumption:cumulant:totallyfinite} corresponds to standard integrability/absolute summability of higher-order cumulant function in the time series case \citep{brillinger1965introduction}, analogously to \cref{assumption:spectra:exists}.

\begin{theorem}[Consistency of the multitaper periodogram]\label{res:consistency}
    If the processes satisfy \cref{assumption:spectra:exists,assumption:cumulant:totallyfinite}
    and the tapers satisfy \cref{assumption:tapers:finitesample,assumption:tapers:concentration,assumption:tapers:orthogonality,assumption:tapers:consistency} and in addition $M_n\rightarrow \infty$ as $n\rightarrow \infty$, then for all $\freq\in\RRd$, for all $p,q\in[P]$,
    $$E\left\{\mtpgram{p,q;n}(\freq)\right\}\rightarrow \asdf{p,q}(\freq), \quad 
    \Var\left\{\mtpgram{p,q;n}(\freq)\right\} \rightarrow 0,$$
    as $n\rightarrow \infty$ .
    \begin{proof}
        See \suppsection{3.4}{proof:res:consistency}.
    \end{proof}
\end{theorem}
As a result, we see that if we grow the number of tapers with appropriate control on their behaviour, then we obtain an estimator which is mean-square consistent, which is not true when using a single taper.


There are two main approaches to establishing asymptotic normality of the tapered Fourier transform.
The first is to use Brillinger mixing conditions \citep{brillinger1982asymptotic}, which assume that all of the higher-order cumulant measures exist and are totally finite.
The second is to use an $\alpha$-mixing condition \citep[for example]{yang2024fourier}, typically with the assumption that at least finitely many moments exist and are finite (see e.g.\ \cite{biscio2016standard} for a discussion of the two assumptions in the point process setting).
In \suppsection{2.1}{supp:asymptotic:normality}, we give a result using the general $\alpha$-mixing central limit theorem from \cite{biscio2019general}.
However, we will use Brillinger mixing here as we already introduced higher-order cumulants.

\begin{assumption}[Brillinger mixing]\label{assumption:mixing:brillinger}
    For all $r\in\NN$, for all $p_1,\ldots,p_r\in[P]$, the joint moment measures of $\sampledmeasure{p_1}, \ldots, \sampledmeasure{p_r}$ up to order $r$ exist and are finite, and $\reducedcumulantmeasure{p_1,\ldots,p_r}^{(\sampledmeasurebase)}$ is totally finite.
\end{assumption}
\Cref{assumption:mixing:brillinger} will be used for asymptotic normality of the tapered Fourier transforms.
Since the tapered Fourier transforms are complex-valued, convergence will be to a complex normal distribution.
We write $\mathcal{CN}(\mu, S, R)$ for a multivariate complex-normal distribution mean vector $\mu$, covariance matrix $S$ and relation matrix $R$.
Furthermore, since we transform real-valued processes, the tapered Fourier transforms are symmetric.
In addition, for certain wavenumbers, the tapered Fourier transforms are real valued (at zero, and multiple of the Nyquist wavenumbers).
This is the same as the time series case, see \cite{brillinger1974timeseries}, with a slightly more complicated formulation due to the different kind of sampling present.
For this reason, we restrict attention to wavenumbers where these issues do not occur.

\begin{theorem}[Asymptotic normality of tapered Fourier transforms]\label{theorem:normality_dft}
    Let the processes satisfy \cref{assumption:spectra:exists,assumption:mixing:brillinger} and the tapers satisfy \cref{assumption:tapers:finitesample,assumption:tapers:concentration,assumption:tapers:orthogonality,assumption:tapers:consistency}.
    Consider $\freq_1,\ldots,\freq_r\in\RRd$ such that, for all $i, j\in[r]$ with $i\neq j$, $2\freq_i,2\freq_j, \freq_i\pm\freq_j \notin \bigcup_{p\in[P]} \Aliasfreq_p$. Then for all $i\in[r]$
    \begin{align*}
        J_{m,n}(\freq_i) = \left[\dft{p}{m,n}(\freq_i)\right]_{1\leq p\leq P} \xrightarrow{d} \mathcal{CN}(0, \tilde{f}(\freq_i), 0)
    \end{align*}
    as $n\rightarrow\infty$.
    In addition, for any two distinct $m,m'$, the vectors $J_{m,n}(\freq_i)$ and $J_{m',n}(\freq_i)$ are asymptotically independent, so the matrix $J_n(\freq_i) = [J_{m,n}(\freq_i)^{T}]_{1\leq m\leq M}$ is asymptotically complex normal with uncorrelated rows.
    Finally, for $i\neq j$, $J_n(\freq_i)$ and $J_n(\freq_j)$ are asymptotically independent.
    \begin{proof}
        See \suppsection{3.5}{proof:theorem:normality_dft}.
    \end{proof}
\end{theorem}

\subsection{Practical construction of tapers on a region of interest}\label{sec:estimation:interpolate}
For a general region which we observe (including non-convex regions) we can construct a family of tapers which satisfy the finite sample conditions in \cref{assumption:tapers:finitesample}, and for which we can check how well asymptotic conditions in the remaining assumptions are satisfied.
We do this by first taking (multidimensional) Slepian sequences on some grid which can be constructed numerically using the methodology of \cite{simons2011spatiospectral}.
Then we multilinearly interpolate these taper sequences to construct a family of continuous tapers on the region of interest.

\begin{proposition}\label{res:general:interpolate}
    Let $g_1,\ldots,g_M$ be a family of discrete space tapers on a grid $\grid=\Delta\circ \ZZ^d+\gridoffset$, only non-zero on the bounded region $\tilde{\region} = \set{\loc \in \region \mid \loc+\Delta \circ [-1,1]^d\subseteq \region}.$
    Then if we construct a family of tapers via multilinear interpolation of $g_1,\ldots,g_m$, then the interpolated family satisfies \cref{assumption:tapers:finitesample}, provided they are correctly normalised.
    Furthermore, their Fourier transform is
    \begin{align*}
        H_{m}(\freq) &= G_m^{(\grid)}(\freq) \prod_{j=1}^d \sinc^2(\pi \Delta_j \freq_j)
    \end{align*}
    where $\freq_j$ and $\Delta_j$ are the $j$\Th components of $\freq$ and $\Delta$ respectively, and $\sinc(x) = \sin(x)/x$.

    \begin{proof}
        See \suppsection{3.6}{proof:res:general:interpolate}.
    \end{proof}
\end{proposition}
The construction on the smaller region $\tilde{\region}$ is to ensure that the interpolated tapers are zero outside $\region$, avoiding linear interpolation across holes not fully covered by the grid, for example.
In order to correctly normalize the tapers, we need their $L^2$ norm, which can be computed exactly with finite sums using the results in \suppsection{7.2}{app:interp:withinterp}.
We can then check the conditions of \cref{assumption:tapers:concentration,assumption:tapers:orthogonality} numerically.

Additional specific details of the taper construction and effect of the linear interpolation are given in \suppsection{7}{app:interp}.
In the case of random fields sampled on a regular grid, the tapered Fourier transforms can be computed efficiently using Fast Fourier Transforms \citep{cooley1965algorithm}, as is well known.
For point patterns, we can use Non-Uniform Fast Fourier Transforms, the properties of which were studied by \cite{dutt1993fast}.
We use the NUFFT implementation from \cite{barnett2019parallel}.

\begin{figure}

\includegraphics{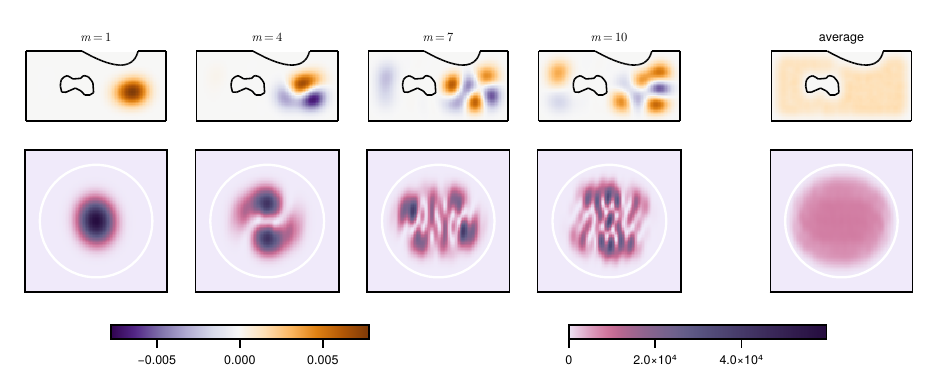}
    \caption{
        Four of the tapers and the total average absolute taper for an irregular region (top), and the corresponding spectral windows (bottom).
        The spatial data has axes 0 to 1000 and 0 to 500, and the wavenumber domain plots range from $-0.01$ to $0.01$ in each dimension.
    }
    \label{fig:bci_tapers}
\end{figure}

A subset of the tapers we will use for the Barro Colorado Island data (see \cref{sec:experiment:application}) is shown in \cref{fig:bci_tapers}.
Whilst any given taper concentrates on a certain region of the spatial domain, the total weight is approximately uniform across the spatial domain, except at the borders (\cref{fig:bci_tapers}, top row).
Similarly, in the wavenumber domain we see that the spectral windows are concentrated on different regions within the bandwidth of zero (denoted by the white circle), but together they cover this region evenly (\cref{fig:bci_tapers}, bottom row).
This points to the major benefit of multitapering, as if we were to use a single taper and kernel smoothing, we would be focusing on a small part of the observational region whereas multitapering uses most of the region evenly.

\subsection{Irregular sampling}\label{sec:estimation:irregular_sampling}
So far, we have only considered the case where the processes are sampled on a regular grid.
However, we may also want to handle situations in which they are not.
This specific case has been considered previously in the literature, for example \cite{masry1978poisson, masry2003alias, matsuda2009fourier}.
In fact, this also fits into the framework we introduce in this paper.
Say we have a mean-zero random field $Y$ which we sample at the now random locations given by the point process $X$.
Then we can regard this sampled process as a mark sum measure, say $\randmeasurebase$, corresponding to the marked point process with marks $Y(x)$ for $x\in X$.
Then, if the marks are independent of the locations, from \cite{daley2003introduction}, page 338, we have
\begin{align*}
    f_{\randmeasurebase\randmeasurebase}(\freq)
    &= \int_{\RRd}f_{YY}(\freq-\freq') f_{XX}(\freq')\de\freq' + \intensity{X} f_{YY}(\freq), \qquad \freq \in \RRd
\end{align*}
where $f_{\randmeasurebase\randmeasurebase}$ is the spectral density function of the marked process, $f_{YY}$ is the spectral density function of the random field, $f_{XX}$ is the spectral density function of the point process, and $\intensity{X}$ is the intensity of the point process.
If the point process used for sampling is Poisson, then
\begin{align*}
    f_{\randmeasurebase\randmeasurebase}(\freq)
    &= \intensity{X}\var{Y(0)} + \intensity{X}^2 f_{YY}(\freq), \qquad \freq \in \RRd.
\end{align*}
The spectra of the random field $f_{YY}$ can then be estimated by plug-in and appropriate rearrangement.
That estimator would be identical to the one we would obtain if the sample locations were uniform over the region and we used the methodology proposed by \cite{matsuda2009fourier}.
For more details, see \suppsection{9}{app:irregular_sampling}.

\subsection{Coherence and partial coherence}
As usual for spectral estimation, we use a plug-in estimator for the magnitude coherence, phase and their partial counterparts. 
This requires the estimated spectral density matrix at a given frequency to be invertible, so we need to have more tapers than processes, a necessary but not sufficient condition \citep{walden2000unified}.
In particular, we cannot use the periodogram as a plug-in estimator for the magnitude partial coherence as the periodogram is not invertible \citep{walden2000unified}.

The probability density function of the absolute value of the sample correlation of a bivariate complex proper normal random vector with absolute correlation $\rho$ and $N$ observations is
\begin{equation}
    f(r; \rho, N) = 2(N-1)(1-\rho^2)^N r (1-r^2)^{N-2} {_2F_1}(N,N,1; \rho^2 r^2) \label{eq:abscoh_dist}
\end{equation}
if $0<r<1$ and zero otherwise \citep{miller1980hypothesis}.
Provided the assumptions of Theorem~\ref{theorem:normality_dft} hold, the matrix $J_n(\freq)$ is asymptotically complex normal.
Therefore, the plug in estimator for $\coh{p,q}$, which is the absolute value of the sample correlation of the matrix $J_n(\freq)$, has an asymptotic distribution with probability density function given by \cref{eq:abscoh_dist}, with $\rho=\acoh{p,q}$ and $N=M$, as in the case of segment averaging \citep{goodman1957joint,carter1973statistics}.

The plug-in estimator for the magnitude partial coherence is the sample partial correlation of $J_n(\freq)$ (with known zero mean).
This partial correlation is the sample correlation of the residuals of the linear regressions of $J_p(\freq)$ and $J_q(\freq)$ with $\set{J_r(\freq) \mid r\in [P]\setminus\set{p,q}}$ \citep{whittaker2009graphical}.
If the number of tapers $M$ is small, this approximation may be poor, however, in the spatial setting the number of tapers is typically large and thus this approximation performs well (see the simulations in \suppsection{12}{app:bci:extrafigures}).

\section{Application and simulation study}
\subsection{Barro Colorado Island data}\label{sec:experiment:application}

The Barro Colorado Island study records the locations of every individual tree of at least 10mm in trunk diameter within a 1000 by 500 metre rectangle of tropical rainforest, along with additional measurements, such as the diameter and species name of each individual tree \citep{condit2019complete}.
In addition, they record topological features and the concentration of various soil chemicals, all sampled on regular grids.

Whilst the study records data on a rectangular domain, it is known that the forest is not homogeneous.
In particular, there is a region near the northern border of the plot known to consist of a much younger forest dominated by the pioneer tree species \textit{Gustavia superba} \citep{hubbell1983diversity,hubbell1986commonness}, which we therefore remove from all of our analyses to make homogeneity a more reasonable assumption.
Similarly, we also remove a swamp region in the centre of the plot where the soil has much higher moisture content and the associated flora is more typical of a riparian habitat \citep{harms2001habitat}.

There are around 200 species present in the plot, and many additional covariates.
To fully analyse such data, we would need to develop methodology to handle that high-dimensional setting.
This is beyond the scope of this paper, so for the sake of illustration we will consider a small subset of the available species in our analysis.
In particular, \cite{harms2001habitat} list five species which are known to be associated with the gradient of the terrain.\footnote{Strictly speaking this list is based partially on the BCI dataset, and therefore this selection criteria is to some extent double dipping, though the list is based on an older census.}
We take the three most abundant species from this list, \textit{Beilschmiedia tovarensis}, \textit{Poulsenia armata} and \textit{Unonopsis pittieri}, and consider the gradient of the terrain as the fourth process.

To test the significance of magnitude (partial) coherence we use the distribution in \cref{eq:abscoh_dist} and a Bonferroni correction \citep{bonferroni1936teoria}.
In particular, we correct as if we perform $K_n P(P-1)/2$ tests, where $P=4$ is the number of processes and $K_n$ are the approximate number of wavenumbers on which the multitaper estimate is uncorrelated.
Approximately, the multitaper estimate is uncorrelated when wavenumbers are spaced two bandwidths apart and when we use only wavenumbers where the first coordinate is non-negative (as there is a symmetry), see \suppsection{2.3}{sec:covariance:wavenumbers} for details.
We display more than $K_n$ wavenumbers for visualization purposes, but the threshold is based on $K_n$, because correcting for all displayed wavenumbers would be overly conservative.
It would be preferable to utilise envelope approaches, e.g.\ \cite{mrkvivcka2023false}, but they require resampling from the null distribution, which is difficult when the region shape is irregular, and is beyond the scope of this paper. 

\begin{figure}

\includegraphics{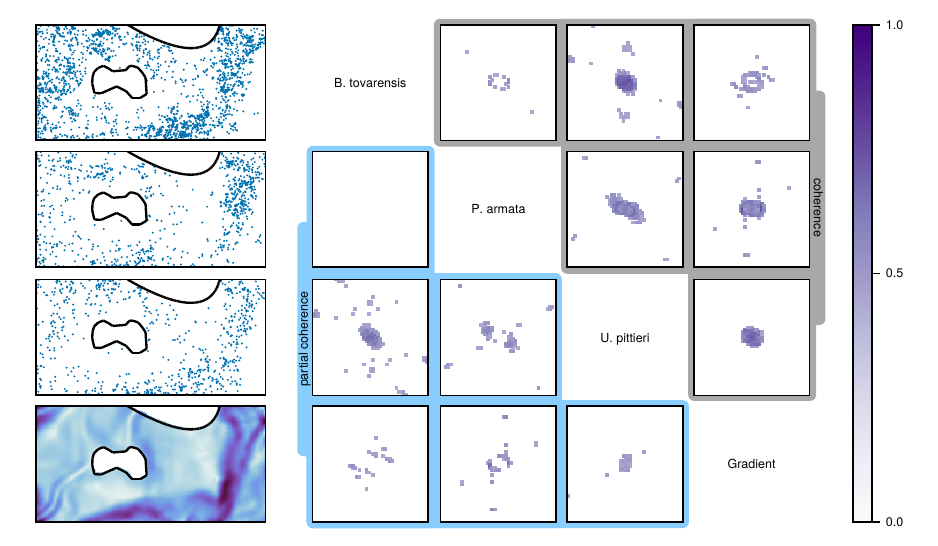}
    \caption{The results of computing the magnitude coherence (upper triangle) and magnitude partial coherence (lower triangle) between \textit{B. tovarensis}, \textit{P. armata}, \textit{U. pittieri} and the gradient of the terrain.
    The data is shown on the left. The spectral plots are shown from -0.05 to 0.05 in each axis, and the data is shown from 0 to 1000 and 0 to 500.}
    \label{fig:bci_results}
\end{figure}

\Cref{fig:bci_results} shows the coherence and partial coherence between each of the four processes from the BCI, alongside the observed data.
We have thresholded the coherence and partial coherence based on the marginal asymptotic null distribution with the correction described above.
As expected, we can see that there is (statistically) significant coherence between each of the three species locations and the gradient (\cref{fig:bci_results}, plot matrix upper-right triangle).
In addition, there is also significant coherence between each of the pairs of species.
When we consider partial coherence, thus correcting to some extent for the other processes, we see that there is still residual coherence between the gradient and each of the species' locations. (\cref{fig:bci_results}, plot matrix lower-left triangle). However, \textit{B. tovarensis} and \textit{P. armata} do not have significant partial coherence whereas \textit{U. pittieri} has significant partial coherence with both other species. From the phase, shown in the \suppsection{12}{app:bci:extrafigures}, we see that all significant associations have a positive sign. Although this example is a preliminary analysis that could change when all other processes and covariates are considered, this example highlights the utility of partial coherence and how it is likely to lead to new inferences compared to existing methods.

\subsection{Simulation study}\label{sec:experiment:simulation}
To investigate the asymptotic properties of the proposed method, we conduct a simulation study, examining the distribution of magnitude coherence and partial coherence between two point processes and a random field, and comparing it to the asymptotic distribution given in \cref{eq:abscoh_dist}.
In the following models, we always have and two point processes (PP1 and PP2) and a random field (RF).
In each case, the random field is a log-Gaussian process, where the Gaussian process has mean -6 and a Mat\'ern covariance function with length scale 20~m, smoothness 3 and variance 1, and is recorded on a grid with spacing every 5~m starting from (0~m, 0~m) as in the case of the gradient data from Barro Colorado Island.
The specific details of the three models are:
\begin{itemize}
    \item Model 1: The two point processes are independent log-Gaussian Cox processes with the Gaussian process described above driving them (independently of the recorded random field).
    \item Model 2: The two point processes are independently Cox processes with the random intensity driven by the field (i.e. they are log-Gaussian Cox processes, see \cite{moller1998log}).
    \item Model 3: The second point process (PP2) is a Cox process with the random intensity driven by the field, and the first point process (PP1) is the result of random clustering around the second point process, where for each point in the first process, there are Poisson(1) offspring, who are placed randomly around the parent with a Normal(0, $5^2$) distribution in each direction.
\end{itemize}

\begin{figure}[h]
    \centering

\includegraphics{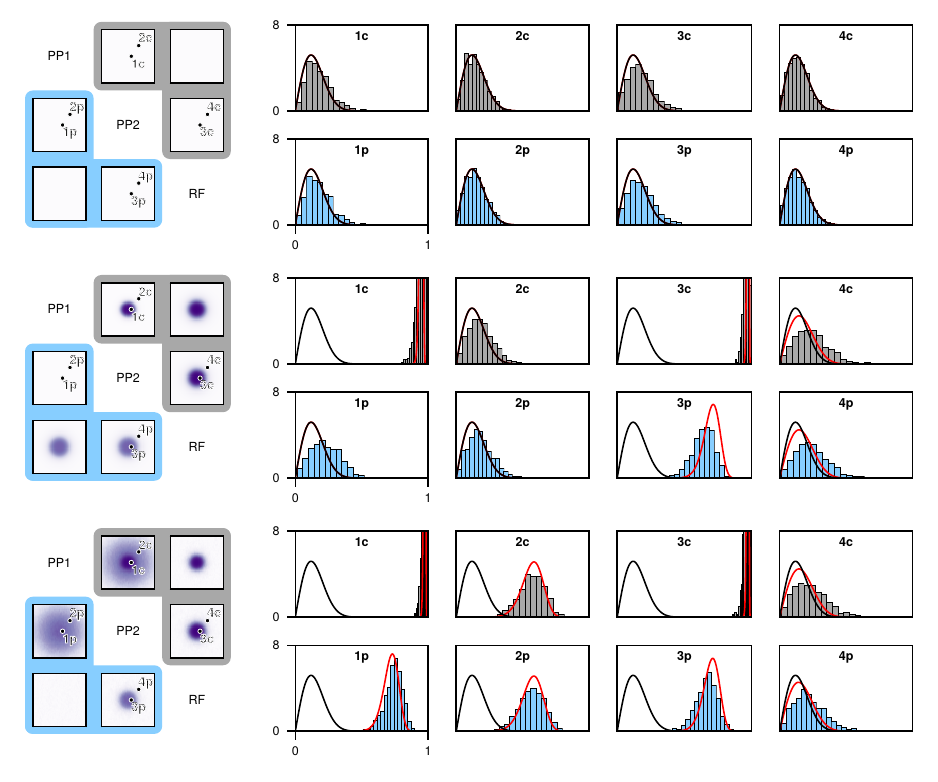}
    \caption{
        Results of the simulation study. 
        The left column shows for each model the true magnitude (partial) coherence in the upper (lower) triangle where both wavenumber components $\freq_1$ and $\freq_2$ range from -0.05 to 0.05.
        The right hand column shows the estimated magnitude (partial) coherence on top (bottom), and asymptotic distribution (red line) and null distribution (black line).
        The rows correspond to models one to three from top to bottom. The result is optimal when the asymptotic (red line) matches the simulation-based histogram.
    }
    \label{fig:simulation_study}
\end{figure}

For each case, we generate 1000 replications, and compute estimates of the spectral density matrix function.
All simulations use the region from the Barro Colorado Island data with the swamp removed.
It is not possible to visualize all of the results, so we show a subset.
In particular, we focus on two wavenumbers which illustrate different areas of the spectrum.
The first is at a low wavenumber on the slope (so we expect smoothing effects to degrade the quality of the approximation), the second is at a high wavenumber, exploring the effects in the tails.
At each of these wavenumbers, we show the empirical and asymptotic distributions of magnitude coherence and magnitude partial coherence between the point processes (processes 1 and 2) and between one of the point processes and the field (processes 2 and 3).
The left hand column of \cref{fig:simulation_study} shows the true spectral matrix for each model, with the magnitude (partial) coherence in the upper (lower) triangle, respectively.
The locations at which we compare the empirical to asymptotic distributions are marked with a point labelled so that it corresponds to the appropriate panel in the right hand column.
For model 1 (the null model), the empirical distribution of the magnitude coherence and magnitude partial coherence align well with the asymptotic distribution (\cref{fig:simulation_study}, top section).
In models 2 and 3, the asymptotic distributions begin to break down in some of the non-null cases (\cref{fig:simulation_study}, middle and bottom sections).
However, we would still detect the effect as being significant when compared to the null asymptotic distribution.
Whilst not perfect, the results are promising and suggest that the asymptotic distribution is a good approximation for the magnitude coherence and magnitude partial coherence in many cases.

\section*{Acknowledgement}
Sofia Olhede would like to thank the European Research Council under Grant CoG 2015-682172NETS, within the Seventh European Union Framework Program.

\setcounter{equation}{0}%
\setcounter{theorem}{0}
\setcounter{lemma}{0}
\setcounter{corollary}{0}
\setcounter{proposition}{0}
\setcounter{definition}{0}
\setcounter{assumption}{0}
\setcounter{remark}{0}
\setcounter{step}{0}
\setcounter{condition}{0}
\setcounter{property}{0}
\setcounter{restrictions}{0}
\setcounter{example}{0}
\setcounter{algo}{0} 
\renewcommand\thetheorem{\text{A}\arabic{theorem}}
\renewcommand\thelemma{\text{A}\arabic{lemma}}
\renewcommand\thecorollary{\text{A}\arabic{corollary}}
\renewcommand\theproposition{\text{A}\arabic{proposition}}
\renewcommand\thedefinition{\text{A}\arabic{definition}}
\renewcommand\theassumption{\text{A}\arabic{assumption}}
\renewcommand\theremark{\text{A}\arabic{remark}}
\renewcommand\thestep{\text{A}\arabic{step}}
\renewcommand\thecondition{\text{A}\arabic{condition}}
\renewcommand\theproperty{\text{A}\arabic{property}}
\renewcommand\therestrictions{\text{A}\arabic{restrictions}}
\renewcommand\theexample{\text{A}\arabic{example}}
\renewcommand\thealgo{\text{A}\arabic{algo}}
\appendix

\label{app:sim:details}
\subsection*{Simulation details}
In the example constructed in \cref{sec:introduction}, each cluster process was generated by placing a cluster at every point in the parent process, where the number of children $N$ and location relative to the parent $X$ were given by $N \sim \mathrm{Poisson}(1),$ $X \sim \mathrm{Normal}(0, 5^2I_2)$ where $I_2$ is the identity matrix.
The intensity for the inhomogeneous Poisson was specified by $\log(\lambda(x)) = 9 Y(x) - 7$ where $Y$ is the gradient of the Barro Colorado Island terrain.
The choice of coefficients was based roughly on fitting a model for the intensity of \textit{P. armata} based on the gradient.

\clearpage
\setcounter{section}{0}
\renewcommand\thesection{S\arabic{section}}
\setcounter{equation}{0}%
\setcounter{theorem}{0}
\setcounter{lemma}{0}
\setcounter{corollary}{0}
\setcounter{proposition}{0}
\setcounter{definition}{0}
\setcounter{assumption}{0}
\setcounter{remark}{0}
\setcounter{step}{0}
\setcounter{condition}{0}
\setcounter{property}{0}
\setcounter{restrictions}{0}
\setcounter{example}{0}
\setcounter{algo}{0} 
\renewcommand\thetheorem{\text{S}\arabic{theorem}}
\renewcommand\thelemma{\text{S}\arabic{lemma}}
\renewcommand\thecorollary{\text{S}\arabic{corollary}}
\renewcommand\theproposition{\text{S}\arabic{proposition}}
\renewcommand\thedefinition{\text{S}\arabic{definition}}
\renewcommand\theassumption{\text{S}\arabic{assumption}}
\renewcommand\theremark{\text{S}\arabic{remark}}
\renewcommand\thestep{\text{S}\arabic{step}}
\renewcommand\thecondition{\text{S}\arabic{condition}}
\renewcommand\theproperty{\text{S}\arabic{property}}
\renewcommand\therestrictions{\text{S}\arabic{restrictions}}
\renewcommand\theexample{\text{S}\arabic{example}}
\renewcommand\thealgo{\text{S}\arabic{algo}}

\thispagestyle{empty}
\vbox{%
\hsize\textwidth
\linewidth\hsize
\vskip 0.1in
\makeatletter\@toptitlebar\makeatother
\centering
{\LARGE\sc Supplementary material for \makeatletter\@title\makeatother\par}
\makeatletter\@bottomtitlebar\makeatother
\vskip 0.1in
}
\renewcommand{\shorttitle}{Supplementary material}


\section{Additional results}
\subsection{Asymptotic normality under $\alpha$-mixing}\label{supp:asymptotic:normality}
To prove asymptotic normality under $\alpha$-mixing, we need an additional assumption on the tapers, and a condition on the mixing of the processes.
\begin{assumption}[Additional taper regularity]\label{assumption:tapers:normality}
    For all $m\in\NN$, $\sup_{n\in\NN}\ell(\region_n) \norm{h_{m,n}}_\infty^2 <\infty,$
    and letting $D_n = \set{x \in \ZZ^d \mid x + [-1/2,1/2]^d \cap \region_n \neq\emptyset}$, $\liminf_{n\rightarrow\infty} \ell(\region_n)\abs{D_n}^{-1} > 0$.
\end{assumption}
The first of these conditions just ensures that the largest value of the taper goes to zero sufficiently fast relative to the area of the observational window.
In the time series case, one usually scales the tapers by the square root of the length of the time series, which satisfies this condition.
The second condition is required because the construction in \cite{biscio2019general} breaks up space into a grid, and the region needs to not become too concentrated around the grid centres.
This condition is also satisfied when we grow a convex region.

To directly utilise the central limit theorem from \cite{biscio2019general}, we need to construct a marked point process from which all statistics can be computed.
In particular define the marked point processes on $\RRd\times\mathcal{M}$ where $\mathcal{M} = [0,\infty)\times[P]$ by $U=\cup_{p\in[P]} U_p$ where $U_{p} = \set{(x,W_p(x),p) \mid x\in X_p}$ if the $p$\Th process is sampled continuously, and $U_p=\set{(x,Y_p(x),p) \mid x\in \grid_p}$ if the $p$\Th process is sampled on a grid.
Now for $A\subset\RRd$ and $B \subset \mathcal{M}$ let $U_{A,B} = U \cap A\times B$. Then the $\alpha$-mixing coefficient from \cite{biscio2019general} is defined as
\begin{align*}
    \alpha^U_{c_1,c_2} = \sup\left\{ \alpha(\sigma(U_{E_1,\mathcal{M}}), \sigma(U_{E_2,\mathcal{M}})) \mid E_1 \subset \RRd, E_2\subset\RRd,\ell(E_1)<c_1,\ell(E_2)<c_2, d(E_1,E_2)>0\right\},
\end{align*}
for $c_1,c_2\geq0$ where $d$ denotes the infimum of a distance between points in the sets and $\alpha(\cdot,\cdot)$ is the mixing coefficient for random variables, see \cite{biscio2019general} for details.

\begin{assumption}[Mixing conditions]\label{assumption:normality}
    Assume $f(\freq)$ is positive definite for all $\freq\in\RRd$.
    Furthermore, there exists $\epsilon>0$ such that
    $
        \sup_{n\in\NN}\alpha^U_{2, \infty}(s) = O(1/s^{d+\epsilon}),
    $
    and there exists $\tau>2d/\epsilon$ such that 
    $$\EE{\left(\sum_{p=1}^P \abs{Z_p}^2\right)^{1+\tau/2}}  < \infty$$
    where
    \begin{align*}
        Z_p &= \begin{cases}
            \randmeasurebase_p([0,1]^d) & \text{if }\randmeasure{p}\text{ corresponds to a point process,}\\
            Y_p(0) & \text{if }\randmeasure{p}\text{ corresponds to a random field.}\\
        \end{cases}
    \end{align*}
\end{assumption}

\begin{theorem}[Asymptotic normality under $\alpha$-mixing]\label{theorem:normality_dft:alpha}
    We can replace the taper \cref{assumption:tapers:consistency} with \cref{assumption:tapers:normality} and the process \cref{assumption:mixing:brillinger} with \cref{assumption:normality} in \cref{theorem:normality_dft}.
    Then the result of \cref{theorem:normality_dft} continues to hold.
    \begin{proof}
        See \cref{proof:theorem:normality_dft:alpha}.
    \end{proof}
\end{theorem}

\subsection{Growing a template region}\label{sec:estimation:template}
One common way to construct a sequence of regions is to start with some bounded convex template region $\region$, and then grow it by scaling all of the dimensions by some factor $l_n\in\RR_{>0}^d$, so that $\region_n = l_n\circ\region$.
In this case, we can be more specific about the construction of the tapers.
Note the convexity assumption is required so that we satisfy the previous nesting assumption that $\region_n\subset \region_{n+1}$.
We will also assume without loss of generality that the template region contains zero for the same reason (one can always shift it to zero and shift it back again).

\begin{proposition}[Growing a template region]\label{res:general:rescale}
    Let $\region$ be some bounded convex region in $\RRd$ which contains zero.
    Let $\set{l_n}_{n\in\NN}$ be a sequence taking elements in $\RR_{>0}^d$ such that $l_n$ is elementwise strictly increasing and tend to infinity as $n\rightarrow\infty$.
    Say we have a family of tapers $\set{h_m}_{m\in\NN}$ on $\region$ that are bounded, continuous and orthonormal.
    Furthermore, for all $m\in\NN$ there exists $\delta>0$ and $C_m>0$ such that for all $x\in\RRd$, 
    $
        \abs{h_m(x)} + \abs{H_m(x)} \leq {C_m}{(1+\norm{x}_2)^{-d-\delta}}.
    $
     These constants scale so $\max_{m\in [M_n]} C_m \elemprod{l_n}^{1/2}  (\min l_n)^{-d-\delta} \rightarrow 0$ and $\max_{m\in [M_n]} C_m \elemprod{l_n}^{-1/2} \rightarrow 0$ as $n\rightarrow\infty$.
    Choose the bandwidth $b_n$ so that $b_n\rightarrow 0$ and
    \begin{align}
        \min_{m\in[M_n]} \int_{\ball{b_n\circ l_n}} \abs{H_m(\freq)}^2 \de\freq &\rightarrow 1,
        \label{eq:growingdomain:concentration}
    \end{align}
    as $n\rightarrow \infty$.
    Then the family of tapers on $\region_n$ given by $h_{m,n}(x) = \elemprod{l_n}^{-1/2} h_m(x\oslash l_n)$ satisfies \cref{assumption:tapers:finitesample,assumption:tapers:concentration,assumption:tapers:orthogonality,assumption:tapers:consistency} and \cref{assumption:tapers:normality}.

    \begin{proof}
        See \cref{proof:res:general:rescale}.
    \end{proof}
\end{proposition}
Here we take $\min l_n$ to mean the smallest element of the vector $l_n$ for a given $n$.
When $M_n=M$ is fixed, then $b_n \min l_n \rightarrow \infty$ as $n\rightarrow\infty$ is sufficient for \cref{eq:growingdomain:concentration}, but \cref{eq:growingdomain:concentration} is slightly stronger when $M_n$ is allowed to grow, which is required for consistency.
In this case, the ordering of the tapers then matters, as we want to introduce new tapers sufficiently slowly relative to the growth of their concentration.
This is discussed explicitly in \cref{res:rect:minbias}.
The condition on the decay of the tapers is required to use aliasing results and ensure that the errors when growing the number of tapers are controlled.
\cref{res:general:rescale} is however reassuring, in that the natural extension of multitapering to a growing region when one does have a reasonable family of tapers will satisfy our various assumptions.

For completeness, we can also consider the case where the template region is rectangular.
In this case, we can construct a family of tapers on this region by taking the outer product of one-dimensional tapers.
This will also give an existence proof for a family of tapers satisfying the conditions of \cref{res:general:rescale}, as we can simply make a family on a rectangle contained in $\region$ (which is possible due to convexity), and then set the family to be zero on the rest of the region.
This is of course not the most efficient way to construct tapers, and is intended only to show that such a family exists. 

\begin{proposition}[Growing a rectangular region]\label{res:rect:outerprod}
    Let $\region=\prod_{j=1}^d [a_j,b_j]$, where $a_j<0<b_j$.
    Say $\set{g_{m}}_{m\in\NN}$ is an orthonormal family tapers supported on a subset of [0,1] that are bounded and continuous.
    Let $\gamma$ be a bijection from $\NN$ to $\NN^d$ so that $\gamma(m)$ maps from a taper index $m$ to a $d$-dimensional vector of indices from which we select the one-dimensional tapers.
    Writing $M_{n,j} = \max_{m\in[M_n]} \gamma(m)_j$ and $l_{n,j}$ for the $j$\Th element of $l_n$, assume that
    \begin{align*}
        \min_{m\in [{M}_{n,j}]}\int_{-b_n l_{n,j}/\sqrt{d}}^{b_n l_{n,j}/\sqrt{d}} \abs{g_{m}(x)}^2 \de x &\rightarrow 1.
    \end{align*}
    Furthermore, for all $m\in\NN$ there exists $\delta>0$ and $C_m>0$ such that for all $x\in\RRd$, 
    $
        \abs{g_m(x)} + \abs{G_m(x)} \leq {C_m}{(1+\abs{x})^{-1-\delta}},
    $
    where $\max_{m\in [{M}_{n,j}]} C_m l_{n,j}^{-1/2} \rightarrow 0$ as $n\rightarrow\infty$.

    Now define
    \begin{align*}
        h_{m}(x) = \prod_{j=1}^d \frac{1}{\sqrt{b_j-a_j}}g_{\gamma(m)_{j}}\left(\tfrac{x_j-a_j}{b_j-a_j}\right)
    \end{align*}
    Then using the family of tapers $\set{h_m}_{m\in\NN}$ in the construction in \cref{res:general:rescale} satisfies \cref{assumption:tapers:finitesample,assumption:tapers:concentration,assumption:tapers:orthogonality,assumption:tapers:consistency} and \cref{assumption:tapers:normality}.

    \begin{proof}
        See \cref{proof:res:rect:outerprod}.
    \end{proof}
\end{proposition}
Note here that the choice of bijection $\gamma$ is not arbitrary, as it determines how quickly ${M}_{n,j}$ grows in each dimension.
Typically, one would use $M = \prod_{j=1}^d M_j$ tapers in practice, using the first $M_j$ tapers in each dimension.
In this separable case, we have a slight relaxation of the condition on the decay of the tapers and their Fourier transforms, as we only require that they decay at a rate of $1+\delta$ in each dimension.
The factor of $1/\sqrt{d}$ modifying the bandwidth is because the previous assumptions were made for concentration on a ball, but now it is easier to work with a box when a separable structure is present. Since $d$ is fixed and $b_n$ is chosen this does not matter.

Finally, we need to check that a univariate family satisfying the required conditions exists.
The minimum bias tapers proposed by \cite{riedel1995minimum}, and used the point process setting by \cite{rajala2023what}, satisfy these conditions.
\begin{proposition}[Minimum bias tapers]\label{res:rect:minbias}
    The minimum bias tapers, i.e. given for all $m\in\NN$ by $g_m(x) = \sqrt{2} \sin(\pi m x) \indicator_{[0,1]}(x),$ $x \in \RR$, satisfy the conditions of \cref{res:rect:outerprod} if for all $n\in\NN$, $b_n {l_{n,j}} \geq M_{n,j}\sqrt{d}$ and there exists some $\delta>0$ such that $M_{n,j}^{1+\delta} l_{n,j}^{-1/2} \rightarrow 0$ as $n\rightarrow\infty$.

    \begin{proof}
        See \cref{proof:res:rect:minbias}.
    \end{proof}
\end{proposition}
The assumption $b_n {l_{n,j}} \geq M_{n,j} \sqrt{d}$ is the usual choice of bandwidth versus number of tapers for the minimum bias tapers \citep{riedel1995minimum}, provided that we view $b_n/\sqrt{d}$ as the bandwidth.
The assumption $M_{n,j}^{1+\delta} l_{n,j}^{-1/2} \rightarrow 0$ is required to ensure that the tapers satisfy \cref{assumption:tapers:consistency}, which in turn is needed for the consistency result in the non-Gaussian setting.
All of these assumptions can be easily satisfied as we are free to choose the bandwidth $b_n$ and the largest taper index in each dimension $M_{n,j}$ (as we choose $M_n$ and $\gamma$).

\subsection{Covariance across wavenumbers}\label{sec:covariance:wavenumbers}
One useful additional result is that fixed $M_n=M$, we have a covariance given by
\begin{align*}
    \Cov\left(\mtpgram{p,q;n}(\freq), \mtpgram{p',q';n}(\freq')\right) 
    = o(1) + \sum_{m=1}^{M_n} \sum_{m'=1}^{M_n}\Bigg\{& \int_{K_{p,p'}} H_{p;m,n}(\freq-\freq'') \conj{H_{p';m',n}(\freq'-\freq'')} \asdf{p,p'}(\freq'')\de\freq''\\
    &\;\;\times \int_{K_{q,q'}} \conj{H_{q;m,n}(\freq'-\freq'')} H_{q';m',n}(\freq'-\freq'') \conj{\asdf{q,q'}(\freq'')}\de\freq''\\ 
    &+ \int_{K_{p,q'}} H_{p;m,n}(\freq-\freq'') H_{q';m',n}(\freq'-\freq'') \asdf{p,q'}(\freq'')\de\freq' \\
    &\;\;\times \int_{K_{p',q}} \conj{H_{q;m,n}(\freq-\freq'') H_{q';m',n}(\freq'-\freq'')} \asdf{q,q'}(\freq'')\de\freq''\Bigg\}.
\end{align*}
where the $o(1)$ term comes from the fourth order cumulants, and is negligible as $n\rightarrow\infty$.
Importantly, provided that the wavenumbers are not too close to each other (modulo some details related to aliasing), this covariance is negligible.
In particular, we should space wavenumbers such that the distance between them is at least $2b_n$, so that the tapers do not overlap in wavenumber space.

\section{Proofs of main results}

\subsection{\texorpdfstring{\cref{res:bias:finite}}{Proposition}}\label{proof:res:bias:finite}
Recall the definitions of $\Aliasfreq_p$ and $w_p$ given in \cref{def:aliased_sdf}.
\begin{lemma}\label{lemma:alias:periodic}
    For all $\freq \in \RRd$, for all $\aliasfreq\in\Aliasfreq_p$,
    \begin{align*}
        H_{p;m,n}(\freq+\aliasfreq) &= H_{p;m,n}(\freq) w_p(\aliasfreq),
    \end{align*}
    and for all $\aliasfreq\in\Aliasfreq_p\cap\Aliasfreq_q$
    \begin{align*}
        H_{p;m,n}(\freq+\aliasfreq)\conj{H_{q;m,n}(\freq+\aliasfreq)}
        &= H_{p;m,n}(\freq)\conj{H_{q;m,n}(\freq)}w_p(\aliasfreq)\conj{w_q(\aliasfreq)}.
    \end{align*}
    \begin{proof}
        For a process sampled on a grid, then for some $\aliasfreq\in\Aliasfreq_p$, recall $\aliasfreq = z'\oslash \Delta_p$ for some $z'\in\ZZ^d$. Then
        \begin{align*}
            H_{p;m,n}(\freq+\aliasfreq) &= \elemprod{\Delta_p} \sum_{\loc\in\grid_p} h_{p;m,n}(\loc)e^{-2\pi i \ip{(\freq+\aliasfreq)}{\loc}} \\
            &= \elemprod{\Delta_p} \sum_{z\in\ZZ^d} h_{p;m,n}(z\circ\Delta_p+\gridoffset_p)e^{-2\pi i \ip{\freq}{(z\circ\Delta_p+\gridoffset_p)}} e^{-2\pi i \ip{\aliasfreq}{z\circ\Delta_p}} e^{-2\pi i \ip{\aliasfreq}{\gridoffset_p}} \\
            &= \elemprod{\Delta_p} \sum_{z\in\ZZ^d} h_{p;m,n}(z\circ\Delta_p+\gridoffset_p)e^{-2\pi i \ip{\freq}{(z\circ\Delta_p+\gridoffset_p)}} e^{-2\pi i \ip{\aliasfreq}{z\circ\Delta_p}} e^{-2\pi i \ip{\aliasfreq}{\gridoffset_p}} \\
            &= H_{p;m,n}(\freq) w_p(\aliasfreq)
        \end{align*}
        where the last line holds because and $w_p(\aliasfreq) = e^{-2\pi i \ip{\aliasfreq}{\gridoffset_p}}$.
        In particular, this means we have $\ip{\aliasfreq}{z\circ\Delta_p} = \ip{z'}{z} \in \ZZ$ and hence $e^{-2\pi i \ip{\aliasfreq}{z\circ\Delta_p}} = 1$.
        If one of the processes is sampled continuously, then $\Aliasfreq_p\cap\Aliasfreq_q = \set{0}$ and $w_p(0) = w_q(0) = 1$, so the second result is trivially true.
        If both processes are sampled on a grid then as $\Aliasfreq_p\cap\Aliasfreq_q$ is contained in $\Aliasfreq_p$ and $\Aliasfreq_q$ the result continues to hold.
    \end{proof}
\end{lemma}

We will prove a more general result than \cref{res:bias:finite} which will also be useful for quantifying covariance over wavenumbers and tapers in \cref{theorem:normality_dft}.
\begin{lemma}\label{lemma:mean:dftcross}
  Given \cref{assumption:spectra:exists,assumption:tapers:finitesample},
  \begin{align}
    \cov{\dft{p}{m,n}(\freq_1), {\dft{q}{m,n}(\freq_2)}} &= \int_{K_{p,q}} H_{p;m,n}(\freq') \conj{H_{q;m',n}(\freq' - \freq_2 + \freq_1)} \sdf{p,q}(\freq_1-\freq')\de\freq'
    \end{align}
    for any $\freq_1,\freq_2\in\RRd$.
  \begin{proof}
    \cref{assumption:tapers:finitesample} implies that we may use \cref{res:cov_process_freq}, setting $\phi_p(\loc) = h_{p;m}(\loc)e^{-2\pi i \ip{\loc}{\freq_1}}$ and $\phi_q(\loc) = h_{q;m'}(\loc)e^{-2\pi i \ip{\loc}{\freq_2}}$. Note
    \begin{align*}
      \Phi_{p}(\freq) &= \int_\RRd h_{p;m}(\loc)e^{-2\pi i \ip{\loc}{\freq_1}} e^{-2\pi i \ip{\loc}{\freq}} \de \freq \\
      &= H_{p;m}(\freq_1+\freq). \\
    \end{align*}

    From \cref{res:cov_process_freq} we have
    \begin{align*}
        \cov{\dft{p}{m,n}(\freq_1), {\dft{q}{m,n}(\freq_2)}} 
        &= \int_\RRd H_{p;m,n}(\freq_1-\freq') \conj{H_{q;m',n}(\freq_2-\freq')} \sdf{p,q}(\freq')\de\freq' \\
        &= \int_\RRd H_{p;m,n}(\freq'') \conj{H_{q;m',n}(\freq''-\freq_1+\freq_2)} \sdf{p,q}(\freq_1-\freq'')\de\freq''.
    \end{align*}
    Now breaking this up
    \begin{align*}
        &\cov{\dft{p}{m,n}(\freq_1), {\dft{q}{m,n}(\freq_2)}} \\
        &= \int_\RRd H_{p;m,n}(\freq') \conj{H_{q;m',n}(\freq'-\freq_1+\freq_2)} \sdf{p,q}(\freq_1-\freq')\de\freq' \\
        &= \sum_{\aliasfreq\in\Aliasfreq_{p}\cap\Aliasfreq_q} \int_{K_{p,q}+\aliasfreq} H_{p;m,n}(\freq') \conj{H_{q;m',n}(\freq'-\freq_1+\freq_2)} \sdf{p,q}(\freq_1-\freq')\de\freq' \\
        &= \sum_{\aliasfreq\in\Aliasfreq_{p}\cap\Aliasfreq_q} \int_{K_{p,q}} H_{p;m,n}(\freq'+\aliasfreq) \conj{H_{q;m',n}(\freq'-\freq_1+\freq_2+\aliasfreq)} \sdf{p,q}(\freq_1-\freq'+\aliasfreq)\de\freq' \\
        &= \sum_{\aliasfreq\in\Aliasfreq_{p}\cap\Aliasfreq_q} \int_{K_{p,q}} H_{p;m,n}(\freq') \conj{H_{q;m',n}(\freq'-\freq_1+\freq_2)} \sdf{p,q}(\freq_1-\freq'+\aliasfreq)w_p(\aliasfreq)\conj{w_q(\aliasfreq)}\de\freq'
    \end{align*}
    from the periodicity result in \cref{lemma:alias:periodic}.
    Finally we have that
    \begin{align*}
        &\int_{K_{p,q}}\sum_{\aliasfreq\in\Aliasfreq_{p}\cap\Aliasfreq_q}  \abs{H_{p;m,n}(\freq') \conj{H_{q;m',n}(\freq'-\freq_1+\freq_2)} \sdf{p,q}(\freq_1-\freq')}\de\freq' \\
        &= \int_{K_{p,q}}  \abs{H_{p;m,n}(\freq') \conj{H_{q;m',n}(\freq'-\freq_1+\freq_2)}}\sum_{\aliasfreq\in\Aliasfreq_{p}\cap\Aliasfreq_q}\abs{\sdf{p,q}(\freq_1-\freq'+\aliasfreq)}\de\freq' \\
        &< \infty
    \end{align*}
    by \cref{lemma:alias:boundedcontinuous} and \cref{assumption:tapers:finitesample}.
    Therefore, we may interchange the sum and integral by Fubini's theorem to obtain
    \begin{align*}
        &\cov{\dft{p}{m,n}(\freq_1), {\dft{q}{m,n}(\freq_2)}} \\
        &= \int_{K_{p,q}} H_{p;m,n}(\freq') \conj{H_{q;m',n}(\freq'-\freq_1+\freq_2)} \sum_{\aliasfreq\in\Aliasfreq_{p}\cap\Aliasfreq_q} \sdf{p,q}(\freq_1-\freq'+\aliasfreq)w_p(\aliasfreq)\conj{w_q(\aliasfreq)}\de\freq' \\
        &= \int_{K_{p,q}} H_{p;m,n}(\freq') \conj{H_{q;m',n}(\freq'-\freq_1+\freq_2)} \asdf{p,q}(\freq_1-\freq') \de\freq'.
    \end{align*}
    as required.
  \end{proof}
\end{lemma}

\begin{proof}[Proof of \cref{res:bias:finite}]
  Follows from \cref{lemma:mean:dftcross}, by setting $k_1=k_2=k$ and $m'=m$.
\end{proof}

\subsection{\texorpdfstring{\cref{res:bias:asymp}}{Theorem}}\label{proof:res:bias:asymp}

In the process of proving \cref{res:bias:asymp} we will prove a stronger proposition which will also be useful for the correlation between the tapered Fourier transforms in \cref{theorem:normality_dft}, and for the consistency results when growing $M_n$.

\begin{proposition}\label{res:strong:bias:asymp}
    Given that the process satisfies \cref{assumption:spectra:exists}, and the tapers satisfy \cref{assumption:tapers:finitesample,assumption:tapers:concentration} we have for any $\freq_1,\freq_2\in\RRd$
    \begin{align*}
        \max_{m,m'\in[M_n]}\abs{\cov{\dft{p}{m,n}(\freq_1), \dft{q}{m',n}(\freq_2)} - Q_{p,q}(m,m',\freq_1,\freq_2)} &\rightarrow 0
    \end{align*}
    as $n\rightarrow\infty$, where
    \begin{align*}
        Q_{p,q}(m,m',\freq_1,\freq_2) &= 
        \begin{cases}
            \asdf{p,q}(\freq_1)\conj{w_{q}(\freq_2-\freq_1)} & \text{if } m=m' \text{ and }\freq_1-\freq_2 \in \Aliasfreq_{q} \\
            \asdf{p,q}(\freq_2)w_{p}(\freq_1-\freq_2) & \text{if } m=m' \text{ and }\freq_1-\freq_2 \in \Aliasfreq_{p}\setminus\Aliasfreq_{q} \\
            0 & \text{otherwise.}
        \end{cases}
    \end{align*}
\begin{proof}    
    Fix some $\freq_1,\freq_2\in\RRd$ and fix some $n\in\NN$ and $m,m'\in[M_n]$.
    We will establish bounds to show the desired convergence.
    We can without loss of generality assume that $\freq_1-\freq_2\notin \Aliasfreq_p \setminus \Aliasfreq_q$, as the case $\freq_1-\freq_2 \in \Aliasfreq_p \setminus \Aliasfreq_q$ can be recovered from conjugate symmetry, because we have
    \begin{align*}
        \cov{\dft{q}{m,n}(\freq_1), \dft{p}{m,n}(\freq_2)} &= \conj{\cov{\dft{p}{m,n}(\freq_2), \dft{q}{m,n}(\freq_1)}} \\
        Q_{p,q}(m,m,\freq_1,\freq_2) &= \conj{Q_{q,p}(m,m,\freq_2,\freq_1)}.
    \end{align*}

    So without loss of generality assume that $\freq_1-\freq_2\notin \Aliasfreq_p$.
    From \cref{lemma:mean:dftcross}
    \begin{align*}
        \cov{\dft{p}{m,n}(\freq_1), \dft{q}{m',n}(\freq_2)}
        &= \int_{K_{p,q}} H_{p;m,n}(\freq') \conj{H_{q;m',n}(\freq'-\freq_1+\freq_2)} \asdf{p,q}(\freq_1-\freq') \de\freq'\\
        &= \int_{\ball{b_n}} H_{p;m,n}(\freq') \conj{H_{q;m',n}(\freq'-\freq_1+\freq_2)} [\asdf{p,q}(\freq_1-\freq') - \asdf{p,q}(\freq_1)]\de\freq' \\
        & \;\; + \int_{K_{p,q}\setminus \ball{b_n}} H_{p;m,n}(\freq') \conj{H_{q;m',n}(\freq'-\freq_1+\freq_2)} \asdf{p,q}(\freq_1-\freq')\de\freq' \\
        & \;\; + \asdf{p,q}(\freq_1)\int_{\ball{b_n}} H_{p;m,n}(\freq') \conj{H_{q;m',n}(\freq'-\freq_1+\freq_2)} \de\freq'.
    \end{align*}
    To consider each of these in turn, define
    \begin{align*}
        T_{p,q;n}^{(1)}(m,m',\freq_1,\freq_2) &= \int_{\ball{b_n}} H_{p;m,n}(\freq') \conj{H_{q;m',n}(\freq'-\freq_1+\freq_2)} [\asdf{p,q}(\freq_1-\freq') - \asdf{p,q}(\freq_1)]\de\freq'  \\
        T_{p,q;n}^{(2)}(m,m',\freq_1,\freq_2) & = \int_{K_{p,q}\setminus \ball{b_n}} H_{p;m,n}(\freq') \conj{H_{q;m',n}(\freq'-\freq_1+\freq_2)} \asdf{p,q}(\freq_1-\freq')\de\freq' \\
        T_{p,q;n}^{(3)}(m,m',\freq_1,\freq_2) &= \asdf{p,q}(\freq)\int_{\ball{b_n}} H_{p;m,n}(\freq') \conj{H_{q;m',n}(\freq'-\freq_1+\freq_2)} \de\freq' - Q_{p,q}(m,m',\freq_1,\freq_2)
    \end{align*}
    then to complete the proof we need only show that
    \begin{align*}
        &\max_{m,m'\in[M_n]}\abs{T_{p,q;n}^{(1)}(m,m',\freq_1,\freq_2)} \rightarrow 0 \\
        &\max_{m,m'\in[M_n]}\abs{T_{p,q;n}^{(2)}(m,m',\freq_1,\freq_2)} \rightarrow 0 \\
        &\max_{m,m'\in[M_n]}\abs{T_{p,q;n}^{(3)}(m,m',\freq_1,\freq_2)} \rightarrow 0
    \end{align*}
    as $n\rightarrow\infty$.

    \textbf{The $T^{(1)}$ term.}

    Firstly
    \begin{align*}
        &\hspace{-2em} \abs{T_{p,q;n}^{(1)}(m,m',\freq_1,\freq_2)}\\
        &\leq \int_{\ball{b_n}} \abs{H_{p;m,n}(\freq') \conj{H_{q;m',n}(\freq'-\freq_1+\freq_2)}} \abs{\asdf{p,q}(\freq_1-\freq') - \asdf{p,q}(\freq_1)} \de\freq' \\
        &\leq \int_{\ball{b_n}} \abs{H_{p;m,n}(\freq') \conj{H_{q;m',n}(\freq'-\freq_1+\freq_2)}} \de\freq' \sup_{\freq''\in\ball{b_n}}\abs{\asdf{p,q}(\freq_1-\freq'') - \asdf{p,q}(\freq_1)}
    \end{align*}
    and furthermore
    \begin{align*}
        &\hspace{-2em} \int_{\ball{b_n}} \abs{H_{p;m,n}(\freq') \conj{H_{q;m',n}(\freq'-\freq_1+\freq_2)}} \de\freq' \\
        &\leq \left(\int_{\ball{b_n}} \abs{H_{p;m,n}(\freq')}^2 \de\freq' \int_{\ball{b_n}} \abs{H_{q;m',n}(\freq'-\freq_1+\freq_2)}^2 \de\freq'\right)^{1/2} \\
        &\leq \left(\int_{K_p} \abs{H_{p;m,n}(\freq')}^2 \de\freq' \int_{K_q} \abs{H_{q;m',n}(\freq')}^2 \de\freq'\right)^{1/2}.
    \end{align*}
    In addition, by the triangle inequality
    \begin{align*}
        & \hspace{-2em} \max_{m\in[M_n]} \int_{K_p} \abs{H_{p;m,n}(\freq')}^2 \de\freq' \\
        &\leq 1 + \max_{m\in[M_n]} \abs{\int_{K_p} \abs{H_{p;m,n}(\freq')}^2 \de\freq'  -1} \\
        & \leq 1 + \max_{m\in[M_n]} \abs{\int_{B_n} \abs{H_{p;m,n}(\freq')}^2 \de\freq'  -1} + \max_{m\in[M_n]} \abs{\int_{K_p\setminus B_n} \abs{H_{p;m,n}(\freq')}^2 \de\freq' }\\
        &= 1 + o(1)
    \end{align*}
    as $n\rightarrow \infty$ by \cref{assumption:tapers:concentration}.
    Putting this together we have
    \begin{align*}
        \max_{m\in[M_n]}{\abs{T_{p,q;n}^{(1)}(m,m',\freq_1,\freq_2)}}
        &\leq \sup_{\freq''\in\ball{b_n}}\abs{\asdf{p,q}(\freq_1-\freq'') - \asdf{p,q}(\freq_1)} ( 1+ o(1)) \rightarrow 0
    \end{align*}
    as $n\rightarrow \infty$ by continuity of $\asdf{p,q}$ from \cref{lemma:alias:boundedcontinuous} and the assumption that $b_n\rightarrow 0$ as $n\rightarrow\infty$.

    \textbf{The $T^{(3)}$ term.}

    We deal with the third term next, as we will need to use the result for processes sampled on grids to prove the second term goes to zero when only one process is sampled on a grid.
    Recall that
    \begin{align*}
        T_{p,q;n}^{(3)}(m,m',\freq_1,\freq_2) &= \asdf{p,q}(\freq_1)\int_{\ball{b_n}} H_{p;m,n}(\freq') \conj{H_{q;m',n}(\freq'-\freq_1+\freq_2)} \de\freq' - Q_{p,q}(m,m',\freq_1,\freq_2).
    \end{align*}
    Firstly, if $\freq_1-\freq_2\notin \Aliasfreq_q$, then we note $Q_{p,q}(m,m',\freq_1,\freq_2)=0$ and
    \begin{align*}
        &\hspace{-2em}\max_{m,m'\in[M_n]}\abs{T_{p,q;n}^{(3)}(m,m',\freq_1,\freq_2)} \\
        &\leq \norm{\asdf{p,q}}_\infty \max_{m,m'\in[M_n]}\abs{\int_{\ball{b_n}} H_{p;m,n}(\freq') \conj{H_{q;m',n}(\freq'-\freq_1+\freq_2)} \de\freq'} \\
        &\leq \norm{\asdf{p,q}}_\infty \left(\max_{m\in[M_n]}\int_{\ball{b_n}} \abs{H_{p;m,n}(\freq')}^2 \de\freq' \max_{m'\in[M_n]}\int_{\ball{b_n}} \abs{H_{q;m',n}(\freq'-\freq_1+\freq_2)}^2 \de\freq'\right)^{1/2}.
    \end{align*}
    Firstly
    \begin{align*}
        \max_{m\in[M_n]}\int_{\ball{b_n}} \abs{H_{p;m,n}(\freq')}^2 \de\freq' \rightarrow 1
    \end{align*}
    by \cref{assumption:tapers:concentration} and the triangle inequality.
    Secondly, take $\aliasfreq\in\Aliasfreq_q$ such that $\freq_1-\freq_2+\aliasfreq\in K_q$.
    Since $\freq_1-\freq_2\notin \Aliasfreq_q$, we have that $\ball{b_n}+\freq_1-\freq_2+\aliasfreq\subseteq K_q \setminus B_n$ for sufficiently large $n$.
    Therefore, by periodicity of $H_{q;m;,n}$
    \begin{align*}
        \max_{m'\in[M_n]}\int_{\ball{b_n}} \abs{H_{q;m',n}(\freq'-\freq_1+\freq_2)}^2 \de\freq' 
        &= \max_{m'\in[M_n]}\int_{\ball{b_n}} \abs{H_{q;m',n}(\freq'-\freq_1+\freq_2+\aliasfreq)}^2 \de\freq' \\
        &= \max_{m'\in[M_n]}\int_{\ball{b_n}-\freq_1+\freq_2+\aliasfreq} \abs{H_{q;m',n}(\freq'')}^2 \de\freq''\\
        &\leq \max_{m'\in[M_n]}\int_{K_q\setminus B_n} \abs{H_{q;m',n}(\freq'')}^2 \de\freq'' \\
        &\rightarrow 0
    \end{align*}
    as $n\rightarrow\infty$ by \cref{assumption:tapers:concentration}.
    Therefore in this case $\max_{m,m'\in[M_n]}\abs{T_{p,q;n}^{(3)}(m,m',\freq_1,\freq_2)} \rightarrow 0$ as $n\rightarrow\infty$.

    Now consider the case where $\freq_1-\freq_2\in \Aliasfreq_q$. In this case from \cref{lemma:alias:periodic}
    \begin{align*}
        T_{p,q;n}^{(3)}(m,m',\freq_1,\freq_2) &= \asdf{p,q}(\freq_1)\int_{\ball{b_n}} H_{p;m,n}(\freq') \conj{H_{q;m',n}(\freq')w_q(-\freq_1+\freq_2)} \de\freq' - Q_{p,q}(m,m',\freq_1,\freq_2).
    \end{align*}
    Now further differentiating the cases based on whether $m=m'$ or $m\neq m'$, we have
    \begin{align*}
        \max_{m,m'\in[M_n]} \abs{T_{p,q;n}^{(3)}(m,m',\freq_1,\freq_2)}
        &\leq \max_{m\in[M_n]} \abs{T_{p,q;n}^{(3)}(m,m,\freq_1,\freq_2)} + \max_{m\in[M_n]}\max_{m'\in[M_n]\setminus\set{m}} \abs{T_{p,q;n}^{(3)}(m,m',\freq_1,\freq_2)} 
    \end{align*}
    Consider the second term first.
    In this case, since $m\neq m'$, $Q_{p,q}(m,m',\freq_1,\freq_2)=0$.
    Therefore, since for sufficiently large $n$, $\ball{b_n}\subseteq K_p\cap K_q$
    \begin{align*}
        \abs{T_{p,q;n}^{(3)}(m,m',\freq_1,\freq_2)}
        &\leq \norm{\asdf{p,q}}_\infty \abs{\int_{\ball{b_n}} H_{p;m,n}(\freq') \conj{H_{q;m',n}(\freq')} \de\freq'} \rightarrow 0
    \end{align*}
    as $n\rightarrow 0$ by \cref{assumption:tapers:orthogonality}, since $m\neq m'$.

    Finally, we have the last term, when $\freq_1-\freq_2\in \Aliasfreq_p$ and $m=m'$.
    In this case,
    \begin{align*}
        T_{p,q;n}^{(3)}(m,m',\freq_1,\freq_2)
        &= \asdf{p,q}(\freq_1)\int_{\ball{b_n}} H_{p;m,n}(\freq') \conj{H_{q;m',n}(\freq')w_q(-\freq_1+\freq_2)} \de\freq' - \asdf{p,q}(\freq_1)\conj{w_q(\freq_2-\freq_1)} \\
        &= \asdf{p,q}(\freq_1)\conj{w_q(\freq_2-\freq_1)}\left(\int_{\ball{b_n}} H_{p;m,n}(\freq') \conj{H_{q;m',n}(\freq')} \de\freq' - 1\right).
    \end{align*}
    Therefore
    \begin{align*}
        \max_{m\in[M_n]}\abs{T_{p,q;n}^{(3)}(m,m',\freq_1,\freq_2)}
        &\leq \norm{\asdf{p,q}}_\infty \max_{m\in[M_n]}\abs{\int_{\ball{b_n}} H_{p;m,n}(\freq') \conj{H_{q;m,n}(\freq')} \de\freq' - 1} \rightarrow 0
    \end{align*}
    as $n\rightarrow\infty$ by \cref{assumption:tapers:concentration}.
    Therefore, we have shown that
    \begin{align*}
        \max_{m,m'\in[M_n]}\abs{T_{p,q;n}^{(3)}(m,m',\freq_1,\freq_2)} \rightarrow 0
    \end{align*}
    as $n\rightarrow\infty$.

    \textbf{The $T^{(2)}$ term.}

    Now for the second term, recall
    \begin{align*}
        T_{p,q;n}^{(2)}(m,m',\freq_1,\freq_2) 
        &= \int_{K_{p,q}\setminus \ball{b_n}} H_{p;m,n}(\freq') \conj{H_{q;m',n}(\freq'-\freq_1+\freq_2)} \asdf{p,q}(\freq_1-\freq')\de\freq'.
    \end{align*}
    Therefore
    \begin{align*}
        \abs{T_{p,q;n}^{(2)}(m,m',\freq_1,\freq_2)}
        &\leq \int_{K_{p,q}\setminus \ball{b_n}} \abs{H_{p;m,n}(\freq') \conj{H_{q;m',n}(\freq'-\freq_1+\freq_2)}}\de\freq' \norm{\asdf{p,q}}_\infty.
    \end{align*}

    This is easiest dealt with by considering the different special cases for the processes $p,q$.
    Begin by assuming that both processes are sampled on grids.
    Potentially, $H_{p;m,n}$ and $H_{q;m,n}$ can repeat within $K_{p,q}$, however, the important observation is that this happens finitely many times, and they never repeat in the same way (so that they are only at most both large at zero), because we assumed that $\freq_1-\freq_2\notin \Aliasfreq_p \setminus \Aliasfreq_{q}$.
    The proof proceeds by splitting the integral into parts where one of $H_{p;m,n}$ or $H_{q;m,n}$ are large, and a remaining set where they are both small.
    
    More formally, note that $\Aliasfreq_p \cap K_{p,q}$ and $\Aliasfreq_q \cap K_{p,q}$ are both finite sets.
    Also, by construction, $\Aliasfreq_p \cap \Aliasfreq_q \cap K_{p,q} = \set{0}$.
    This means that for sufficiently large $n$ (and thus small $b_n$) we can partition $K_{p,q}\setminus \ball{b_n}$ into the finite union of $\bigcup\limits_{\aliasfreq \in \Aliasfreq_p\cap K_{p,q}\setminus\set{0}} \ball{b_n}+\aliasfreq$, $\bigcup\limits_{\aliasfreq \in \Aliasfreq_q\cap K_{p,q}\setminus\set{0}} \ball{b_n}+\aliasfreq$ and $K_{p,q}\setminus \left(\bigcup\limits_{\aliasfreq\in\Aliasfreq_p\cup\Aliasfreq_q} \ball{b_n}+\aliasfreq\right)$.
    This means
    \begin{align*}
        \abs{T_{p,q;n}^{(2)}(m,m',\freq_1,\freq_2)}
        &\leq \norm{\asdf{p,q}}_\infty\sum_{\aliasfreq\in\Aliasfreq_p\cap K_{p,q}\setminus\set{0}} \int_{\ball{b_n}+\aliasfreq} \abs{H_{p;m,n}(\freq') \conj{H_{q;m',n}(\freq'-\freq_1+\freq_2)}} \de\freq' \\
        &+ \norm{\asdf{p,q}}_\infty\sum_{\aliasfreq\in\Aliasfreq_q\cap K_{p,q}\setminus\set{0}} \int_{\ball{b_n}+\aliasfreq} \abs{H_{p;m,n}(\freq') \conj{H_{q;m',n}(\freq'-\freq_1+\freq_2)}} \de\freq' \\
        &+ \norm{\asdf{p,q}}_\infty \int_{K_{p,q}\setminus \left(\bigcup_{\aliasfreq\in\Aliasfreq_p\cup\Aliasfreq_q} \ball{b_n}+\aliasfreq\right)} \abs{H_{p;m,n}(\freq') \conj{H_{q;m',n}(\freq'-\freq_1+\freq_2)}} \de\freq'
    \end{align*}
    The two sums are finite, so if we can bound the maximum over $m,m'$ of their summands we are done.
    In particular, for some $\aliasfreq\in\Aliasfreq_p\cap K_{p,q}\setminus\set{0}$ we have
    \begin{align*}
        &\hspace{-2em} \int_{\ball{b_n}+\aliasfreq} \abs{H_{p;m,n}(\freq') \conj{H_{q;m',n}(\freq'-\freq_1+\freq_2)}} \de\freq' \\
        &\leq \left(\int_{\ball{b_n}+\aliasfreq} \abs{H_{p;m,n}(\freq')}^2\de\freq' \int_{\ball{b_n}+\aliasfreq} \abs{H_{q;m',n}(\freq'-\freq_1+\freq_2)}^2\de\freq'\right)^{1/2} \\
        &\leq \left(\int_{\underbrace{K_p\setminus \ball{b_n}}_{(a)}} \abs{H_{p;m,n}(\freq')}^2\de\freq' \int_{\underbrace{K_q+\aliasfreq}_{(b)}} \abs{H_{q;m',n}(\freq'-\freq_1+\freq_2)}^2\de\freq'\right)^{1/2}
    \end{align*}
    for sufficiently large $n$, where (a) comes from $\ball{b_n}+\aliasfreq \cap \ball{b_n} = \emptyset$ and (b) follows from $\ball{b_n} \subseteq K_q$ (which is independent of $m,m'$).
    By periodicity, 
    \begin{align*}
        \int_{\ball{b_n}+\aliasfreq} \abs{H_{p;m,n}(\freq') \conj{H_{q;m',n}(\freq'-\freq_1+\freq_2)}} \de\freq'
        &\leq \left(\int_{K_p\setminus \ball{b_n}} \abs{H_{p;m,n}(\freq')}^2\de\freq' \int_{K_q} \abs{H_{q;m',n}(\freq')}^2\de\freq'\right)^{1/2}
    \end{align*}
    as $n\rightarrow \infty$ by the taper assumptions.
    Now to finish we see that
    \begin{align*}
        \max_{m\in[M_n]} \int_{K_p\setminus \ball{b_n}} \abs{H_{p;m',n}(\freq')}^2\de\freq' \int_{K_q} \abs{H_{q;m,n}(\freq')}^2\de\freq' &= o(1)(1+o(1)) \\
        &= o(1)
    \end{align*}
    as $n\rightarrow\infty$ by \cref{assumption:tapers:concentration}.
    The argument for the second term is similar, except that we note either $\freq_1-\freq_2\in\Aliasfreq_q$, in which case we use periodicity, or $\freq_1-\freq_2\notin\Aliasfreq_q$, in which case we have that $\ball{b_n}+\aliasfreq-\freq_1+\freq_2 \cap \ball{b_n} = \emptyset$ for sufficiently large $n$.
    Then the argument is the same.

    Finally then
    \begin{align*}
        &\int_{K_{p,q}\setminus \left(\bigcup_{\aliasfreq\in\Aliasfreq_p\cup\Aliasfreq_q} \ball{b_n}+\aliasfreq\right)} \abs{H_{p;m,n}(\freq') \conj{H_{q;m',n}(\freq'-\freq_1+\freq_2)}} \de\freq' \\
        &\leq \left(\int_{K_{p,q}\setminus \left(\bigcup_{\aliasfreq\in\Aliasfreq_p\cup\Aliasfreq_q} \ball{b_n}+\aliasfreq\right)} \abs{H_{p;m,n}(\freq')}^2 \de\freq' \int_{K_{p,q}\setminus \left(\bigcup_{\aliasfreq\in\Aliasfreq_p\cup\Aliasfreq_q} \ball{b_n}+\aliasfreq\right)} \abs{H_{q;m',n}(\freq'-\freq_1+\freq_2)}^2 \de\freq'\right)^{1/2}.
    \end{align*}
    Consider the first of these terms. We will increase the domain of integration by removing only certain balls. In particular,
    \begin{align*}
        \int_{K_{p,q}\setminus \left(\bigcup_{\aliasfreq\in\Aliasfreq_p\cup\Aliasfreq_q} \ball{b_n}+\aliasfreq\right)} \abs{H_{p;m,n}(\freq')}^2 \de\freq'
        &\leq \int_{K_{p,q}\setminus \left(\bigcup_{\aliasfreq\in\Aliasfreq_p} \ball{b_n}+\aliasfreq\right)} \abs{H_{p;m,n}(\freq')}^2 \de\freq' \\
        &= \sum_{\aliasfreq'\in\Aliasfreq_p \cap K_{p,q}}\int_{(K_{p}+\aliasfreq') \setminus \left(\bigcup_{\aliasfreq\in\Aliasfreq_p} \ball{b_n}+\aliasfreq\right)} \abs{H_{p;m,n}(\freq')}^2 \de\freq' \\
        &\leq \sum_{\aliasfreq'\in\Aliasfreq_p \cap K_{p,q}}\int_{(K_{p}+\aliasfreq') \setminus \left(\ball{b_n}+\aliasfreq'\right)} \abs{H_{p;m,n}(\freq')}^2 \de\freq' \\
        &= \sum_{\aliasfreq'\in\Aliasfreq_p \cap K_{p,q}}\int_{K_{p} \setminus \ball{b_n}} \abs{H_{p;m,n}(\freq'+\aliasfreq')}^2 \de\freq' \\
        &= \sum_{\aliasfreq'\in\Aliasfreq_p \cap K_{p,q}}\int_{K_{p} \setminus \ball{b_n}} \abs{H_{p;m,n}(\freq')}^2 \de\freq'
    \end{align*}
    by periodicity. 
    Because $\Aliasfreq_p \cap K_{p,q}$ is a finite set and by \cref{assumption:tapers:concentration},
    \begin{align*}
    \max_{m\in [M_n]}\int_{K_{p,q}\setminus \left(\bigcup_{\aliasfreq\in\Aliasfreq_p\cup\Aliasfreq_q} \ball{b_n}+\aliasfreq\right)} \abs{H_{p;m,n}(\freq')}^2 \de\freq' \rightarrow 0
    \end{align*}
    as $n\rightarrow\infty$.
    Now the second term satisfies
    \begin{align*}
        \max_{m'\in [M_n]} \int_{K_{p,q}\setminus \left(\bigcup_{\aliasfreq\in\Aliasfreq_p\cup\Aliasfreq_q} \ball{b_n}+\aliasfreq\right)} \abs{H_{q;m',n}(\freq'-\freq_1+\freq_2)}^2 \de\freq'
        &\leq  \max_{m'\in [M_n]} \int_{K_{p,q}} \abs{H_{q;m',n}(\freq')}^2 \de\freq' \\
        &\rightarrow \abs{\Aliasfreq_q\cap K_{p,q}} < \infty
    \end{align*}
    as $n\rightarrow\infty$ by \cref{assumption:tapers:concentration}, the triangle inequality (as we detailed before), and periodicity.
    Therefore, when both processes are sampled on grids
    \begin{align*}
        \max_{m,m'\in[M_n]}\abs{T_{p,q;n}^{(2)}(m,m',\freq_1,\freq_2)} \rightarrow 0
    \end{align*}
    as $n\rightarrow\infty$.
    
    When at least one of the processes, say $p$ (wlog) is recorded continuously but the other is not, then $K_{p,q}=\RRd$, $\asdf{p,q}=\sdf{p,q}$ and $H_{p;m,n}=H_{m,n}$ so
    \begin{align*}
        \abs{T_{p,q;n}^{(2)}(m,m',\freq_1,\freq_2)}
        &= \abs{ \int_{\RRd\setminus \ball{b_n}} H_{m,n}(\freq') \conj{H_{q;m',n}(\freq'-\freq_1+\freq_2)}\sdf{p,q}(\freq_1-\freq')\de\freq'} \\
        &\leq \int_{\RRd\setminus \ball{b_n}} \abs{H_{m,n}(\freq') \conj{H_{q;m',n}(\freq'-\freq_1+\freq_2)}\sdf{p,q}(\freq_1-\freq')}\de\freq' \\
        &\leq \left(\int_{\RRd\setminus \ball{b_n}} \abs{H_{m,n}(\freq')}^2 \de\freq' \int_{\RRd\setminus \ball{b_n}} \abs{\conj{H_{q;m',n}(\freq'-\freq_1+\freq_2)}\sdf{p,q}(\freq_1-\freq')}^2\de\freq'\right)^{1/2}
    \end{align*}
    the maximum over $m,m'$ of which tends to zero provided that the second term converges to something finite.
    Indeed by Cauchy–Schwarz
    \begin{align*}
        &\hspace{-4em} \max_{m'\in[M_n]}\int_{\RRd\setminus \ball{b_n}} \abs{\conj{H_{q;m',n}(\freq'-\freq_1+\freq_2)}\sdf{p,q}(\freq_1-\freq')}^2\de\freq'\\
        &\leq \max_{m'\in[M_n]}\int_{\RRd\setminus \ball{b_n}} \abs{\conj{H_{q;m',n}(\freq'-\freq_1+\freq_2)}}^2\sdf{q,q}(\freq_1-\freq')\sdf{p,p}(\freq_1-\freq')\de\freq' \\
        &\leq \max_{m'\in[M_n]}\int_{\RRd} \abs{{H_{q;m',n}(\freq'-\freq_1+\freq_2)}}^2\sdf{q,q}(\freq_1-\freq')\de\freq' \norm{\sdf{p,p}}_\infty \\
        &= \max_{m'\in[M_n]}\int_{K_q} \abs{{H_{q;m',n}(\freq'-\freq_1+\freq_2)}}^2\asdf{q,q}(\freq_1-\freq')\de\freq' \norm{\sdf{p,p}}_\infty \;\;{\small(\text{by Definition 1})}\\
        &\leq \max_{m'\in[M_n]}\int_{K_q} \abs{{H_{q;m',n}(\freq')}}^2\de\freq'\norm{\asdf{q,q}}_\infty \norm{\sdf{p,p}}_\infty \\
        &\rightarrow \norm{\asdf{q,q}}_\infty \norm{\sdf{p,p}}_\infty \\
        &<\infty
    \end{align*}
    by \cref{assumption:tapers:concentration}.

    Finally, when both processes are sampled continuously
    \begin{align*}
        &\hspace{-2em} \max_{m,m'\in[M_n]} \abs{T_{p,q;n}^{(2)}(m,m',\freq_1,\freq_2)} \\
        &\leq \norm{\sdf{p,q}}_\infty \max_{m,m'\in[M_n]} \int_{\RRd\setminus \ball{b_n}} \abs{H_{m,n}(\freq') \conj{H_{m',n}(\freq')}}\de\freq'  \\
        &\leq \norm{\sdf{p,q}}_\infty \left(\max_{m\in[M_n]} \int_{\RRd\setminus \ball{b_n}} \abs{H_{m,n}(\freq')}^2 \de\freq'\right)^{1/2} \left(\max_{m'\in[M_n]} \int_{\RRd\setminus \ball{b_n}} \abs{H_{m',n}(\freq')}^2 \de\freq'\right)^{1/2} \\
        &\rightarrow 0
    \end{align*}
    as $n\rightarrow\infty$ by \cref{assumption:tapers:concentration}.
\end{proof}
\end{proposition}

\begin{proof}[Proof of \cref{res:bias:asymp}]
    Noting we assume that $\lambda_p$ is known, we have
    \begin{align*}
        \EE{\pgram{p,q}{m,n}(\freq)}
        &= \cov{\dft{p}{m,n}(\freq), \dft{q}{m,n}(\freq)}
    \end{align*}
    and so
    \begin{align*}
        \abs{\EE{\pgram{p,q}{m,n}(\freq)} - \asdf{p,q}(\freq)}
        &= \abs{\cov{\dft{p}{m,n}(\freq), \dft{q}{m,n}(\freq)} - \asdf{p,q}(\freq)} \\
        &\leq \max_{m\in[M_n]} \abs{\cov{\dft{p}{m,n}(\freq), \dft{q}{m,n}(\freq)} - \asdf{p,q}(\freq)} \\
        &\rightarrow 0
    \end{align*}
    as $n\rightarrow\infty$ by \cref{res:strong:bias:asymp} with $\freq_1=\freq_2=\freq$.
\end{proof}

\subsection{\texorpdfstring{\cref{prop:reduced_cumulant_measure}}{Proposition}}\label{proof:prop:reduced_cumulant_measure}
\begin{proof}[Proof of \cref{prop:reduced_cumulant_measure}]
    If all of the processes are sampled continuously, then this is simply Proposition 12.6.III of \cite{daley2007introductionV2}, and so we consider the case when the last process is not sampled continuously, and is instead sampled on $\grid=\Delta\circ\ZZ^d$.
    Differently from \cite{daley2007introductionV2}, we put the lags in the first arguments, not the last, but this is really just a cosmetic change here because the groups we use for the shift invariance are abelian.
    
    The argument is essentially the same as Lemma 12.1.IV of \cite{daley2007introductionV2}, except that we need to handle the grid sampling.
    In particular, the original argument in \cite{daley2007introductionV2} uses a shift invariance to diagonal shifts on the group $D^{(r)}_x$, for $r \in \NN$ and $x\in\RR^d$ where 
    \begin{align*}
        D_x^{(r)}(y_1,\ldots,y_r) = (y_1+x,\ldots,y_r+x)
    \end{align*}
    for $y_1,\ldots,y_r\in\RR^d$.
    In particular, that the $r$\Th order cumulant measure is invariant to such shifts.

    However, when at least one process is sampled on a grid this is no longer the case, as the grid is fixed, and thus not invariant to shifts, even when the underlying process is homogeneous (as a shift could change the number of included grid points).
    However, we still do have an invariance to diagonal shifts on the grid, provided all processes are sampled on the same grid (or can be embedded on the same grid).
    In particular, define the diagonal grid shifts by $\tilde{D}_x^{(r)}$ for $x\in\grid$, where
    \begin{align*}
        \tilde{D}_x^{(r)}(y_1,\ldots,y_r) = (y_1+x,\ldots,y_r+x)
    \end{align*}
    for $y_1,\ldots,y_r\in\RR^d$.

    Consider the space $\mathcal{Y} = \RR^{d(r-1)} \times \grid$.
    Using an essentially identical argument to that in Lemma 12.1.IV of \cite{daley2007introductionV2}, we consider the function from $\RR^{d(r-1)} \times \grid$ given by $((y_1,\ldots,y_{r-1}, x)) \mapsto (x+y_1, \ldots, x+y_{r-1}, x)$.
    This is a bijection from $\mathcal{Y}$ to itself, and continuous, so measurable.
    This means we can use Lemma A2.7.II of \cite{daley2003introduction}, so that we have for any bounded measurable function $g:\RR^{d(r-1)}\times\grid\rightarrow\CC$,
    \begin{align*}
        &\int_{\RR^{d(r-1)}\times \grid} g(y_1, \ldots, y_r) \cumulantmeasure{p_1,\ldots p_r}^{(\sampledmeasurebase)}(\de y_1 \times \cdots \times \de y_r) \\
        &= \int_{\grid} \int_{\RR^{d(r-1)}} g(x+u_1, \ldots, x+u_{r-1}, x) \reducedcumulantmeasure{p_1,\ldots p_r}^{(\sampledmeasurebase)}(\de u_1 \times \cdots \times \de u_{r-1}) \ell_{\grid}(\de x),
    \end{align*} 
    where $\ell_{\grid}$ is the counting measure on the grid $\grid$.
    Now because the last process is sampled on the grid, we only have mass at the points in the grid in the last coordinate of $\cumulantmeasure{p_1,\ldots p_r}^{(\sampledmeasurebase)}$, and similarly with $\ell_{\grid}$, and so
    \begin{align*}
        &\int_{\RR^{dr}} g(y_1, \ldots, y_r) \cumulantmeasure{p_1,\ldots p_r}^{(\sampledmeasurebase)}(\de y_1 \times \cdots \times \de y_r) \\
        &= \int_{\RR^{d(r-1)}\times \grid} g(y_1, \ldots, y_r) \cumulantmeasure{p_1,\ldots p_r}^{(\sampledmeasurebase)}(\de y_1 \times \cdots \times \de y_r) \\
        &= \int_{\grid} \int_{\RR^{d(r-1)}} g(x+u_1, \ldots, x+u_{r-1}, x) \reducedcumulantmeasure{p_1,\ldots p_r}^{(\sampledmeasurebase)}(\de u_1 \times \cdots \times \de u_{r-1}) \ell_{\grid}(\de x) \\
        &= \int_{\RRd} \int_{\RR^{d(r-1)}} g(x+u_1, \ldots, x+u_{r-1}, x) \reducedcumulantmeasure{p_1,\ldots p_r}^{(\sampledmeasurebase)}(\de u_1 \times \cdots \times \de u_{r-1}) \ell_{\grid}(\de x).
    \end{align*}
    This establishes the required result.
\end{proof}

\subsection{\texorpdfstring{\cref{res:consistency}}{Proposition}}\label{proof:res:consistency}
\begin{proof}[Proof of \cref{res:consistency}]
    Begin with the bias result.
    Noting that we assume $\lambda_p$ is known, we have
    \begin{align*}
        \abs{\mtpgram{p,q;n}(\freq) - \asdf{p,q}(\freq)} &= \abs{\frac{1}{M_n}\sum_{m=1}^{M_n} \pgram{p,q}{m,n}(\freq)-\asdf{p,q}(\freq)} \\
        &\leq \max_{m\in[M_n]} \abs{\pgram{p,q}{m,n}(\freq) - \asdf{p,q}(\freq)} \\
        &= \max_{m\in[M_n]} \abs{\cov{\dft{p}{m,n}(\freq), \dft{q}{m,n}(\freq)} - \asdf{p,q}(\freq)} \\
        &\rightarrow 0
    \end{align*}
    as $n\rightarrow\infty$ by \cref{res:strong:bias:asymp} with $\freq_1=\freq_2=\freq$.

    Now for the variance, firstly, for any $p,q\in[P]$ and $\freq\in\RRd$,
    \begin{align*}
        \var{\mtpgram{p,q;n}(\freq)} &= \frac{1}{M_n^2}\sum_{m=1}^{M_n}\sum_{m'=1}^{M_n} \cov{\pgram{p,q}{m,n}(\freq), \pgram{p,q}{m',n}(\freq)}.
    \end{align*}
    Now using the relations between moments and cumulants for mean-zero random variables
    \begin{align*}
        \cov{\pgram{p,q}{m,n}(\freq), \pgram{p,q}{m',n}(\freq')} 
        &= \EE{\pgram{p,q}{m,n}(\freq)\conj{\pgram{p,q}{m',n}(\freq')}} - \EE{\pgram{p,q}{m,n}(\freq)}\EE{\conj{\pgram{p,q}{m',n}(\freq')}} \\
        &= \EE{\dft{p}{m,n}(\freq) \conj{\dft{q}{m,n}(\freq)} \conj{\dft{p}{m',n}(\freq')} \dft{q}{m',n}(\freq')} \\ &\quad - \EE{\pgram{p,q}{m,n}(\freq)}\EE{\conj{\pgram{p,q}{m',n}(\freq')}} \\
        &= \cum{\dft{p}{m,n}(\freq), \conj{\dft{q}{m,n}(\freq)}, \conj{\dft{p}{m',n}(\freq')}, \dft{q}{m',n}(\freq')} 
        \\ &\quad + \EE{\dft{p}{m,n}(\freq)\conj{\dft{p}{m',n}(\freq')}}\EE{\conj{\dft{q}{m,n}(\freq)}\dft{q}{m',n}(\freq')} 
        \\ &\quad + \EE{\dft{p}{m,n}(\freq)\dft{q}{m',n}(\freq')}\EE{\conj{\dft{q}{m,n}(\freq)\dft{p}{m',n}(\freq')}},
    \end{align*}
    where $\cum{\cdot}$ is the joint cumulant.
    Therefore we can split the variance into multiple parts, in particular
    \begin{align*}
        \var{\mtpgram{p,q;n}(\freq)} 
        &= \frac{1}{M_n^2}\sum_{m=1}^{M_n}\sum_{m'=1}^{M_n} \cum{\dft{p}{m,n}(\freq), \conj{\dft{q}{m,n}(\freq)}, \conj{\dft{p}{m',n}(\freq)}, \dft{q}{m',n}(\freq)} \\
        &+ \frac{1}{M_n^2}\sum_{m=1}^{M_n}\sum_{m'=1}^{M_n} \EE{\dft{p}{m,n}(\freq)\conj{\dft{p}{m',n}(\freq)}}\EE{\conj{\dft{q}{m,n}(\freq)}\dft{q}{m',n}(\freq)} \\
        &+ \frac{1}{M_n^2}\sum_{m=1}^{M_n}\sum_{m'=1}^{M_n} \EE{\dft{p}{m,n}(\freq)\dft{q}{m',n}(\freq)}\EE{\conj{\dft{q}{m,n}(\freq)\dft{p}{m',n}(\freq)}}.
    \end{align*}
    Again we name each term
    \begin{align*}
        S_n^{(1)} &= \frac{1}{M_n^2}\sum_{m=1}^{M_n}\sum_{m'=1}^{M_n} \cum{\dft{p}{m,n}(\freq), \conj{\dft{q}{m,n}(\freq)}, \conj{\dft{p}{m',n}(\freq)}, \dft{q}{m',n}(\freq)} \\
        S_n^{(2)} &= \frac{1}{M_n^2}\sum_{m=1}^{M_n}\sum_{m'=1}^{M_n} \EE{\dft{p}{m,n}(\freq)\conj{\dft{p}{m',n}(\freq)}}\EE{\conj{\dft{q}{m,n}(\freq)}\dft{q}{m',n}(\freq)} \\
        S_n^{(3)} &= \frac{1}{M_n^2}\sum_{m=1}^{M_n}\sum_{m'=1}^{M_n} \EE{\dft{p}{m,n}(\freq)\dft{q}{m',n}(\freq)}\EE{\conj{\dft{q}{m,n}(\freq)\dft{p}{m',n}(\freq)}}.
    \end{align*}
    Firstly then
    \begin{align*}
        S_n^{(2)} 
        &= \frac{1}{M_n^2}\sum_{m=1}^{M_n}\sum_{m'=1}^{M_n} \EE{\dft{p}{m,n}(\freq)\conj{\dft{p}{m',n}(\freq)}}\EE{\conj{\dft{q}{m,n}(\freq)}\dft{q}{m',n}(\freq)} \\
        &= \frac{1}{M_n^2}\sum_{m=1}^{M_n}\sum_{m'=1}^{M_n} \abs{\cov{\dft{p}{m,n}(\freq),\dft{p}{m',n}(\freq)}}^2.
    \end{align*}
    For notational convenience, write the following
    \begin{align*}
        C_n(m,m') &= \cov{\dft{p}{m,n}(\freq),\dft{p}{m',n}(\freq)}, \\
        Q(m,m') &= \asdf{p,q}(\freq)\delta_{m,m'}.
    \end{align*}
    then note that
    \begin{align*}
        \abs{C_n(m,m')-Q(m,m')}^2 = \abs{C_n(m,m')}^2 - 2\Re{C_n(m,m')Q(m,m')} + \abs{Q(m,m')}^2
    \end{align*}
    so rearranging
    \begin{align*}
        \abs{C_n(m,m')}^2 &= \abs{C_n(m,m')-Q(m,m')}^2 + 2\Re{C_n(m,m')Q(m,m')} - \abs{Q(m,m')}^2 \\
        &\leq \abs{C_n(m,m')-Q(m,m')}^2 + 2\abs{C_n(m,m')}\abs{Q(m,m')} + \abs{Q(m,m')}^2 \\
        &\leq \abs{C_n(m,m')-Q(m,m')}^2 + 2\abs{C_n(m,m')-Q(m,m')}\abs{Q(m,m')} + 2\abs{Q(m,m')}^2.
    \end{align*}
    Therefore,
    \begin{align*}
        S_n^{(2)} &= \frac{1}{M_n^2}\sum_{m=1}^{M_n}\sum_{m'=1}^{M_n} \abs{C_n(m,m')}^2 \\
        &\leq \frac{1}{M_n^2}\sum_{m=1}^{M_n}\sum_{m'=1}^{M_n} \abs{C_n(m,m')-Q(m,m')}^2 \\
        &\quad + 2\frac{1}{M_n^2}\sum_{m=1}^{M_n}\sum_{m'=1}^{M_n} \abs{C_n(m,m')-Q(m,m')}\abs{Q(m,m')} \\
        &\quad + 2\frac{1}{M_n^2}\sum_{m=1}^{M_n}\sum_{m'=1}^{M_n} \abs{Q(m,m')}^2 \\
        &\leq \max_{m,m'\in[M_n]} \abs{C_n(m,m')-Q(m,m')}^2 \\
        &\quad + \max_{m,m'\in[M_n]} \abs{C_n(m,m')-Q(m,m')} \norm{\asdf{p,q}}_\infty \\
        &\quad + \frac{\norm{\asdf{p,q}}_\infty}{M_n} \\
        &\rightarrow 0
    \end{align*}
    as $n\rightarrow \infty$ by \cref{res:strong:bias:asymp} with $\freq_1=\freq_2=\freq$ and the assumption that $M_n\rightarrow\infty$.

    Now 
    \begin{align*}
        S_n^{(3)} &= \frac{1}{M_n^2}\sum_{m=1}^{M_n}\sum_{m'=1}^{M_n} \EE{\dft{p}{m,n}(\freq)\dft{q}{m',n}(\freq)}\EE{\conj{\dft{q}{m,n}(\freq)\dft{p}{m',n}(\freq)}} \\
        &= \frac{1}{M_n^2}\sum_{m=1}^{M_n}\sum_{m'=1}^{M_n} \cov{\dft{p}{m,n}(\freq),\dft{q}{m',n}(-\freq)}\cov{\dft{q}{m,n}(-\freq),\dft{p}{m',n}(\freq)} \\
    \end{align*}
    by the same argument we used for $S_n^{(2)}$ setting $Q(m,m') = \asdf{p,q}(\freq)\delta_{m,m'}$ if $2\freq \in \Aliasfreq_p\cup \Aliasfreq_q$ and zero otherwise, we get
    \begin{align*}
        S_n^{(3)} &\rightarrow 0
    \end{align*}
    as $n\rightarrow\infty$.

    Finally then we need to deal with the last term.
    \begin{align*}
        S_n^{(1)} &= \frac{1}{M_n^2}\sum_{m=1}^{M_n}\sum_{m'=1}^{M_n} \cum{\dft{p}{m,n}(\freq), \conj{\dft{q}{m,n}(\freq)}, \conj{\dft{p}{m',n}(\freq)}, \dft{q}{m',n}(\freq)}.
    \end{align*}
    In particular, let us study the term
    \begin{align*}
        \cum{\dft{p}{m,n}(\freq), \conj{\dft{q}{m,n}(\freq)}, \conj{\dft{p}{m',n}(\freq)}, \dft{q}{m',n}(\freq)}.
    \end{align*}
    Recall that we defined the grid $\grid$ to be the grid which contains all of the other grids of interest.
    Therefore, when the $p$\Th process is sampled on a grid, we can write $g_{p;m,n} = \elemprod{\Delta_p} h_{m,n}\indicator_{\grid_p}$, and think about the grid sampled transforms as sums of $g_{p;m,n}$ over the grid $\grid$.
    For convenience, when the $p$\Th process is sampled continuously, we can set $g_{p;m,n} = h_{m,n}$, which is just the continuous transform.

    For further notational convenience, define
    \begin{align*}
        \phi_{1,n}(x) &= g_{p;m,n}(x)e^{-2\pi i \ip{\freq}{x}}, \\
        \phi_{2,n}(x) &= g_{q;m,n}(x)e^{2\pi i \ip{\freq}{x}}, \\
        \phi_{3,n}(x) &= g_{p;m',n}(x)e^{2\pi i \ip{\freq}{x}}, \\
        \phi_{4,n}(x) &= g_{q;m',n}(x)e^{-2\pi i \ip{\freq}{x}}.
    \end{align*}
    
    Again assume without loss of generality that if at least one process is sampled on a grid, then the $q$\Th process is sampled on a grid.
    Since the functions $h_{m,n}$ are bounded and have bounded support, from \cref{prop:reduced_cumulant_measure} we have that
    \begin{align*}
        & \cum{\dft{p}{m,n}(\freq), \conj{\dft{q}{m,n}(\freq)}, \conj{\dft{p}{m',n}(\freq)}, \dft{q}{m',n}(\freq)} \\
        &= \cum{\int_\RRd \phi_{1,n}(x)\randmeasurecenter{p}(\de x),  \int_\RRd \phi_{2,n}(x)\randmeasurecenter{q}, \int_\RRd \phi_{3,n}(x)\randmeasurecenter{p}(\de x), \int_\RRd \phi_{4,n}(x)\randmeasurecenter{q}(\de x)} \\
        &= \int_{\RR^{4d}} \phi_{1,n}(x_1) \phi_{2,n}(x_2) \phi_{3,n}(x_3) \phi_{4,n}(x_4) \cumulantmeasure{p,q,p,q}^{(\sampledmeasurebase)}(\de x_1 \times \de x_2 \times \de x_3 \times \de x_4) \\
        &= \int_\RRd \int_{\RR^{3d}} \phi_{1,n}(\lag_1+x) \phi_{2,n}(\lag_2+x) \phi_{3,n}(\lag_3+x) \phi_{4,n}(x) \reducedcumulantmeasure{p,q,p,q}^{(\sampledmeasurebase)}(\de \lag_1 \times \de \lag_2 \times \de \lag_3) \ell_q(\de x).
    \end{align*}
    Now notice that the functions $\phi_{1,n},\ldots,\phi_{4,n}$ are all bounded with bounded support, and therefore the integral is well-defined. 
    Furthermore, they are continuous functions, with continuous and $L^1$ Fourier transforms, meaning that we can write them as their Fourier inverse transforms.
    In particular, we have
    \begin{align*}
        & \cum{\dft{p}{m,n}(\freq), \conj{\dft{q}{m,n}(\freq)}, \conj{\dft{p}{m',n}(\freq)}, \dft{q}{m',n}(\freq)} \\
        &= \int_\RRd \phi_{4,n}(x) \int_{\RR^{3d}} \prod_{j=1}^3 \int_\RRd \Phi_{j,n}(\freq_j)e^{2\pi i \ip{\freq_j}{(\lag_1+x)}} \de\freq_j \reducedcumulantmeasure{p,q,p,q}^{(\sampledmeasurebase)}(\de \lag_1 \times \de \lag_2 \times \de \lag_3) \ell_q(\de x).
    \end{align*}
    We will need to interchange the order of intergration, for which we will need to use Fubini's theorem.
    To check this is legitimate, we see
    \begin{align*}
        &\hspace{-10em} \int_\RRd \abs{\phi_{4,n}(x)} \int_{\RR^{3d}} \prod_{j=1}^3 \int_\RRd \abs{\Phi_{j,n}(\freq_j)} \de\freq_j \abs{\reducedcumulantmeasure{p,q,p,q}^{(\sampledmeasurebase)}}(\de \lag_1 \times \de \lag_2 \times \de \lag_3) \ell_q(\de x) \\
        &= \prod_{j=1}^3 \norm{\Phi_{j,n}}_1  \sum_{x\in\grid} \abs{\phi_{4,n}(x)} \abs{\reducedcumulantmeasure{p,q,p,q}^{(\sampledmeasurebase)}}\left(\RR^{3d}\right) \\
        &<\infty
    \end{align*}
    where $\abs{\reducedcumulantmeasure{p,q,p,q}^{(\sampledmeasurebase)}}$ is the absolute value measure defined from the reduce cumulant measure using the Jordan-Hahn decomposition, and the integral is finite because the functions $\Phi_{j,n}$ are $L^1$ functions, $\phi_{1,n}$ is bounded and has bounded support, so that sum/integral (depending on wether the $q$\Th process is grid-sampled) is finite, and $\reducedcumulantmeasure{p,q,p,q}^{(\sampledmeasurebase)}$ is a totally finite measure.

    Therefore, we can interchange the order of integration and write
    \begin{align*}
        & \cum{\dft{p}{m,n}(\freq), \conj{\dft{q}{m,n}(\freq)}, \conj{\dft{p}{m',n}(\freq)}, \dft{q}{m',n}(\freq)} \\
        &= \int_\RRd \int_\RRd \int_\RRd \Phi_{1,n}(\freq_1) \Phi_{2,n}(\freq_2) \Phi_{3,n}(\freq_3)
        \int_\RRd \phi_{4,n}(x) e^{2\pi i \ip{\freq_1+\freq_2+\freq_3}{x}} \de x \\
        & \hspace{2em} \cdot e^{2\pi i (\ip{\freq_1}{\lag_1}+\ip{\freq_2}{\lag_2}+\ip{\freq_3}{\lag_3})} \reducedcumulantmeasure{p,q,p,q}^{(\sampledmeasurebase)}(\de \lag_1 \times \de \lag_2 \times \de \lag_3) \de\freq_1 \de\freq_2 \de\freq_3 \\
        &= \int_\RRd \int_\RRd \int_\RRd \Phi_{1,n}(\freq_1) \Phi_{2,n}(\freq_2) \Phi_{3,n}(\freq_3)
        \Phi_{4,n}(-\freq_1-\freq_2-\freq_3) \\
        & \hspace{2em} \cdot e^{2\pi i (\ip{\freq_1}{\lag_1}+\ip{\freq_2}{\lag_2}+\ip{\freq_3}{\lag_3})} \reducedcumulantmeasure{p,q,p,q}^{(\sampledmeasurebase)}(\de \lag_1 \times \de \lag_2 \times \de \lag_3) \de\freq_1 \de\freq_2 \de\freq_3 \\
        &\leq \int_\RRd \int_\RRd \int_\RRd \abs{\Phi_{1,n}(\freq_1) \Phi_{2,n}(\freq_2) \Phi_{3,n}(\freq_3)
        \Phi_{4,n}(-\freq_1-\freq_2-\freq_3)} \de\freq_1 \de\freq_2 \de\freq_3 \abs{\reducedcumulantmeasure{p,q,p,q}^{(\sampledmeasurebase)}}\left(\RR^{3d}\right).
    \end{align*}
    Now we consider two cases. Firstly, if the $p$\Th process is sampled continuously, then $\Phi_{4,n}(\freq') = H_{m',n}(\freq'-\freq)$ which is in $L^1$.
    Therefore from \cref{res:taper:conv_bound} we have
    \begin{align*}
        & \hspace{-2em} \cum{\dft{p}{m,n}(\freq), \conj{\dft{q}{m,n}(\freq)}, \conj{\dft{p}{m',n}(\freq)}, \dft{q}{m',n}(\freq)} \\
        &\leq \prod_{j=1}^4 \norm{\Phi_j}_2^{\frac{1}{4}} \norm{\Phi_j}_1^{\frac{1}{2}}\abs{\reducedcumulantmeasure{p,q,p,q}^{(\sampledmeasurebase)}}\left(\RR^{3d}\right) \\
        &= \norm{H_{m,n}}_2^{\frac{1}{4}} \norm{H_{m,n}}_1^{\frac{1}{2}} \norm{H_{m',n}}_2^{\frac{1}{4}} \norm{H_{m',n}}_1^{\frac{1}{2}}\abs{\reducedcumulantmeasure{p,q,p,q}^{(\sampledmeasurebase)}}\left(\RR^{3d}\right) \\
        &= \norm{H_{m,n}}_2^{\frac{1}{4}} \norm{H_{m,n}}_1^{\frac{1}{2}} \norm{H_{m',n}}_2^{\frac{1}{4}} \norm{H_{m',n}}_1^{\frac{1}{2}}\abs{\reducedcumulantmeasure{p,q,p,q}^{(\sampledmeasurebase)}}\left(\RR^{3d}\right).
    \end{align*}
    Therefore,
    \begin{align*}
        S_n^{(1)} 
        &= \frac{1}{M_n^2}\sum_{m=1}^{M_n}\sum_{m'=1}^{M_n} \cum{\dft{p}{m,n}(\freq), \conj{\dft{q}{m,n}(\freq)}, \conj{\dft{p}{m',n}(\freq)}, \dft{q}{m',n}(\freq)} \\
        &\leq \abs{\reducedcumulantmeasure{p,q,p,q}^{(\sampledmeasurebase)}}\left(\RR^{3d}\right)\max_{m \in [M_n]} \norm{H_{m,n}}_2^{\frac{1}{2}} \norm{H_{m,n}}_1\max_{m\in[M_n]}\norm{H_{m',n}}_2^{\frac{1}{2}} \norm{H_{m',n}}_1 \\
        &= \abs{\reducedcumulantmeasure{p,q,p,q}^{(\sampledmeasurebase)}}\left(\RR^{3d}\right)\max_{m \in [M_n]} \norm{H_{m,n}}_1\max_{m\in[M_n]} \norm{H_{m',n}}_1 \\
        &\rightarrow 0
    \end{align*}
    as $n\rightarrow\infty$.
    
    Now if the $p$\Th process is sampled on a grid, then we have $\Phi_{4,n}(\freq') = H_{q;m',n}(\freq'-\freq)$, which repeats, so needs different treatment.
    In particular, we have that for $\aliasfreq\in\Aliasfreq_q$, $\abs{\Phi_{4,n}(\freq'+\aliasfreq)} = \abs{\Phi_{4,n}(\freq')}$. 

    Now
    \begin{align*}
        & \int_\RRd \int_\RRd \int_\RRd \abs{\Phi_{1,n}(\freq_1) \Phi_{2,n}(\freq_2) \Phi_{3,n}(\freq_3) \Phi_{4,n}(-\freq_1-\freq_2-\freq_3)} \de\freq_1 \de\freq_2 \de\freq_3 \\
        &= \sum_{\aliasfreq_1,\aliasfreq_2,\aliasfreq_3\in\Aliasfreq_{q}} \int_{K_p+\aliasfreq_1} \int_{K_q+\aliasfreq_2} \int_{K_q+\aliasfreq_3} \abs{\Phi_{1,n}(\freq_1) \Phi_{2,n}(\freq_2) \Phi_{3,n}(\freq_3)
        \Phi_{4,n}(-\freq_1-\freq_2-\freq_3)} \de\freq_1 \de\freq_2 \de\freq_3 \\
        &= \sum_{\aliasfreq_1,\aliasfreq_2,\aliasfreq_3\in\Aliasfreq_{q}} \int_{K_q} \int_{K_q} \int_{K_q} \abs{\Phi_{1,n}(\freq_1+\aliasfreq_1) \Phi_{2,n}(\freq_2+\aliasfreq_2) \Phi_{3,n}(\freq_3+\aliasfreq_3)
        \Phi_{4,n}(-\freq_1-\freq_2-\freq_3)} \de\freq_1 \de\freq_2 \de\freq_3 \\
        &\leq \left(\int_{K_q} \abs{\Phi_{4,n}(\freq)}^{\frac{4}{3}}\de\freq\right)^{3/4} \prod_{j=1}^{3} \sum_{\aliasfreq_j\in\Aliasfreq_{q}}\int_{K_q} \left(\abs{\Phi_{j,n}(\freq_j+\aliasfreq_j)}^{\frac{4}{3}} \de\freq_j\right)^{3/4} \\
        &\leq \left(\int_{K_q} \abs{\Phi_{4,n}(\freq)}^{2}\de\freq\right)^{1/4}\left(\int_{K_q} \abs{\Phi_{4,n}(\freq)}\de\freq\right)^{1/2} \prod_{j=1}^{3} \sum_{\aliasfreq_j\in\Aliasfreq_{q}} \int_{K_q} \left(\abs{\Phi_{j,n}(\freq_j+\aliasfreq_j)}^{2} \de\freq_j\right)^{\frac{1}{2}} \ell(K_q)^{\frac{1}{4}} \\
        &\leq \left(\int_{K_q} \abs{\Phi_{4,n}(\freq)}^{2}\de\freq\right)^{1/4}\left(\int_{K_q} \abs{\Phi_{4,n}(\freq)}\de\freq\right)^{1/2} \ell(K_q)^{1/4} \prod_{j=1}^{3} \sum_{\aliasfreq_j\in\Aliasfreq_{q}} \left(\int_{K_q+\aliasfreq_j} \abs{\Phi_{j,n}(\freq_j)}^{2} \de\freq_j\right)^{1/2},
    \end{align*}
    where we used \cref{lemma:taper:hos:altbound}.
    Therefore,
    \begin{align*}
        S_n^{(1)} 
        &= \frac{1}{M_n^2}\sum_{m=1}^{M_n}\sum_{m'=1}^{M_n} \cum{\dft{p}{m,n}(\freq), \conj{\dft{q}{m,n}(\freq)}, \conj{\dft{p}{m',n}(\freq)}, \dft{q}{m',n}(\freq)} \\
        &\leq \abs{\reducedcumulantmeasure{p,q,p,q}^{(\sampledmeasurebase)}}\left(\RR^{3d}\right) \ell(K_q)^{1/4} \max_{m \in [M_n]} \left(\int_{K_q} \abs{H_{q;m,n}(\freq)}^{2}\de\freq\right)^{1/4}\left(\int_{K_q} \abs{H_{q;m,n}(\freq)}\de\freq\right)^{1/2} \\
        &\hspace{4em} \times \left\{\max_{m\in[M_n]} \sum_{\aliasfreq \in\Aliasfreq_{q}} \left(\int_{K_q+\aliasfreq} \abs{H_{m,n}(\freq)}^{2} \de\freq\right)^{1/2}\right\}^3 \\
        &\rightarrow 0
    \end{align*}
    as $n\rightarrow\infty$ by \cref{assumption:tapers:concentration,assumption:tapers:consistency}.
    This completes the proof of the variance result.
\end{proof}

\begin{lemma}\label{res:taper:conv_bound}
    For $r\geq 3$, consider some functions $A_j:\RRd\rightarrow\CC$ for $1\leq j\leq r$ and $n\in\NN$ such that $\norm{A_j}_1<\infty$ and $\norm{A_j}_2<\infty$, then
    \begin{align*}
        \int_\RRd\cdots\int_\RRd \abs{\left\{\prod_{j=1}^{r-1} A_j(\freq_j)\right\} A_r\left(-\sum_{j=1}^{r-1} \freq_j\right)} \de\freq_1\cdots\de\freq_{r-1} &\leq \prod_{j=1}^{r}\left(\int_\RRd \abs{A_j(\freq)}^{\frac{r}{r-1}}\de\freq\right)^{\frac{r-1}{r}}
    \end{align*}
    \begin{proof}
    Firstly we have
    \begin{align*}
        \norm{A_j}_{\frac{r}{r-1}} &\leq \left(\int_\RRd \abs{A_j(\freq)}^{\frac{r}{r-1}}\de\freq\right)^{\frac{r-1}{r}} \\
        &= \left(\norm{A_j^\frac{2}{r-1} A_j^\frac{r-2}{r-1}}_{1}\right)^{\frac{r-1}{r}} \\
        &\leq \left(\norm{A_j^\frac{2}{r-1}}_{r-1} \norm{A_j^\frac{r-2}{r-1}}_{\frac{r-1}{r-2}}\right)^{\frac{r-1}{r}} \\
        &= \left(\int_\RRd \abs{A_j(\freq)}^2 \de\freq\right)^{\frac{1}{r}} \left(\int_\RRd \abs{A_j(\freq)} \de\freq\right)^{\frac{r-2}{r}} \\
        &= \norm{A_j}_2^{\frac{1}{r}} \norm{A_j}_1^{\frac{r-2}{r}} \\
        &< \infty
    \end{align*}
    from H\"older's inequality and by the assumption of finiteness of $\norm{A_j}_1$ and $\norm{A_j}_2$.
    
    Following the approach of \cite{brillinger1982asymptotic}, we begin by making the substitution
    \begin{align*}
        k_1'&=k_1 \\
        k_j'&=k_j+k_{j-1}' \qquad \text{for  } 2\leq j\leq r-1,
    \end{align*}
    then we get
    \begin{align*}
        &\int_\RRd\cdots\int_\RRd \abs{ A_1(\freq_1')\left\{\textstyle\prod_{j=2}^{r-1} A_j(\freq_j'-\freq_{j-1}')\right\} A_r(-\freq_{r-1}') }\de\freq_1'\cdots\de\freq_{r-1}' \\
        &= \int_\RRd \abs{\phi_{r-1}(\freq) A_r(-\freq)} \de\freq \\
        &= \norm{\phi_{r-1} A_r}_1,
    \end{align*}
    where
    \begin{align*}
        \phi_{1}(\freq) &= A_{1}(\freq), \\
        \phi_{j}(\freq) &= \int_\RRd \abs{A_{j}(\freq-\freq') \phi_{j-1}(\freq')} \de\freq,
    \end{align*}
    for $2\leq j\leq r-1$.

    Now we have by H\"older's inequality
    \begin{align*}
        \norm{\phi_{r-1} A_{r}}_1 & \leq \norm{A_{r}}_{\frac{r}{r-1}} \norm{\phi_{r-1}}_r
    \end{align*}
    and by definition and Young's inequality
    \begin{align*}
        \norm{\phi_{1}}_{\frac{r}{r-1}} &= \norm{A_{1}}_{\frac{r}{r-1}} \\
        \norm{\phi_{j}}_{\frac{r}{r-j}} &= \norm{A_{j}\ast\phi_{j-1}}_{\frac{r}{r-1}} 
        \\ & \leq \norm{A_{j}}_{\frac{r}{r-1}} \norm{\phi_{j-1}}_{\frac{r}{r-(j-1)}} 
    \end{align*}
    for $2\leq j\leq r-1$.
    Here Young's inequality applies because, $\norm{A_{j}}_{\frac{r}{r-1}}$ is finite, and $\norm{\phi_{j-1}}_{\frac{r}{r-(j-1)}}$ is therefore finite by induction.
    Applying this recursively we obtain
    \begin{align*}
        \norm{\phi_{r-1} A_r}_1 &\leq \prod_{j=1}^r \norm{A_{j}}_{\frac{r}{r-1}}
    \end{align*}
    giving the result.
    \end{proof}
\end{lemma}

\begin{lemma}\label{lemma:taper:hos:altbound}
    For any $r>0$, if $h:\RRd \rightarrow\CC$ for $1\leq j\leq r$ is such that $\norm{h}_1<\infty$ and $\norm{h}_2<\infty$, then
    \begin{align*}
        \norm{h \indicator_K}_{\frac{r}{r-1}} &\leq \norm{h \indicator_K}_2 \ell(K)^{\frac{r-2}{2r}}.
    \end{align*}
    \begin{proof}
        We have
        \begin{align*}
            \norm{h \indicator_K}_{\frac{r}{r-1}}
            &=\norm{(h\indicator_K)^{\frac{r}{r-1}} \indicator_K}_1^{\frac{r-1}{r}} \\
            &\leq \left(\norm{(h\indicator_K)^{\frac{r}{r-1}}}_{\frac{2(r-1)}{r}} \norm{\indicator_K}_{\frac{2(r-1)}{r-2}} \right)^{\frac{r-1}{r}} \\
            &= \left(\left(\int_K \abs{h(x)}^2 \de x\right)^{\frac{r}{2(r-1)}} \ell(K)^{\frac{r-2}{2(r-1)}}\right)^{\frac{r-1}{r}} \\
            &= \norm{h \indicator_K}_2 \ell(K)^{\frac{r-2}{2r}}
        \end{align*}
        where in the second line we used H\"older's inequality with $p=\frac{2(r-1)}{r}$ and $q=\frac{2(r-1)}{r-2}$.
    \end{proof}
\end{lemma}

\subsection{\texorpdfstring{\cref{theorem:normality_dft}}{Theorem}}\label{proof:theorem:normality_dft}
\begin{proof}[Proof of \cref{theorem:normality_dft}]
    To establish this result, we use a similar argument to \cite{brillinger1982asymptotic}.
    In particular, we need to show that the mean is asymptotically zero, the variance and relation converges to a finite value, and that the cumulants of order greater than two converge to zero.
    The mean is asymptotically zero by construction.
    Recall we have restricted ourselves to the case where $\freq_1,\ldots,\freq_r$ are such that $\freq_i\pm\freq_j\not\in\cup_{p\in[P]}\Aliasfreq_p$.
    Now from \cref{res:strong:bias:asymp}, for $j\in[r]$
    \begin{align*}
        \cov{\dft{p}{m,n}(\freq_j), \dft{q}{m,n}(\freq_j)}
        &\rightarrow \asdf{p,q}(\freq_j) \conj{w_q(0)} \\
        &= \asdf{p,q}(\freq_j)
    \end{align*}
    as $n\rightarrow\infty$ and have relation
    \begin{align*}
        \mathrm{Rel}(\dft{p}{m,n}(\freq_j), \dft{q}{m,n}(\freq_j))
        &= \cov{\dft{p}{m,n}(\freq_j), \conj{\dft{q}{m,n}(\freq_j)}} \\
        &= \cov{\dft{p}{m,n}(\freq_j), \dft{q}{m,n}(-\freq_j)} \\
        &\rightarrow 0
    \end{align*}
    as $n\rightarrow\infty$ by conjugate symmetry of $\dft{q}{m,n}$ and because by assumption $\freq_j-(-\freq_j)=2\freq_j \notin \Aliasfreq_p \cup \Aliasfreq_q$.

    Finally for two different frequencies where $i\neq j$,
    \begin{align*}
        \cov{\dft{p}{m,n}(\freq_i), \dft{q}{m,n}(\freq_j)}
        &\rightarrow 0
    \end{align*}
    as $n\rightarrow\infty$ and
    \begin{align*}
        \mathrm{Rel}(\dft{p}{m,n}(\freq_i), \dft{q}{m,n}(\freq_j))
        &= \cov{\dft{p}{m,n}(\freq_i), \dft{q}{m,n}(-\freq_j)} \\
        &\rightarrow 0
    \end{align*}
    as $n\rightarrow\infty$ again from the assumption that $\freq_i\pm\freq_j\notin \Aliasfreq_p \cup \Aliasfreq_q$.
    The higher-order cumulants tending to zero follows the same argument used for the proof of consistency in \cref{proof:res:consistency}, modified to the arbitrary order cumulants.

    This means we have asymptotic normality and uncorrelatedness across $\freq_i$ and $\freq_j$, and therefore we have asymptotic independence across $\freq_i$ and $\freq_j$.
\end{proof}

\subsection{\texorpdfstring{\cref{res:general:interpolate}}{Proposition}}\label{proof:res:general:interpolate}
\begin{proof}[Proof of \cref{res:general:interpolate}]
    The restriction to the set $\tilde{\region} = \set{\loc \in \region \mid \loc+\Delta \circ [-1,1]^d\subseteq \region}$ ensures that when we linearly interpolate the tapers, we do not leave the region $\region$.
    This holds because the contribution to the linear interpolation from the discrete tapers we make at a point $\loc\in\grid$ only impacts the values of the taper at points within $\loc+\Delta \circ [-1,1]^d$, so if we only include such boxes contained in $\region$, the resultant tapers are supported on a subset of $\region$.

    Now, by construction the tapers are bounded and continuous.
    Using \cref{supp:res:ip_interp}, we can standardise the tapers to have unit $L^2$ norm.
    Finally, from \cref{res:taper_ft} we have
    \begin{align*}
        H_{m}(\freq) &= G_m^{(\grid)}(\freq) \prod_{j=1}^d \sinc^2(\pi \Delta_j \freq_j)
    \end{align*}
    where $G_m^{(\grid)}$ is the Fourier transform of the taper $g_m^{(\grid)}$.
    The Fourier transform $G_m^{(\grid)}$  is a finite sum because the region $\region$ is bounded, and therefore it is bounded.
    As a result, $\norm{H_{m}}_1<\infty$ since $\sinc^2$ is integrable.
\end{proof}

\section{Proofs of additional results}

\subsection{Proof of \texorpdfstring{\cref{theorem:normality_dft:alpha}}{Theorem}}\label{proof:theorem:normality_dft:alpha}
\begin{proof}[Proof of \cref{theorem:normality_dft:alpha}]
    We utilise the central limit theorem result from \cite{biscio2019general}.
    There are four assumptions which must be verified to use this result, which they name $(\mathcal{H}1)-(\mathcal{H}4)$.
    For their full statement, see \cite{biscio2019general}.
    We will address each of these in turn.
    Note that the result in \cite{biscio2019general} is for real valued statistics, and ours are complex valued.
    We handle this with the standard isomorphism between $\CC$ and $\RR^2$, so that we can treat the real and imaginary parts separately.
    In particular, write
    \begin{align*}
        \phi:\CC \rightarrow \RR^2, \quad \phi(x) = (\Re(x), \Im(x))^\top.
    \end{align*}

    The statistic in question whose asymptotic normality we wish to establish is the vector given by
    \begin{align*}
        T_n = \ell(\region_n)^{1/2} \mathrm{vec}\; [\phi(J_{p;m,n}(\freq_j))]_{p\in[P], m\in [M], j\in[r]},
    \end{align*}
    which is a vector in $\RR^{2PMr}$, recalling $P$ is the number of processes, $M$ then number of tapers and $r$ the number of wavenumbers.

    Firstly, starting with $(\mathcal{H}1)$.
    We require that $\region_1 \subset \region_2 \subset \cdots$ and $\ell(\bigcup_{n=1}^\infty \region_n) = \infty$, which we satisfy by assumption.
    Furthermore, we need to be able to decompose the statistic $T_n$ into an additive form,
    \begin{align*}
        T_n = \sum_{l\in\ZZ^d\cap\region_n} Z_n(l)
    \end{align*}
    where $Z_n(l)$ is some function of the general process $U$, that only depends on $U$ in the region $C(l)=l+(-1/2,1/2]^d$.
    In \cite{biscio2019general}, they consider more generality than this, but we do not require such generality.
    In particular, we can consider a fixed grid and need only consider the boxes of the form $l+(-1/2,1/2]^d$.
    So in the notation of \cite{biscio2019general}, we set $\eta=0$ and $R=0$ (see \cite{biscio2019general} for their definition).
    Our tapered Fourier transforms certainly can be written in this form, as we can simply set
    \begin{align*}
        J_{p;m,n}(\freq, l) = \int_{C(l)} h_{p;m,n}(x) e^{-2\pi i \ip{\freq}{x}} \randmeasurecenter{p}(\de x).
    \end{align*}
    and then
    \begin{align*}
        Z_n(l) = \ell(\region_n)^{1/2} \mathrm{vec}\; [\phi(J_{p;m,n}(\freq_j, l))]_{p\in[P], m\in [M], j \in [r]}.
    \end{align*}

    Now to satisfy $(\mathcal{H}2)$, since we set $\eta=0$, we just require that the process is $\alpha$-mixing.
    In other words, the first part of \cref{assumption:normality}.

    Now to satisfy $(\mathcal{H}3)$, we need to show a moment condition.
    In particular, we require that the exists some $\tau > 2d/\epsilon$ such that
    \begin{align*}
        \sup_{n\in\NN}\sup_{l\in\ZZ^d\cap\region_n} \EE{\abs{Z_n(l) - \EE{Z_n(l)}}^{2+\tau}} < \infty,
    \end{align*}
    where $\abs{\cdot}$ here is the Frobenius norm.
    Note that $\EE{Z_n(l)}=0$, because we assumed $\lambda_p$ is known.
    Furthermore, we have
    \begin{align*}
        \EE{\abs{Z_n(l) - \EE{Z_n(l)}}^{2+\tau}}
        &= \EE{\abs{Z_n(l)}^{2+\tau}} \\
         &= \ell(\region_n)^{1+\tau/2} \EE{\left(\sum_{p=1}^P\sum_{m=1}^M \sum_{j=1}^{r} \Re{J_{p;m,n}(\freq_j, l)}^2+\Im{J_{p;m,n}(\freq_j, l)}^2\right)^{1+\tau/2}} \\
        &= \ell(\region_n)^{1+\tau/2} \EE{\left(\sum_{p=1}^P\sum_{m=1}^M \sum_{j=1}^{r} \abs{J_{p;m,n}(\freq_j, l)}^2\right)^{1+\tau/2}}.
    \end{align*}
    Now define
    \begin{align*}
        Z_p &= \begin{cases}
            \randmeasurebase_p^0([0,1]^d) & \text{if }\randmeasure{p}\text{ corresponds to a point process,}\\
            Y_p(\Delta_p) & \text{if }\randmeasure{p}\text{ corresponds to a random field.}\\
        \end{cases}
    \end{align*}
    In the case of a point process we have
    \begin{align*}
        \abs{J_{p;m,n}(\freq_j, l)} &= \abs{\int_{C(l)} h_{p;m,n}(x) e^{-2\pi i \ip{\freq}{x}} \randmeasurecenter{p}(\de x)} \\
        &= \abs{\int_{C(l)} h_{m,n}(x) e^{-2\pi i \ip{\freq}{x}} \randmeasurecenter{p}(\de x)}\\
        &= \abs{\int_{C(l)} h_{m,n}(x) e^{-2\pi i \ip{\freq}{x}} \randmeasure{p}(\de x) - \lambda_p\int_{C(l)} h_{m,n}(x) e^{-2\pi i \ip{\freq}{x}}\de x}\\
        & \leq \norm{h_{m,n}}_\infty \left\{\randmeasure{p}(C(l)) + \lambda_p\right\}.\\
    \end{align*}
    because both $\randmeasure{p}(C(l))$ and $\lambda_p$ are non-negative and $\ell(C(l))=1$.
    In the case of a random field we have
    \begin{align*}
        \abs{J_{p;m,n}(\freq_j, l)} &= \abs{\int_{C(l)} h_{p;m,n}(x) e^{-2\pi i \ip{\freq}{x}} \randmeasurecenter{p}(\de x)} \\
        &= \left(\prod_{j=1}^d \Delta_{p;j}\right) \sum_{u \in \grid_p \cap C(l)} h_{m,n}(u) e^{-2\pi i \ip{\freq}{u}} [Y_p(u)-\lambda_p] \\
        & \leq \left(\prod_{j=1}^d \Delta_{p;j}\right) \norm{h_{m,n}}_\infty \sum_{u \in \grid_p \cap C(l)} \abs{Y_p(u)-\lambda_p}. \\
    \end{align*}
    W.l.o.g we can consider the case when at most one such point is present in $\grid_p \cap C(l)$, as one could always just rescale the grid $\ZZ^d$ by a constant (or rescale space).
    Clearly then the moments in question will be larger when there is a point (else $|Z_n(l)|$ is zero).
    Therefore, by stationarity we can choose to consider the variables around zero, so
    \begin{align*}
        \EE{\abs{Z_n(l) - \EE{Z_n(l)}}^{2+\tau}} 
        &\leq C \ell(\region_n)^{1+\tau/2} \max_{m\in[M]}\norm{h_{m,n}}_\infty^{2+\tau} \EE{\left(\sum_{p=1}^P\sum_{m=1}^M \sum_{j=1}^{r} \abs{Z_{p}-\EE{Z_p}}^2\right)^{1+\tau/2}}
    \end{align*}
    where the constant $C$ encodes nuisance constants like the products of $\Delta_p$, and
    \begin{align*}
        Z_p &= \begin{cases}
            \randmeasurebase_p([0,1]^d) & \text{if }\randmeasure{p}\text{ corresponds to a point process,}\\
            Y_p(0) & \text{if }\randmeasure{p}\text{ corresponds to a random field.}\\
        \end{cases}
    \end{align*}
    Therefore,
    \begin{align*}
        \sup_{n\in\NN}\sup_{l\in\ZZ^d\cap\region_n} & \EE{\abs{Z_n(l) - \EE{Z_n(l)}}^{2+\tau}} \\
        &\leq C \cdot \left(\sum_{p=1}^P\sum_{m=1}^M \sum_{j=1}^{r} \abs{Z_{p}-\EE{Z_p}}^2\right)^{1+\tau/2} \left[\max_{m\in[M]}\sup_{n\in\NN} \ell(\region_n) \norm{h_{m,n}}_\infty^{2}\right]^{1+\tau/2}
    \end{align*}
    which is finite by \cref{assumption:tapers:normality,assumption:normality}.

    Finally, need to show that $(\mathcal{H}4)$ holds. In particular, that
    \begin{align*}
        \liminf_{n\rightarrow\infty}\lambda_{\min} \left(\frac{\Sigma_n}{\abs{D_n}}\right) > 0
    \end{align*}
    where $\Sigma_n = \var{T_n}$ and $\lambda_{\min}$ is the smallest eigenvalue of a matrix.
    Now we have that
    \begin{align*}
        \Sigma_n = \ell(\region_n) S_n
    \end{align*}
    where $S_n$ is a matrix formed by appropriate combinations of $\EE{\pgram{p,q}{m,n}(\freq_j)}$ and zero.
    We have that for any $n\in\NN$
    \begin{align*}
        \lambda_{\min} \left(\frac{\Sigma_n}{\abs{D_n}}\right) = \frac{\ell(\region_n)}{\abs{D_n}}\lambda_{\min} \left(S_n\right).
    \end{align*}
    Now $\frac{\ell(\region_n)}{\abs{D_n}}$ is bounded by 1, and $\lambda_{\min} \left(S_n\right)$ converges, therefore
    \begin{align*}
        \liminf_{n\rightarrow\infty}\lambda_{\min} \left(\frac{\Sigma_n}{\abs{D_n}}\right)
        &= \left(\lim_{n\rightarrow} \lambda_{\min} \left(S_n\right)\right) \left(\liminf_{n\rightarrow\infty} \frac{\ell(\region_n)}{\abs{D_n}}\right).
    \end{align*}
    By \cref{assumption:normality} the second part is strictly positive.
    For the other part we see
    \begin{align*}
        \lim_{n\rightarrow\infty} \lambda_{\min} \left(S_n\right) = \lambda_{\min} \left(\lim_{n\rightarrow\infty} S_n\right).
    \end{align*}
    Now $\lim_{n\rightarrow\infty} S_n$ is 
    a block-diagonal matrix comprised of $\tilde{f}(\freq_j)$ by \cref{lemma:mean:dftcross}.
    
    Finally, if $f(\freq)$ is positive definite for all $\freq$, then $\tilde{f}(\freq)$ is also positive definite by \cref{lemma:posdef:alias}.
    Therefore, $\lambda_{\min} \left(\lim_{n\rightarrow\infty} S_n\right) > 0$ since block diagonal matrices have eigenvalues given by the union of the eigenvalues of their blocks.
    Therefore, from \cite{biscio2019general}
    \begin{align*}
         \ell(\region_n)^{-1/2} T_n \xrightarrow{d} \mathcal{N}(0, I_{2PMr})
    \end{align*}
    as $n\rightarrow\infty$. Meaning that
    \begin{align*}
        S_n^{-1/2} \mathrm{vec}\; [\phi(J_{p;m,n}(\freq_j))]_{p\in[P], m\in [M], j\in[r]} \xrightarrow{d} \mathcal{N}(0, I_{2PMr}).
    \end{align*}
    Writing $S$ to be the limit of $S_n$ (which is a block diagonal matrix with blocks $\tilde{f}(\freq_j)$), we have that
    \begin{align*}
        \mathrm{vec}\; [\phi(J_{p;m,n}(\freq_j))]_{p\in[P], m\in [M], j\in[r]} \xrightarrow{d} \mathcal{N}(0, S)
    \end{align*}
    as $n\rightarrow\infty$.
    This can then be appropriately converted back to complex notation and unpacked to give the desired result.
\end{proof}

\subsection{\texorpdfstring{\cref{res:general:rescale}}{Proposition}}\label{proof:res:general:rescale}
\begin{proof}[Proof of \cref{res:general:rescale}]
    Recall that the region is of the form
    \begin{align*}
        \region_n = \region \circ l_n
    \end{align*}
    for some template region $\region$ and we have set for $x\in\RRd$ and all $m,n\in\NN$
    \begin{align*}
        h_{m,n}(x) = \elemprod{l_n}^{-1/2} h_m\left(x\oslash l_n\right),
    \end{align*}
    where $\set{h_m}_{m\in\NN}$ is an orthonormal family of tapers on $\region$ that are continuous, orthonormal, and the bandwidth $b_n$ so that $b_n\rightarrow 0$ and
    \begin{align*}
        \min_{m\in[M_n]} \int_{\ball{b_n\circ l_n}} \abs{H_m(\freq)}^2 \de\freq &\rightarrow 1,\\
    \end{align*}
   
    Beginning with checking \cref{assumption:tapers:finitesample}, by construction $h_m$ is bounded and continuous, and therefore $h_{m,n}$ is also bounded and continuous for any $n\in\NN$, since we scale by a decreasing sequence. 
    Furthermore, if $x\notin\region_n$ then $x\oslash l_n \not\in\region$, and so $h_{m,n}(x)=0$.

    Next we have
    \begin{align*}
        \ipalt{h_{m,n}}{h_{m',n}} &= \int_{\RRd} h_{m,n}(x)\conj{h_{m',n}(x)} \de x \\
        &= \frac{1}{\elemprod{l_n}} \int_{\RRd} h_{m}(x \oslash l_n)\conj{h_{m'}(x \oslash l_n)} \de x \\
        &= \frac{\elemprod{l_n}}{\elemprod{l_n}} \int_{\RRd} h_{m}(u)\conj{h_{m'}(u)} \de u \\
        &= \ipalt{h_{m}}{h_{m'}}.
    \end{align*}
    Since $h_m$ is orthonormal, we have that $\ipalt{h_{m,n}}{h_{m',n}} = \delta_{m,m'}$.

    We also have that for all $n\in\NN$
    \begin{align*}
        \int_{\RRd} \abs{H_{m,n}(\freq)} \de \freq 
        &= \elemprod{l_n}^{1/2} \int_{\RRd} \abs{H_{m}(\freq \circ l_n)} \de \freq \\
        &= \elemprod{l_n}^{-1/2} \int_{\RRd} \abs{H_{m}(\freq')} \de \freq' \\
        &\leq C_m \elemprod{l_n}^{-1/2} \int_{\RRd} (1+\norm{x}_2)^{-d-\delta} \de \freq' \\
        &<\infty.
    \end{align*}

    As for \cref{assumption:tapers:concentration}, we need to prove that
    \begin{align*}
        \max_{m\in[M_n]} \abs{\int_{\ball{b_n}} H_{p;m,n}(\freq) \conj{H_{q;m,n}(\freq)} \de\freq-1} &\rightarrow 0, \\
        \max_{m\in[M_n]} \abs{\int_{K_{p}\setminus\ball{b_n}} \abs{H_{p;m,n}(\freq)}^2 \de\freq} &\rightarrow 0,
    \end{align*}
    as $n\rightarrow\infty$.

    Now we will need to utilise the aliasing relationship.
    In particular, we have that
    \begin{align*}
        H_{p;m,n}(\freq) &= \sum_{\aliasfreq\in\Aliasfreq_p} H_{m,n}(\freq+\aliasfreq) w_p(\aliasfreq)
    \end{align*}
    which holds given the conditions by \cref{lemma:alias:tapers}.
    
    Therefore, we have
    \begin{align*}
        & \int_{\ball{b_n}} H_{p;m,n}(\freq) \conj{H_{q;m,n}(\freq)} \de\freq \\
        &= \sum_{\aliasfreq_p\in\Aliasfreq_p} \sum_{\aliasfreq_q\in\Aliasfreq_q} w_p(\aliasfreq_p) \conj{w_q(\aliasfreq_q)} \int_{\ball{b_n}} H_{m,n}(\freq+\aliasfreq_p) \conj{H_{m,n}(\freq+\aliasfreq_q)} \de\freq \\
        &= \sum_{\aliasfreq_p\in\Aliasfreq_p\setminus\set{0}} \sum_{\aliasfreq_q\in\Aliasfreq_q\setminus\set{0}} w_p(\aliasfreq_p) \conj{w_q(\aliasfreq_q)} \int_{\ball{b_n}} H_{m,n}(\freq+\aliasfreq_p) \conj{H_{m,n}(\freq+\aliasfreq_q)} \de\freq \\
        &\hspace{2em} + \sum_{\aliasfreq_p\in\Aliasfreq_p\setminus\set{0}} w_p(\aliasfreq_p) \int_{\ball{b_n}} H_{m,n}(\freq+\aliasfreq_p) \conj{H_{m,n}(\freq)} \de\freq \\
        &\hspace{2em} + \sum_{\aliasfreq_q\in\Aliasfreq_q\setminus\set{0}} \conj{w_q(\aliasfreq_q)} \int_{\ball{b_n}} H_{m,n}(\freq) \conj{H_{m,n}(\freq+\aliasfreq_q)} \de\freq \\
        &\hspace{2em} + \int_{\ball{b_n}} \abs{H_{m,n}(\freq)}^2 \de\freq.
    \end{align*}
    Therefore
    \begin{align*}
        &\max_{m\in[M_n]} \abs{\int_{\ball{b_n}} H_{p;m,n}(\freq) \conj{H_{q;m,n}(\freq)} \de\freq-1} \\
        &\leq \max_{m\in[M_n]} \abs{\sum_{\aliasfreq_p\in\Aliasfreq_p\setminus\set{0}} \sum_{\aliasfreq_q\in\Aliasfreq_q\setminus\set{0}} w_p(\aliasfreq_p) \conj{w_q(\aliasfreq_q)} \int_{\ball{b_n}} H_{m,n}(\freq+\aliasfreq_p) \conj{H_{m,n}(\freq+\aliasfreq_q)} \de\freq} \\
        &\hspace{2em} + \max_{m\in[M_n]} \abs{\sum_{\aliasfreq_p\in\Aliasfreq_p\setminus\set{0}} w_p(\aliasfreq_p) \int_{\ball{b_n}} H_{m,n}(\freq+\aliasfreq_p) \conj{H_{m,n}(\freq)} \de\freq} \\
        &\hspace{2em} + \max_{m\in[M_n]} \abs{\sum_{\aliasfreq_q\in\Aliasfreq_q\setminus\set{0}} \conj{w_q(\aliasfreq_q)} \int_{\ball{b_n}} H_{m,n}(\freq) \conj{H_{m,n}(\freq+\aliasfreq_q)} \de\freq} \\
        &\hspace{2em} + \max_{m\in[M_n]} \abs{\int_{\ball{b_n}} \abs{H_{m,n}(\freq)}^2 \de\freq - 1}.
    \end{align*}
    Note that
    \begin{align*}
        0\leq \int_{\ball{b_n}} \abs{H_{m,n}(\freq)}^2 \de\freq \leq \int_{\RRd} \abs{H_{m,n}(\freq)}^2 \de\freq = 1
    \end{align*}
    and therefore
    \begin{align*}
        \max_{m\in[M_n]} \abs{\int_{\ball{b_n}} \abs{H_{m,n}(\freq)}^2 \de\freq-1}
        &= 1 - \min_{m\in[M_n]} \int_{\ball{b_n}} \abs{H_{m,n}(\freq)}^2 \de\freq \\
        &= 1 - \min_{m\in[M_n]}{\elemprod{l_n}} \int_{B_{b_n}} \abs{H_{m}(\freq \circ l_n)}^2 \de\freq \\
        &= 1 - \min_{m\in[M_n]}\frac{\elemprod{l_n}}{\elemprod{l_n}} \int_{B_{b_n}\circ l_n} \abs{H_{m}(\freq')}^2 \de\freq' \\
        &= 1 - \min_{m\in[M_n]} \int_{B_{b_n}\circ l_n} \abs{H_{m}(\freq)}^2 \de\freq \\
        &\rightarrow 0
    \end{align*}
    because by assumption, we have that $\min_{m\in[M_n]} \int_{B_{b_n}\circ l_n} \abs{H_{m}(\freq)}^2 \de\freq \rightarrow 1$ as $n\rightarrow\infty$.

    Now recall that 
    \begin{align*}
        \abs{H_m(\freq)} \leq \frac{C_m}{\left(1 + \min_{x\in\ball{b_n}}\norm{\freq}_2\right)^{d+\delta}}
    \end{align*}
    and therefore
    \begin{align*}
        \abs{H_{m,n}(\freq)} \leq \frac{C_m \elemprod{l_n}^{1/2}}{\left(1 + \min_{x\in\ball{b_n}}\norm{\freq\circ l_n}_2\right)^{d+\delta}}.
    \end{align*}

    Now consider the first term. In particular, we have
    \begin{align*}
        &\abs{\int_{\ball{b_n}} H_{m,n}(\freq+\aliasfreq_p) \conj{H_{m,n}(\freq+\aliasfreq_q)} \de\freq} \\
        &\leq \int_{\ball{b_n}} \abs{H_{m,n}(\freq+\aliasfreq_p) \conj{H_{m,n}(\freq+\aliasfreq_q)}} \de\freq \\
        &\leq \ell(\ball{b_n}) \frac{C_m^2 \elemprod{l_n}}{\left(1 + \min_{x\in\ball{b_n}}\norm{(\aliasfreq_q+x)\circ l_n}_2\right)^{d+\delta}\left(1 + \min_{x\in\ball{b_n}}\norm{(\aliasfreq_q+x)\circ l_n}_2\right)^{d+\delta}} \\
        &\leq \frac{\ell(\ball{b_n})\elemprod{l_n}}{\min l_n^{2(d+\delta)}} \frac{C_m^2}{\left(1/\min l_n + \min_{x\in\ball{b_n}}\norm{\aliasfreq_p+x}_2\right)^{d+\delta}\left(1/\min l_n + \min_{x\in\ball{b_n}}\norm{\aliasfreq_q+x}_2\right)^{d+\delta}}
    \end{align*}
    therefore
    \begin{align*}
        & \max_{m\in[M_n]} \abs{\sum_{\aliasfreq_p\in\Aliasfreq_p\setminus\set{0}} \sum_{\aliasfreq_q\in\Aliasfreq_q\setminus\set{0}} w_p(\aliasfreq_p) \conj{w_q(\aliasfreq_q)} \int_{\ball{b_n}} H_{m,n}(\freq+\aliasfreq_p) \conj{H_{m,n}(\freq+\aliasfreq_q)} \de\freq} \\
        &\leq \sum_{\aliasfreq_p\in\Aliasfreq_p\setminus\set{0}} \sum_{\aliasfreq_q\in\Aliasfreq_q\setminus\set{0}} \max_{m\in[M_n]} \abs{\int_{\ball{b_n}} H_{m,n}(\freq+\aliasfreq_p) \conj{H_{m,n}(\freq+\aliasfreq_q)} \de\freq} \\
        &\leq \sum_{\aliasfreq_p\in\Aliasfreq_p\setminus\set{0}} \sum_{\aliasfreq_q\in\Aliasfreq_q\setminus\set{0}} \max_{m\in[M_n]} \frac{\ell(\ball{b_n})}{\min l_n^2} \frac{C_m^2}{\left(1/\min l_n + \min_{x\in\ball{b_n}}\norm{\aliasfreq_p+x}_2\right)^{d+\delta}\left(1/\min l_n + \min_{x\in\ball{b_n}}\norm{\aliasfreq_q+x}_2\right)^{d+\delta}} \\
        &\leq { \max_{m\in[M_n]} C_m^2 } \frac{\ell(\ball{b_n})\elemprod{l_n}}{\min l_n^{2(d+\delta)}} \sum_{\aliasfreq_p\in\Aliasfreq_p\setminus\set{0}} \frac{1}{\left(\min_{x\in\ball{b_n}}\norm{\aliasfreq_p+x}_2\right)^{d+\delta}} \sum_{\aliasfreq_q\in\Aliasfreq_q\setminus\set{0}} \frac{1}{\left(\min_{x\in\ball{b_n}}\norm{\aliasfreq_p+x}_2\right)^{d+\delta}} \\
        &\rightarrow 0
    \end{align*}
    as $n\rightarrow\infty$ because by assumption $\ell(\ball{b_n})\rightarrow 0$ and $\max_{m\in[M_n]}C_m \elemprod{l_n}^{1/2}(\min l_n)^{-d-\delta}\rightarrow 0$  as $n\rightarrow\infty$.
    The second and third terms are symmetric, so we treat only one of them.
    In particular, note that
    \begin{align*}
        \int_{\ball{b_n}} H_{m,n}(\freq) \conj{H_{m,n}(\freq+\aliasfreq_q)} \de\freq
        &\leq \left(\int_{\ball{b_n}} \abs{H_{m,n}(\freq)}^2 \de\freq\right)^{1/2} \left(\int_{\ball{b_n}} \abs{H_{m,n}(\freq+\aliasfreq_q)}^2 \de\freq\right)^{1/2} \\
        &\leq \left(\int_{\ball{b_n}} \abs{H_{m,n}(\freq+\aliasfreq_q)}^2 \de\freq\right)^{1/2} \\
        &\leq \frac{\ell(\ball{b_n})^{1/2}\elemprod{l_n}^{1/2}}{\min l_n^{d+\delta}} \frac{C_m}{\left(1/\min l_n + \min_{x\in\ball{b_n}}\norm{\aliasfreq_q+x}_2\right)^{d+\delta}}
    \end{align*}
    by similar arguments as above.
    Therefore
    \begin{align*}
        &\max_{m\in[M_n]} \abs{\sum_{\aliasfreq_p\in\Aliasfreq_p\setminus\set{0}} w_p(\aliasfreq_p) \int_{\ball{b_n}} H_{m,n}(\freq+\aliasfreq_p) \conj{H_{m,n}(\freq)} \de\freq} \\
        &\leq \sum_{\aliasfreq_p\in\Aliasfreq_p\setminus\set{0}} \max_{m\in[M_n]} \abs{\int_{\ball{b_n}} H_{m,n}(\freq+\aliasfreq_p) \conj{H_{m,n}(\freq)} \de\freq} \\
        &\leq \max_{m\in[M_n]} C_m\frac{\ell(\ball{b_n})^{1/2}\elemprod{l_n}^{1/2}}{\min l_n^{d+\delta}}\sum_{\aliasfreq_p\in\Aliasfreq_p\setminus\set{0}}  \frac{1}{\left(\min_{x\in\ball{b_n}}\norm{\aliasfreq_p+x}_2\right)^{d+\delta}} \\
        &\rightarrow 0
    \end{align*}
    as $n\rightarrow\infty$ by the same arguments as above.
    This completes the proof.

    The second condition, namely
    \begin{align*}
        \max_{m\in[M_n]} \abs{\int_{K_{p}\setminus\ball{b_n}} \abs{H_{p;m,n}(\freq)}^2 \de\freq} \rightarrow 0
    \end{align*}
    as $n\rightarrow \infty$ follows the exact same argument as above (as we do not use the fact that $\ell{(\ball{b_n})}\rightarrow 0$) with the only difference being the term where there is no aliasing.
    In particular, we have
    \begin{align*}
        \max_{m\in[M_n]} \abs{\int_{K_{p}\setminus\ball{b_n}} \abs{H_{p;m,n}(\freq)}^2 \de\freq}
        &\leq \max_{m\in[M_n]} \abs{\sum_{\aliasfreq\in\Aliasfreq_p\setminus\set{0}} \sum_{\aliasfreq'\in\Aliasfreq_p\setminus\set{0}} \hspace{-1em} w_p(\aliasfreq) \conj{w_p(\aliasfreq')} \int_{K_{p}\setminus\ball{b_n}} H_{m,n}(\freq+\aliasfreq) \conj{H_{m,n}(\freq+\aliasfreq')} \de\freq} \\
        &\hspace{2em} + 2\max_{m\in[M_n]} \abs{\sum_{\aliasfreq_p\in\Aliasfreq_p\setminus\set{0}} w_p(\aliasfreq_p) \int_{K_{p}\setminus\ball{b_n}} H_{m,n}(\freq+\aliasfreq_p) \conj{H_{m,n}(\freq)} \de\freq} \\
        &\hspace{2em} + \max_{m\in[M_n]} \abs{\int_{K_{p}\setminus\ball{b_n}} \abs{H_{m,n}(\freq)}^2 \de\freq}.
    \end{align*}

    The first two terms are handled in the same way as above. The last term
    \begin{align*}
        \max_{m\in[M_n]} \abs{\int_{K_{p}\setminus\ball{b_n}} \abs{H_{m,n}(\freq)}^2 \de\freq}
        & \leq \max_{m\in[M_n]} \abs{\int_{K_{p}} \abs{H_{m,n}(\freq)}^2 \de\freq - 1} + \max_{m\in[M_n]} \abs{1-\int_{\ball{b_n}} \abs{H_{m,n}(\freq)}^2 \de\freq} \\
        & \leq \max_{m\in[M_n]} \abs{\int_{K_{p} \circ l_n} \abs{H_{m}(\freq)}^2 \de\freq - 1} + \max_{m\in[M_n]} \abs{1-\int_{\ball{b_n} \circ l_n} \abs{H_{m}(\freq)}^2 \de\freq} \\
        &\leq 2 \left(1-\min_{m\in[M_n]} \int_{\ball{b_n}\circ l_n} \abs{H_{m}(\freq)}^2 \de\freq\right) \\
        &\rightarrow 0
    \end{align*}
    as $n\rightarrow\infty$ by assumption.

    Moving on to \cref{assumption:tapers:orthogonality}, we firstly have $\ipalt{h_{m,n}}{h_{m',n}}=0$ for any $m\neq m'$ by assumption.
    Then by a similar argument to the above, again have
    \begin{align*}
        &\hspace{-2em} \max_{m\in[M_n]} \max_{m'\in[M_n]\setminus \set{m}} \abs{\int_{\ball{b_n}} H_{p;m,n}(\freq) \conj{H_{q;m',n}(\freq)} \de\freq} \\
        &\leq \max_{m\in[M_n]} \max_{m'\in[M_n]\setminus \set{m}} \abs{\sum_{\aliasfreq_p\in\Aliasfreq_p} \sum_{\aliasfreq_q\in\Aliasfreq_q} w_p(\aliasfreq_p) \conj{w_q(\aliasfreq_q)} \int_{\ball{b_n}} H_{m,n}(\freq+\aliasfreq_p) \conj{H_{m',n}(\freq+\aliasfreq_q)} \de\freq} \\
        &\hspace{2em} + \max_{m\in[M_n]} \max_{m'\in[M_n]\setminus \set{m}} \abs{\sum_{\aliasfreq_p\in\Aliasfreq_p} w_p(\aliasfreq_p) \int_{\ball{b_n}} H_{m,n}(\freq+\aliasfreq_p) \conj{H_{m',n}(\freq)} \de\freq} \\
        &\hspace{2em} + \max_{m\in[M_n]} \max_{m'\in[M_n]\setminus \set{m}} \abs{\sum_{\aliasfreq_q\in\Aliasfreq_q} \conj{w_q(\aliasfreq_q)} \int_{\ball{b_n}} H_{m,n}(\freq) \conj{H_{m',n}(\freq+\aliasfreq_q)} \de\freq} \\
        &\hspace{2em} + \max_{m\in[M_n]} \max_{m'\in[M_n]\setminus \set{m}} \abs{\int_{\ball{b_n}} H_{m,n}(\freq) \conj{H_{m',n}(\freq)} \de\freq}.
    \end{align*}
    By the same arguments used above we need only consider the last term.
    In particular, we have (because $H_{m},H_{m'}$ is orthogonal when $m \neq m'$)
    \begin{align*}
        &\hspace{-2em} \max_{m\in[M_n]} \max_{m'\in[M_n]\setminus \set{m}} \abs{\int_{\ball{b_n}} H_{m,n}(\freq) \conj{H_{m',n}(\freq)} \de\freq} \\
        &= \max_{m\in[M_n]} \max_{m'\in[M_n]\setminus \set{m}} \abs{\int_{\ball{b_n}\circ l_n} H_{m}(\freq) \conj{H_{m'}(\freq)} \de\freq} \\
        &= \max_{m\in[M_n]} \max_{m'\in[M_n]\setminus \set{m}} \abs{\int_{\RRd \setminus \ball{b_n}\circ l_n} H_{m}(\freq) \conj{H_{m'}(\freq)} \de\freq - \int_{\RRd} H_{m}(\freq) \conj{H_{m'}(\freq)} \de\freq} \\
        &\leq \max_{m\in[M_n]} \max_{m'\in[M_n]\setminus \set{m}} \left(\int_{\RRd \setminus  \ball{b_n}\circ l_n} \abs{H_{m}(\freq)}^2 \int_{\RRd \setminus \ball{b_n}\circ l_n}\abs{H_{m'}(\freq)}^2 \de\freq\right)^{1/2} \\
        &\leq \max_{m\in[M_n]} \int_{\RRd \setminus  \ball{b_n}\circ l_n} \abs{H_{m}(\freq)}^2 \\
        &\rightarrow 0
    \end{align*}
    as $n\rightarrow\infty$ as already established.

    Now for \cref{assumption:tapers:consistency}.
    In particular, we need to show that
    \begin{align*}
        \max_{m\in[M_n]}{\int_{K_p} \abs{H_{p;m,n}(\freq)} \de\freq} &\rightarrow 0, \\
        \max_{m\in[M_n]} \sum_{\aliasfreq \in\Aliasfreq_{p}} \left(\int_{K_p+\aliasfreq} \abs{H_{m,n}(\freq)}^{2} \de\freq\right)^{1/2} &\rightarrow C,
    \end{align*}
    as $n\rightarrow\infty$.

    Starting with the first condition, we have
    \begin{align*}
        \int_{K_p} \abs{H_{p;m,n}(\freq)} \de\freq 
        &= \int_{K_p} \abs{\sum_{\aliasfreq\in\Aliasfreq_p} H_{m,n}(\freq) w_p(\aliasfreq)} \de\freq \\
        &\leq \sum_{\aliasfreq\in\Aliasfreq_p} \int_{K_p} \abs{H_{m,n}(\freq+\aliasfreq)} \de\freq \\ 
        &= \int_\RRd \abs{H_{m,n}(\freq)} \de\freq \\
        &= \elemprod{l_n}^{1/2} \int_\RRd \abs{H_{m}(\freq \circ l_n)} \de\freq \\
        &= \elemprod{l_n}^{-1/2} \int_\RRd \abs{H_{m}(\freq')} \de\freq' \\
        &= \elemprod{l_n}^{-1/2} \norm{H_{m}}_1 \\
        &\leq \elemprod{l_n}^{-1/2} C_m C
    \end{align*}
    where $C = \int_\RRd (1+\norm{x}_2)^{-d-\delta} \de x < \infty$.
    Therefore the result holds because by assumption
    \begin{align*}
        \max_{m\in[M_n]} C_m \elemprod{l_n}^{-1/2} \rightarrow 0
    \end{align*}
    as $n\rightarrow\infty$.
    For the second condition, we have
    \begin{align*}
        \max_{m\in[M_n]} \sum_{\aliasfreq \in\Aliasfreq_{p}} \left(\int_{K_p+\aliasfreq} \abs{H_{m,n}(\freq)}^{2} \de\freq\right)^{1/2}
        &= \max_{m\in[M_n]}\left(\int_{K_p} \abs{H_{m,n}(\freq)}^{2} \de\freq\right)^{1/2}  + \sum_{\aliasfreq \in\Aliasfreq_{p}\setminus\set{0}} \left(\int_{K_p+\aliasfreq} \abs{H_{m,n}(\freq)}^{2} \de\freq\right)^{1/2} \\
        &= 1 + \max_{m\in[M_n]} \sum_{\aliasfreq \in\Aliasfreq_{p}\setminus\set{0}} \left(\int_{K_p+\aliasfreq} \abs{H_{m,n}(\freq)}^{2} \de\freq\right)^{1/2} \\
        &\leq 1 + \max_{m\in[M_n]} C_m \elemprod{l_n}^{1/2} \ell(K_p)^{1/2} \sum_{\aliasfreq \in\Aliasfreq_{p}\setminus\set{0}} \max_{\freq\in K_p+\aliasfreq} (1+\norm{\freq\circ l_n})^{-d-\delta} \\
        &\leq 1 + \max_{m\in[M_n]} C_m \elemprod{l_n}^{1/2} \ell(K_p)^{1/2} \min(l_n)^{-d-\delta} \sum_{\aliasfreq \in\Aliasfreq_{p}\setminus\set{0}} \max_{\freq\in K_p+\aliasfreq} (\norm{\aliasfreq})^{-d-\delta} \\
        &\rightarrow 1
    \end{align*}
    as $n\rightarrow\infty$, by similar arguments as above around pulling out $\min(l_n)$.

    Finally to establish \cref{assumption:tapers:normality}.
    For any fixed $m$, for any $n\in\NN$ such that $m\in[M_n]$,
    \begin{align*}
        \ell(\region_n) \norm{h_{m,n}}_\infty^2 = \ell(\region) \elemprod{l_n} \frac{1}{\elemprod{l_n}}\norm{h_m}_\infty^2 = \ell(\region) \norm{h_m}_\infty^2 < \infty.
    \end{align*}

    Finally consider $\ell(\region_n)\abs{D_n}^{-1}$.
    Note that the region $\region$ is bounded, and therefore contained in a bounding box $\region\subseteq B=\prod_{j=1}^d[a_j,b_j]$.
    Therefore,
    \begin{align*}
        D_n 
        &\subseteq \set{x \in \ZZ^d : B\circ l_n \cap x+[-1/2,1/2]^d \neq \emptyset}\\
        &\subseteq \set{x \in \ZZ^d : l_{n;j}a_j-1 < x_j < l_{n;j}b_j+1}
    \end{align*}
    where $l_{n;j}$ denotes the $j$\Th element of $l_n$.
    Therefore we have that $\abs{D_n} \leq \prod_{j=1}^d (l_{n;j}(b_j-a_j)+2)$.
    So
    \begin{align*}
        \frac{\ell(\region_n)}{\abs{D_n}}
        &\geq \frac{\ell(\region)}{\prod_{j=1}^d (l_{n;j}(b_j-a_j)+2)} \elemprod{l_n} \\
        &= \frac{\ell(\region)}{\prod_{j=1}^d ((b_j-a_j)+2/l_{n;j})} \\
        &\geq \frac{\ell(\region)}{\prod_{j=1}^d ((b_j-a_j)+2/l_{1;j})} \\
        &> 0,
    \end{align*}
    and so $\liminf_{n\rightarrow\infty}\ell(\region_n) \abs{D_n}^{-1}>0$, completing the proof.
\end{proof}

\subsection{\texorpdfstring{\cref{res:rect:outerprod}}{Proposition}}\label{proof:res:rect:outerprod}

\begin{proof}[Proof of \cref{res:rect:outerprod}]
    In order to verify the assumptions, we will reuse the results from \cref{res:general:rescale}, with appropriate modifications.
    Note that the assumptions on the decay of the tapers in \cref{res:general:rescale} are no longer satistified by these tapers, so parts of the proof requiring these need modification.
    
    Recall that we are interested in the case $\region=\prod_{j=1}^d [a_j,b_j]$, where
    \begin{align*}
        h_{m}(x) = \prod_{j=1}^d \frac{1}{\sqrt{b_j-a_j}}g_{\gamma(m)_{j}}\left(\tfrac{x_j-a_j}{b_j-a_j}\right).
    \end{align*}
    
    Starting with \cref{assumption:tapers:finitesample}, clearly the tapers $h_m$ are bounded and continuous.
    Now if $x \notin \region$, then $(x-a)\oslash(b-a)\notin [0,1]^d$. 
    Therefore by the assumption that $g_m(y)$ is zero when $y\notin[0,1]$, we have $h_m(x)=0$.
    Next we have that for $m,m'\in\NN$
    \begin{align*}
        \ipalt{h_m}{h_{m'}}
        &= \prod_{j=1}^d \frac{1}{b_j-a_j} \int_{\RR} g_{\gamma(m)_j}\left(\frac{x_j-a_j}{b_j-a_j}\right)g_{\gamma(m')_j}\left(\frac{x_j-a_j}{b_j-a_j}\right) \de x_j \\
        &= \prod_{j=1}^d \int_{\RR} g_{\gamma(m)_j}(y)g_{\gamma(m')_j}(y) \de y \\
        &= \prod_{j=1}^d \ipalt{g_{\gamma(m)_j}}{g_{\gamma(m')_j}}.
    \end{align*}
    This is equal to 1 if $\gamma(m)=\gamma(m')$, and zero otherwise.
    Since $\gamma(m)=\gamma(m')$ if and only if $m=m'$, this completes the proof of orthonormality.
    Next we have
    \begin{align*}
        H_m(\freq) &= \prod_{j=1}^d \frac{1}{\sqrt{b_j-a_j}} \int_{\RR} g_{\gamma(m)_j}\left(\frac{x_j-a_j}{b_j-a_j}\right) e^{-2\pi i \freq_j x_j} \de x_j \\
        &= \prod_{j=1}^d \sqrt{b_j-a_j} \int_{\RR} g_{\gamma(m)_j}(y) e^{-2\pi i \freq_j (y(b_j-a_j)+a_j)} \de y \\
        &= \prod_{j=1}^d \sqrt{b_j-a_j} G_{\gamma(m)_j}(\freq_j[b_j-a_j]) e^{-2\pi i \freq_j a_j}
    \end{align*}
    and so
    \begin{align*}
        \norm{H_m}_1 
        &= \int_{\RRd} \abs{H_m(\freq)} \de \freq \\
        &= \prod_{j=1}^d {\sqrt{b_j-a_j}} \int_{\RR} \abs{G_{\gamma(m)_j}\left(\freq_j[b_j-a_j]\right)}\de\freq_j \\
        &= \prod_{j=1}^d \frac{1}{\sqrt{b_j-a_j}} \int_{\RR} \abs{G_{\gamma(m)_j}\left(\freq_j\right)}\de\freq_j \\
        &\leq  \left(\int_{\RR} (1+\abs{x})^{-1-\delta} \de x \right)^d \prod_{j=1}^d C_{\gamma(m)_j} \frac{1}{\sqrt{b_j-a_j}} \\
        &< \infty.
    \end{align*}

    Now we need to verify the asymptotic assumptions (\cref{assumption:tapers:concentration,assumption:tapers:orthogonality,assumption:tapers:consistency}).
    In particular, for any $p\in[P]$, and any $n,m$ we can always write
    \begin{align*}
        H_{p;m,n}(\freq) = \prod_{j=1}^d \tilde{G}_{p;m,j,n}(\freq_j)
    \end{align*}
    where $\freq_j$ is the $j$\Th element of $\freq$, and $\tilde{G}_{p;m,j,n}(\freq_j)$ is Fourier transform of the taper $\tilde{g}_{p;m,j,n}(\freq_j)$, which is either grid sampled on not depending on $p$, from
    \begin{align*}
        \tilde{g}_{m,j,n}(u) &= g_{\gamma(m)_j}(u/l_{n,j}) / l_{n,j}^{1/2}
    \end{align*}
    where $l_{n,j}$ is the $j$\Th element of $l_n$.

    We need the second condition of \cref{assumption:tapers:concentration}, in order to prove the first, and so we start with that.
    For the second condition of \cref{assumption:tapers:concentration}, the nuance is that the integral is over the non-separable set $K_p\setminus\ball{b_n}$, even though the function is separable.
    However, noting that $[-b_n/\sqrt{d},b_n/\sqrt{d}]^d \subset \ball{b_n}$, we can bound the desired integral as
    \begin{align*}
        \int_{K_p\setminus\ball{b_n}} \abs{H_{p;m,n}(\freq)}^2 \de\freq
        &\leq \int_{K_p\setminus \left[-b_n/\sqrt{d},b_n/\sqrt{d}\right]^d} \abs{H_{p;m,n}(\freq)}^2 \de\freq \\
        &= \prod_{j=1}^d \int_{-b_n/\sqrt{d}}^{b_n/\sqrt{d}} \abs{\tilde{G}_{p;m,j,n}(\freq_j)}^2 \de\freq_j
    \end{align*}
    and therefore
    \begin{align*}
        \max_{m\in[M_n]} \abs{\int_{K_p\setminus\ball{b_n}} \abs{H_{p;m,n}(\freq)}^2 \de\freq}
        &\leq \prod_{j=1}^d \max_{m\in[M_{n,j}]} \int_{-b_n/\sqrt{d}}^{b_n/\sqrt{d}} \abs{G_{p;m,n}(\freq_j)}^2 \de\freq_j
    \end{align*}
    where we recall ${M}_{n,j} = \max_{m\in[M_n]}{\gamma(m)_j}$.\footnote{Note here we use $G_{p;m,n}$, which is the Fourier transform of the taper made on the $n$\Th region with the $p$\Th grid, but specifically using the $m$\Th taper (not the $\gamma(m)_j$\Th taper).}
    This upper bound holds because $M_{n,j}$ is such that we include all utilised one-dimensional tapers in the $j$\Th dimension (plus potentially some extras, meaning we have an upper bound).

    Now we can apply the results of \cref{res:general:rescale}, with one-dimensional tapers.
    In particular, we apply \cref{res:general:rescale} to the tapers $\set{g_{m}}_{m\in\NN}$ with $M_n' = {M}_{n,j}$ and $l_n'=l_{n,j}$ and $b_n' = b_n/\sqrt{d}$.
    Note because this is one dimensional, the condition
    $\max_{m\in [M_n']} C_m \elemprod{l_n'}^{1/2}  (\min l_n')^{-d-\delta} \rightarrow 0$ reduces to
    $\max_{m\in[M_{n,j}]} C_m l_{n,j}^{-1/2-\delta} \rightarrow 0$ which is already satisfied by assumption as $l_{n,j}^{-1/2-\delta} \leq l_{n,j}^{-1/2}$ and we assume $\max_{m\in[M_{n,j}]} C_m l_{n,j}^{-1/2} \rightarrow 0$ as $n\rightarrow\infty$.
    Therefore, we have
    \begin{align*}
        \max_{m\in[M_{n,j}]} \int_{-b_n/\sqrt{d}}^{b_n/\sqrt{d}} \abs{G_{p;m,n}(\freq_j)}^2 \de\freq_j
        \rightarrow 0
    \end{align*}
    as $n\rightarrow\infty$ for each $j\in[d]$.

    For this first condition of \cref{assumption:tapers:concentration}, we need to show that
    \begin{align*}
        \max_{m\in[M_n]} \abs{\int_{\ball{b_n}} H_{p;m,n}(\freq) \conj{H_{q;m,n}} \de\freq - 1} \rightarrow 0.
    \end{align*}
    Firstly, to make this separable
    \begin{align*}
        \max_{m\in[M_n]} \abs{\int_{\ball{b_n}} H_{p;m,n}(\freq) \conj{H_{q;m,n}} \de\freq - 1}
        &\leq \max_{m\in[M_n]} \abs{\int_{\left[-b_n/\sqrt{d},b_n/\sqrt{d}\right]^d} H_{p;m,n}(\freq) \conj{H_{q;m,n}} \de\freq - 1} \\
        &+ \max_{m\in[M_n]} \abs{\int_{\ball{b_n}\setminus\left[-b_n/\sqrt{d},b_n/\sqrt{d}\right]^d} H_{p;m,n}(\freq) \conj{H_{q;m,n}} \de\freq}.
    \end{align*}
    Considering the second of these terms
    \begin{align*}
        &\max_{m\in[M_n]} \abs{\int_{\ball{b_n}\setminus\left[-b_n/\sqrt{d},b_n/\sqrt{d}\right]^d} H_{p;m,n}(\freq) \conj{H_{q;m,n}} \de\freq} \\
        &\leq \max_{m\in[M_n]} \left(\int_{\ball{b_n}\setminus\left[-b_n/\sqrt{d},b_n/\sqrt{d}\right]^d} \abs{H_{p;m,n}(\freq)}^2 \de\freq \int_{\ball{b_n}\setminus\left[-b_n/\sqrt{d},b_n/\sqrt{d}\right]^d} \abs{H_{q;m,n}(\freq)}^2 \de\freq\right)^{1/2} \\
        &\leq \max_{m\in[M_n]} \left(\int_{K_{p}\setminus\left[-b_n/\sqrt{d},b_n/\sqrt{d}\right]^d} \abs{H_{p;m,n}(\freq)}^2 \de\freq \int_{K_{q}\setminus\left[-b_n/\sqrt{d},b_n/\sqrt{d}\right]^d} \abs{H_{q;m,n}(\freq)}^2 \de\freq\right)^{1/2} \\
        &\rightarrow 0
    \end{align*}
    as we just established.

    Finally then, note that for any $a_1,\ldots,a_d\in\CC$, and writing for convenience $a_0=1$
    \begin{align*}
        \abs{\prod_{j=1}^d a_j-1} \leq \sum_{j=1}^d \abs{a_j-1}\prod_{j'=0}^{j-1} a_{j'}
    \end{align*} 
    which follows from induction and the relation
    \begin{align*}
        \abs{ab-1} \leq \abs{a-1}\abs{b} + \abs{b-1}
    \end{align*}
    which itself follows from the triangle inequality.

    Setting $a_{m,j} = \int_{-b_n/\sqrt{d}}^{b_n/\sqrt{d}} \tilde{G}_{p;m,j,n}(\freq_j)\conj{\tilde{G}_{q;m,j,n}(\freq_j)} \de\freq_j$, and $a_{m,0}=1$ we see that 
    \begin{align*}
        \max_{m\in[M_n]} \abs{\int_{\left[-b_n/\sqrt{d},b_n/\sqrt{d}\right]^d} H_{p;m,n}(\freq) \conj{H_{q;m,n}} \de\freq - 1}
        &\leq \max_{m\in[M_n]} \sum_{j=1}^d \abs{a_{m,j}-1}\prod_{j'=0}^{j-1} a_{m,j'} \\
        &\leq \sum_{j=1}^d \max_{m\in[M_n]} \abs{a_{m,j}-1} \prod_{j'=0}^{j-1} \max_{m\in[M_n]} \abs{a_{m,j'}} 
    \end{align*}
    so provided
    \begin{align}
        \max_{m \in [M_{n}]} \abs{a_{m,j}-1} \rightarrow 0 \label{eq:proof:res:rect:outerprod:makingseparable}
    \end{align}
    as $n\rightarrow \infty$ the result follows, because \cref{eq:proof:res:rect:outerprod:makingseparable} implies
    \begin{align*}
        \max_{m \in [M_{n}]} \abs{a_{m,j}} \leq \max_{m \in [M_{n}]} \abs{a_{m,j}-1} + 1 \rightarrow 1
    \end{align*}
    as $n\rightarrow \infty$, meaning we can use algebra of limits.
    \Cref{eq:proof:res:rect:outerprod:makingseparable} holds because
    \begin{align*}
        \max_{m \in [M_{n}]} \abs{a_{m,j}-1} \leq \max_{m\in[M_{n,j}]} \abs{\int_{-b_n/\sqrt{d}}^{b_n/\sqrt{d}} G_{p;m,n}(\freq_j)\conj{G_{q;m,n}(\freq_j)} \de\freq_j-1} \rightarrow 0
    \end{align*}
    where we apply \cref{res:general:rescale} in one dimension to obtain the result as above.
    
    Now for \cref{assumption:tapers:orthogonality}, we have again
    \begin{align*}
        &\hspace{-2em} \max_{m\in[M_n]} \max_{m'\in[M_n]\setminus\set{m}} \abs{\int_{\ball{b_n}} H_{p;m,n}(\freq) \conj{H_{q;m',n}} \de\freq} \\
        &\leq \max_{m\in[M_n]} \max_{m'\in[M_n]\setminus\set{m}} \abs{\int_{\left[-b_n/\sqrt{d},b_n/\sqrt{d}\right]^d} H_{p;m,n}(\freq) \conj{H_{q;m',n}} \de\freq} \\
        &+ \max_{m\in[M_n]} \max_{m'\in[M_n]\setminus\set{m}} \abs{\int_{\ball{b_n}\setminus\left[-b_n/\sqrt{d},b_n/\sqrt{d}\right]^d} H_{p;m,n}(\freq) \conj{H_{q;m',n}} \de\freq}.
    \end{align*}
    The second term converges to zero as we have already established.
    The first term is now separable, and so we again apply \cref{res:general:rescale} in one dimension.

    Finally for assumption \cref{assumption:tapers:consistency}, we note for the first part we have separability because
    \begin{align*}
        \max_{m\in[M_n]} \int_{K_p} \abs{H_{p;m,n}(\freq)} \de\freq
        &= \max_{m\in[M_n]} \prod_{j=1}^d \int_{-b_n/\sqrt{d}}^{b_n/\sqrt{d}} \abs{\tilde{G}_{p;m,j,n}(\freq_j)} \de\freq_j \\
        &\leq \prod_{j=1}^d \max_{m\in[M_{n,j}]} \int_{-b_n/\sqrt{d}}^{b_n/\sqrt{d}} \abs{\tilde{G}_{p;m,j,n}(\freq_j)} \de\freq_j,
    \end{align*}
    so again we apply \cref{res:general:rescale} in one dimension.
    Then for the last condition, writing $\Aliasfreq_p = \prod_{j=1}^{d} \Aliasfreq_{p;j}$ and $K_p = \prod_{j=1}^{d} K_{p;j}$, we have
    \begin{align*}
        \max_{m\in[M_n]} \sum_{\aliasfreq \in\Aliasfreq_{p}} \left(\int_{K_p+\aliasfreq} \abs{H_{p;m,n}(\freq)}^{2} \de\freq\right)^{1/2}
        &= \max_{m\in[M_n]} \sum_{\aliasfreq_1 \in\Aliasfreq_{p;1}}\cdots \sum_{\aliasfreq_d \in\Aliasfreq_{p;d}} \prod_{j=1}^d \left(\int_{K_{p;j}+\aliasfreq_j} \abs{\tilde{G}_{p;m,j,n}(\freq_j)}^{2} \de\freq_j\right)^{1/2}\\
        &= \max_{m\in[M_n]} \prod_{j=1}^d \sum_{\aliasfreq_j \in\Aliasfreq_{p;j}} \left(\int_{K_{p;j}+\aliasfreq_j} \abs{\tilde{G}_{p;m,j,n}(\freq_j)}^{2} \de\freq_j\right)^{1/2}\\
        &\leq \prod_{j=1}^d \sum_{\aliasfreq_j \in\Aliasfreq_{p;j}} \max_{m\in[M_{n,j}]} \left(\int_{K_{p;j}+\aliasfreq_j} \abs{G_{p;m,n}(\freq_j)}^{2} \de\freq_j\right)^{1/2}
    \end{align*}
    which converges to 1 by again applying \cref{res:general:rescale} in one dimension.

    \Cref{assumption:tapers:normality} is then satisfied by an identical argument to that in \cref{res:general:rescale}.
\end{proof}

\subsection{\texorpdfstring{\cref{res:rect:minbias}}{Proposition}}\label{proof:res:rect:minbias}
\begin{proof}[Proof of \cref{res:rect:minbias}]
    Recall that
    \begin{align*}
        g_m(x) = \sqrt{2} \sin(\pi m x)\indicator_{[0,1]}(x),
    \end{align*}
    and we assume that for all $n\in\NN$, $b_n {l_{n,j}} \geq M_{n,j}\sqrt{d}$ and there exists some $\delta>0$ such that $M_{n,j}^{1+\delta} l_{n,j}^{-1/2} \rightarrow 0$ as $n\rightarrow\infty$.

    Firstly, we aim to prove that $\set{g_{m}}_{m\in\NN}$ is an orthonormal family tapers supported on a subset of [0,1] that are bounded and continuous.
    Secondly that
    \begin{align*}
        \min_{m\in [{M}_{n,j}]}\int_{-b_n l_{n,j}/\sqrt{d}}^{b_n l_{n,j}/\sqrt{d}} \abs{g_{m}(x)}^2 \de x &\rightarrow 1.
    \end{align*}
    And finally, that for all $m\in\NN$ there exists $\delta>0$ and $C_m>0$ such that for all $x\in\RRd$, 
    $
        \abs{g_m(x)} + \abs{G_m(x)} \leq {C_m}{(1+\abs{x})^{-1-\delta}},
    $
    where $\max_{m\in [{M}_{n,j}]} C_m l_{n,j}^{-1/2} \rightarrow 0$ as $n\rightarrow\infty$.

    The minimum bias tapers are an orthonormal family of functions \citep{riedel1995minimum}.
    Clearly $g_m$ are continuous and $\norm{g_m}_\infty = \sqrt{2} < \infty$ so they are bounded. 
    This establishes the first part.

    For the second part, note\footnote{Note there is a slight typo in \cite{riedel1995minimum} in the statement of the Fourier transform of $g_m$. We have corrected it here.}
    \begin{align*}
        G_m(\freq) = \frac{1}{\sqrt{2}} e^{-\pi i (\freq-\frac{m-1}{2}) } \frac{m \sinc(\pi[\freq-m/2])}{k+m/2}.
    \end{align*}
    Note, $\abs{G_m(\freq)}$ is an even function of frequency.
    Now, we have for all $m\in[M_n]$
    \begin{align*}
        \int_{M_n}^R \abs{G_m(\freq)}^2 \de\freq
        &= \frac{m^2}{2} \int_{M_n}^R \frac{\sinc^2(\pi[k-m/2])}{(k+m/2)^2} \de\freq \\
        &\leq \frac{m^2}{2\pi} \int_{M_n}^R \frac{1}{(k+m/2)^2(k-m/2)^2} \de\freq \\
        &= \frac{m^2}{2\pi} \int_{M_n-m/2}^R \frac{1}{(k+m)^2 k^2} \de\freq \\
        &= \frac{1}{2\pi m}\Bigg[\frac{m}{M_n+m/2}+\frac{m}{M_n-m/2}-2\log\left(\frac{M_n-m/2}{M_n+m/2}\right)\\ &\qquad\qquad -\left(\frac{M_n}{R+m}+\frac{m}{R}\right)+2\log\left(\frac{R+m}{R}\right)\Bigg] \\
        &\rightarrow \frac{1}{2\pi m}\left[\frac{m}{M_n+m/2}+\frac{m}{M_n-m/2}-2\log\left(\frac{M_n-m/2}{M_n+m/2}\right)\right]
    \end{align*}
    as $R \rightarrow \infty$.
    So we have for any $m\in[M_n]$
    \begin{align*}
        \int_{M_n}^\infty \abs{G_m(\freq)}^2 \de\freq
        &\leq \frac{1}{2\pi}\left[\frac{4M_n}{4M_n^2+m^2} + \frac{2}{m}\log\left(\frac{M_n+m/2}{M_n-m/2}\right) \right] \\
        &= \frac{1}{2\pi}\left[\frac{4M_n}{4M_n^2+m^2} + \frac{2}{m}\log\left(1+\frac{2}{2M_n/m-1}\right) \right]\\
        &\leq \frac{1}{2\pi}\left[\frac{1}{4M_n} + \frac{2}{2M_n-m} \right] \\
        &\leq \frac{1}{2\pi}\left[\frac{1}{4M_n} + \frac{2}{M_n} \right] \\
        &= \frac{9}{8\pi M_n}.
    \end{align*}
    So we have
    \begin{align*}
        \max_{m\in[M_n]}\int_{M_n}^\infty \abs{G_m(\freq)}^2 \de\freq 
        \leq \frac{9}{8\pi M_n}.
    \end{align*}
    As a result, we have that
    \begin{align*}
        \max_{m\in[M_n]}\int_{M_n}^\infty \abs{G_m(\freq)}^2 \de\freq \rightarrow 0
    \end{align*}
    as $M_n\rightarrow\infty$.
    Thus
    \begin{align*}
        \min_{m\in[M_n]} \int_{-M_n}^{M_n} \abs{G_m(\freq)}^2 \de\freq
        &= \min_{m\in[M_n]} 1-2\int_{M_n}^\infty \abs{G_m(\freq)}^2 \de\freq \\
        &= 1 - 2\max_{m\in[M_n]} \int_{M_n}^\infty \abs{G_m(\freq)}^2 \de\freq \\
        &\rightarrow 1
    \end{align*}
    as $M_n\rightarrow\infty$.
    Now by assumption $M_{n,j} \leq b_n l_{n,j}/\sqrt{d}$ and so with some relabelling
    \begin{align*}
        \min_{m\in[M_{n,j}]} \int_{-b_n l_{n,j}/\sqrt{d}}^{b_n l_{n,j}/\sqrt{d}} \abs{G_m(\freq)}^2 \de\freq 
        &\leq \min_{m\in[M_{n,j}]} \int_{-M_{n,j}}^{M_{n,j}} \abs{G_m(\freq)}^2 \de\freq \\
        &\rightarrow 1
    \end{align*}
    as $n\rightarrow\infty$.
    
    The final condition is to check the tail decay, more specifically for all $m\in\NN$ there exists $\delta>0$ and $C_m>0$ such that for all $x\in\RRd$, 
    $
        \abs{g_m(x)} + \abs{G_m(x)} \leq {C_m}{(1+\abs{x})^{-1-\delta}},
    $
    where $\max_{m\in [{M}_{n,j}]} C_m l_{n,j}^{-1/2} \rightarrow 0$ as $n\rightarrow\infty$.
    Since the tapers decay like $x^{-2}$ in wavenumber, for some $0<\delta'<1$ we just need to find appropriate choices for $C_m$.
    Recall
    \begin{align*}
        \abs{g_m(x)} 
        &= \sqrt{2} \abs{\sin(\pi m x)}\indicator_{[0,1]}(x) \\
        &\leq \sqrt{2}.
    \end{align*}
    So since our bounding function $C_m(1+\abs{x})^{-1-\delta'}$ is decreasing, we have
    \begin{align*}
        \abs{g_m(x)} \leq C_m(1+\abs{x})^{-1-\delta'}
    \end{align*}
    if
    \begin{align*}
        \sqrt{2} \leq C_m 2^{-1-\delta'} \Leftrightarrow C_m \geq 2^{1+\delta'+1/2},
    \end{align*}
    as we must cover the corner of the box at least.
    Now we need to also consider the Fourier transform.
    Recall that
    \begin{align*}
        \abs{G_m(\freq)} 
        &= \frac{1}{\sqrt{2}} \abs{\frac{m \sinc(\pi[\freq-m/2])}{k+m/2}}.
    \end{align*}
    Since this is symmetric, we can consider the case $\freq\geq 0$.
    We further decompose this into two cases, $\freq \in [0,m]$ and $\freq \in [m,\infty)$.
    If we can find a lower bound for $C_m$ in each case, then taking $C_m$ to be the maximum of these two bounds and the previous bound will suffice.
    Firstly, when $\freq \in [0,m]$, we have
    \begin{align*}
        \abs{G_m(\freq)} 
        &\leq \frac{m}{\sqrt{2}} \abs{\frac{1}{k+m/2}} \\
        &\leq \sqrt{2}.
    \end{align*}
    In particular, again we have a constant bound we need to cover with $C_m(1+\abs{\freq})^{-1-\delta'}$.
    So we have for $\freq \in [0,m]$
    \begin{align*}
        \abs{G_m(\freq)} \leq C_m(1+\abs{\freq})^{-1-\delta'}
    \end{align*}
    if
    \begin{align*}
        \sqrt{2} \leq C_m (1+m)^{-1-\delta'} \Leftrightarrow C_m \geq \sqrt{2} (1+m)^{1+\delta'}.
    \end{align*}
    Since $m \in \NN$, we have
    \begin{align*}
        \sqrt{2} (1+m)^{1+\delta'} \geq 2^{1+\delta'+1/2}
    \end{align*}
    meaning we subsume the previous bound.
    Now consider the case $\freq \in [m,\infty)$.
    In this case, we have
    \begin{align*}
        \abs{G_m(\freq)} 
        &= \frac{1}{\sqrt{2}} \abs{\frac{m \sinc(\pi[\freq-m/2])}{k+m/2}} \\
        &\leq \frac{1}{\sqrt{2}} \frac{m}{\freq^2 - m^2/4}.
    \end{align*}
    So, we now need to find a $C_m$ such that
    \begin{align*}
        \frac{1}{\sqrt{2}} \frac{m}{\freq^2 - m^2/4} \leq C_m(1+\abs{\freq})^{-1-\delta'} \Leftrightarrow C_m \geq \frac{m}{\sqrt{2}} \frac{(1+\abs{\freq})^{1+\delta'}}{\freq^2 - m^2/4}.
    \end{align*}
    So we need only find an upper bound on the last fraction.
    In particular, because $\freq \geq m$ and $m\geq 1$,
    \begin{align*}
        \frac{m}{\sqrt{2}} \frac{(1+\abs{\freq})^{1+\delta'}}{\freq^2 - m^2/4}
        & \leq \frac{m}{\sqrt{2}} \frac{(1+\abs{\freq})^2}{\freq^2 - m^2/4} \\
        & = \frac{1}{\sqrt{2}} \left(m + \frac{1+2x+m^2/4}{x^2-m^2/4}\right) \\
        & = \frac{1}{\sqrt{2}} \left(m + \frac{1}{x+m/2}\left( 2 + \frac{m^2/4+m+1}{x-m/2}\right)\right) \\
        &\leq \frac{1}{\sqrt{2}} \left(m + \frac{1}{3m/2}\left( 2 + \frac{m^2/4+m+1}{m/2}\right)\right) \\
        &= \frac{1}{\sqrt{2}} \left(m + \frac{4}{3m} + \frac{1}{3} + \frac{4}{3m} + \frac{4}{3m^2}\right) \\
        &\leq \frac{1}{\sqrt{2}} \left(m + 4 + \frac{1}{3}\right) \\
        &\leq \frac{1}{\sqrt{2}} \left(5 + m\right).
    \end{align*}
    Therefore, we set
    \begin{align*}
        C_m = \max\set{\frac{1}{\sqrt{2}} \left(5 + m\right), \sqrt{2} (1+m)^{1+\delta'}}.
    \end{align*}
    Importantly, this means
    \begin{align*}
        \max_{m\in[M_{n,j}]} C_m  l_{n,j}^{-1/2} &= \max\set{\frac{1}{\sqrt{2}} \left(5 + M_{n,j}\right) l_{n,j}^{-1/2}, \sqrt{2} (1+M_{n,j})^{1+\delta'}  l_{n,j}^{-1/2}}
        \rightarrow 0
    \end{align*}
    as $n\rightarrow\infty$ because the second case dominates asymptotically and by assumption for some $\delta>0$, $M_{n,j}^{1+\delta}l_{n,j}^{-1/2} \rightarrow 0$ as $n\rightarrow\infty$.
    Therefore, choosing $0<\delta<1$ such that $\delta' < \delta$ we have $\max_{m\in[M_n]} C_m l_{n,j}^{-1/2} $ is dominated by
    \begin{align*}
        M_{n,j}^{1+\delta'}l_{n,j}^{-1/2} = M_{n,j}^{1+\delta}l_{n,j}^{-1/2} \rightarrow 0
    \end{align*}
    as $n\rightarrow\infty$.
\end{proof}

\section{Lemmas for main theorems}

\subsection{Aliasing of tapers}

\begin{lemma}\label{lemma:alias:tapers}
    Consider a function $g:\RRd \rightarrow \CC$ and a grid $\grid$ on which it will be sampled.
    Say that $g$ is continuous and bounded, with an $L^1$ Fourier transform $G$.
    Furthermore, assume that there are some constants $C>0$ and $\delta>0$ such that for all $x\in\RRd$
    \begin{align*}
        \abs{g(x)} + \abs{G(x)} \leq C \left(1+\norm{x}_2\right)^{-d-\delta}.
    \end{align*}
    Then for all $\freq\in\RRd$
    \begin{align*}
        G_m^{(\grid)}(\freq) 
        &= \elemprod{\Delta} \sum_{u\in\grid} g(u) e^{-2\pi i \ip{\freq}{u}} \\
        &= \sum_{z\in\ZZ^d} G_m\left(\freq+z\oslash\Delta \right) e^{-2\pi i \gridoffset \cdot (z\oslash\Delta)}.
    \end{align*}
    \begin{proof}
        We will need the Poisson summation formula \citep{stein1971introduction}.
        There are a variety of conditions for this, but the most convenient here is the conditions given in Lemma 4 of \cite{grochenig1996uncertainty}, namely that for a function $\phi:\RRd\rightarrow \CC$ with Fourier transform $\Phi$, both $\phi$ and $\Phi$ are continuous and satisfy
        \begin{align*}
            \sum_{x\in\ZZ^d} \max_{u\in[0,1]^d} \abs{\phi(x+u)} < \infty,
            \quad 
            \sum_{x\in\ZZ^d} \max_{u\in[0,1]^d} \abs{\Phi(x+u)} < \infty
        \end{align*}
        then
        \begin{align*}
            \sum_{u\in\ZZ^d} \phi(u) e^{-2\pi i \ip{\freq}{u}} = \sum_{z\in\ZZ^d} \Phi(\freq+z).
        \end{align*}
        
        Therefore, we just need to apply this to certain transformations of $g$.
        In particular, given our grid $\grid$, we set
        \begin{align*}
            \phi(x) = \elemprod{\Delta} g(x \circ \Delta + \gridoffset), \qquad x\in\RRd.
        \end{align*}
        Then the Fourier transform is given by
        \begin{align*}
            \Phi(\freq) &= \int_\RRd \phi(x) e^{-2\pi i \ip{\freq}{x}} \de x \\
            &= \elemprod{\Delta} \int_\RRd g(x \circ \Delta + \gridoffset) e^{-2\pi i \ip{\freq}{x}} \de x \\
            &= \frac{\elemprod{\Delta}}{\elemprod{\Delta}} \int_\RRd g(u) e^{-2\pi i \ip{\freq}{(u-\gridoffset)\oslash \Delta}} \de u \\
            &= G(\freq\oslash \Delta) e^{2\pi i \ip{\freq\oslash \Delta}{\gridoffset}}.
        \end{align*}
        Given the conditions on $g$ and $G$, we see that the condition for the Poisson summation holds for $\phi$ and $\Phi$.
        Firstly we have
        \begin{align*}
            \abs{\phi(x)} 
            &= \abs{\elemprod{\Delta} g(x \circ \Delta + \gridoffset)} \\
            &\leq \elemprod{\Delta} C (1+\norm{x \circ \Delta + \gridoffset}_2)^{-d-\delta},
        \end{align*}
        and
        \begin{align*}
            \abs{\Phi(x)}
            &= \abs{G(x\oslash \Delta)} \\
            &\leq C (1+\norm{x \oslash \Delta}_2)^{-d-\delta},
        \end{align*}
        both of which decay sufficiently fast to satisfy the required condition for Poisson summation.        
        Therefore we have for all $\freq\in\RRd$
        \begin{align*}
            H_m^{(\grid)}(\freq) 
            &= \elemprod{\Delta} \sum_{u\in\grid} h(u) e^{-2\pi i \ip{\freq}{u}} \\
            &= \sum_{z\in\ZZ^d} \phi(z) e^{-2\pi i \ip{\freq}{(z\circ \Delta+\gridoffset)}} \\
            &= e^{-2\pi i \ip{\freq}{\gridoffset}} \sum_{z\in\ZZ^d} \phi(z) e^{-2\pi i \ip{(\freq\circ \Delta)}{z}} \\
            &= e^{-2\pi i \ip{\freq}{\gridoffset}} \sum_{z\in\ZZ^d} \Phi(\freq\circ\Delta+z) \\
            &= e^{-2\pi i \ip{\freq}{\gridoffset}} \sum_{z\in\ZZ^d} G((\freq\circ\Delta+z)\oslash \Delta) e^{2\pi i \ip{((\freq+z)\oslash \Delta)}{\gridoffset}} \\
            &= \sum_{z\in\ZZ^d} G(\freq + z\oslash \Delta) e^{2\pi i \ip{(z\oslash \Delta)}{\gridoffset}}
        \end{align*}
        which yields the result.
    \end{proof}
\end{lemma}

\subsection{Covariance of spatial processes}
The following Lemma enables us to work with weighted integrals and sums of the spatial processes.

\begin{lemma}\label{res:cov_process_freq}
  Given \cref{assumption:spectra:exists} hold.
  Without loss of generality, let $\randmeasure{1},\randmeasure{2}$ be either marked point processes or integrated random fields, and $\randmeasure{3},\randmeasure{4}$ be integrated random fields.
  Let $\phi_1,\phi_2,\phi_3,\phi_4$ be integrable functions with bounded support and integrable Fourier transforms, and let $\grid_3,\grid_4$ be a regular grids with sampling intervals $\Delta_3,\Delta_4$ and offset $\gridoffset_3,\gridoffset_4$.
  Then
  \begin{align*}
      \cov{\int_\RRd \phi_1(\loc) \randmeasure{1}(\de\loc), \int_\RRd \phi_2(\loc)\randmeasure{2}(\de\loc)} 
      &= \int_\RRd \Phi_1(-\freq) \conj{\Phi_2(-\freq)} \sdf{1,2}(\freq)\de\freq, \\
      \cov{\int_\RRd \phi_1(\loc) \randmeasure{1}(\de\loc), \int_\RRd \phi_3^{(\grid_3)}(\loc)\randmeasure{3}(\de\loc)} &= \int_\RRd \Phi_1(-\freq) \conj{\Phi_3^{(\grid_3)}(-\freq)} \sdf{1,3}(\freq)\de\freq, \\
      \cov{\int_\RRd \phi_3^{(\grid_3)}(\loc) \randmeasure{3}(\de\loc), \int_\RRd \phi_4^{(\grid_4)}\randmeasure{4}(\de\loc)} &= \int_\RRd \Phi_3^{(\grid_3)}(-\freq) \conj{\Phi_4^{(\grid_4)}(-\freq)} \sdf{1,4}(\freq)\de\freq.
  \end{align*}
  where $\Phi_j$ is the Fourier transform of $\phi_j$, i.e. $\Phi_j(\freq)=\int_\RRd \phi_j(\lag) e^{-2\pi i \ip{\lag}{\freq}} \de\lag$, $\freq\in\RRd$.

  Therefore if $\randmeasure{p},\randmeasure{q}$ are either marked point processes or integrated random fields and $\phi_p,\phi_q$ are integrable functions with bounded support and integrable Fourier transforms, or sampled versions of such functions if the corresponding process is an integrated random field, then from the previous cases
  \begin{align*}
    \cov{\int_\RRd \phi_p(\loc) \randmeasure{p}(\de\loc), \int_\RRd \phi_q(\loc')\randmeasure{q}(\de\loc')} &= \int_\RRd \Phi_p(-\freq) \conj{\Phi_q(-\freq)} \sdf{p,q}(\freq)\de\freq.
  \end{align*}

  \begin{proof}
    We have that for $j\in\set{1,\ldots,4}$, $\phi_j(u)=\int_\RRd \Phi_j(\freq) e^{2\pi i \ip{\freq}{\loc}}\de\freq$.
    Begin with the first case,
    \begin{align*}
        LHS &= \cov{\int_\RRd \phi_1(\loc) \randmeasure{1}(\de\loc), \int_\RRd \phi_2(\loc)\randmeasure{2}(\de\loc)} \\
        &= \int_\RRd \int_\RRd \phi_1(\loc+\lag) \conj{\phi_2(\loc)} \ell(\de \loc) \reducedcumulantmeasure{1,2}(\de\lag) \\
        &= \int_\RRd \int_\RRd \int_\RRd \Phi_1(-\freq) e^{-2\pi i \ip{\freq}{(\loc+\lag)}}\de\freq \conj{\phi_2(\loc)} \ell(\de \loc) \reducedcumulantmeasure{1,2}(\de\lag) \\
        &= \int_\RRd \Phi_1(-\freq) \int_\RRd \conj{\phi_2(\loc)}  e^{-2\pi i \ip{\freq}{\loc}} \ell(\de \loc)\int_\RRd  \reducedcumulantdens{1,2}(\lag) e^{-2\pi i \ip{\freq}{\lag}}  \reducedcumulantmeasure{1,2}(\de\lag) \de\freq \\
        &= \int_\RRd \Phi_1(-\freq) \conj{\Phi_2(-\freq)} \sdf{1,2}(\freq)\de\freq
    \end{align*}
    where the interchange of limits is justified because $\Phi_1$, $\phi_2$ are assumed to be integrable; and $\reducedcumulantmeasure{1,2}$ is totally finite from \cref{assumption:spectra:exists}.

    Now in the second case, $\reducedcumulantmeasure{1,3}$ has a density $\reducedcumulantdens{1,3}$ because one of the processes is an integrated random field, so
    \begin{align*}
      \cov{\int_\RRd \phi_1(u) \randmeasure{1}(\de u), Y_3(s)} 
      &= \int_{\RRd} \phi_1(u) \reducedcumulantdens{1,3}(u-s) \de u \\
      &= \int_{\RRd} \int_{\RRd} \Phi_1(-\freq) e^{-2\pi i \ip{\freq}{u}} \de\freq \reducedcumulantdens{1,3}(u-s) \de u \\
      &= \int_{\RRd} \Phi_1(-\freq) \int_{\RRd} \reducedcumulantdens{1,3}(u-s) e^{-2\pi i \ip{\freq}{u-s}}\de u e^{-2\pi i \ip{\freq}{s}} \de\freq\\
      &= \int_{\RRd} \Phi_1(-\freq) \sdf{1,3}(\freq) e^{-2\pi i \ip{\loc}{\freq}} \de\freq, \\
    \end{align*}
    where the interchange is justified as both $\Phi_1$ and $\reducedcumulantdens{1,3}$ are integrable.
    Therefore
    \begin{align*}
      LHS &= \cov{\int_\RRd \phi_1(u) \randmeasure{1}(\de u), \int_\RRd \conj{\phi_3^{(\grid_3)}(u)} \randmeasure{3}(\de u)} \\ 
      &= \elemprod{\Delta_3} \sum_{\loc \in \grid} \conj{\phi_3^{(\grid_3)}(u)} \int_{\RRd} \Phi_1(-\freq) \sdf{1,3}(\freq) e^{-2\pi i \ip{\loc}{\freq}} \de\freq \\
      &= \int_{\RRd} \Phi_1(-\freq) \sdf{1,3}(\freq) \elemprod{\Delta_3} \sum_{\loc \in \grid} \conj{\phi_3^{(\grid_3)}(u)} e^{-2\pi i \ip{\loc}{\freq}} \de\freq \\
      &= \int_{\RRd} \Phi_1(-\freq) \conj{\Phi_3^{(\grid)}(-\freq)} \sdf{p,q}(\freq)\de\freq,
    \end{align*}
    where the interchange is justified because the summands are only non-zero for a finite subset of $\grid_3$ (as $\phi_3$ has bounded support).

    For the final case, $\reducedcumulantdens{3,4}$ is continuous by assumption, so we have the inverse relation
    \begin{equation*}
      \reducedcumulantdens{3,4}(\lag) = \int_{\RRd} \sdf{3,4}(\freq) e^{2\pi i \ip{\lag}{\freq}} \de\freq.
    \end{equation*}
    Therefore,
    \begin{align*}
      LHS &= \cov{\int_\RRd \phi_3^{(\grid_3)}(u) \randmeasure{3}(\de u), \int_\RRd \phi_4^{(\grid_4)}(u) \randmeasure{4}(\de u)} \\
      &= \elemprod{\Delta_3}\elemprod{\Delta_4} \sum_{u\in\grid_3} \sum_{\loc\in\grid_4} \phi_3(u)\conj{\phi_4(\loc)} \reducedcumulantdens{3,4}(u-\loc) \\
      &= \elemprod{\Delta_3}\elemprod{\Delta_4} \sum_{u\in\grid_3} \sum_{\loc\in\grid_4} \phi_3(u)\conj{\phi_4(\loc)} \int_{\RRd} \sdf{3,4}(\freq) e^{2\pi i \ip{(u-\loc)}{\freq}} \de\freq \\
      &= \int_\RRd \Phi_3^{(\grid_3)}(-\freq) \conj{\Phi_4^{(\grid_4)}(-\freq)} \sdf{3,4}(\freq) \de\freq,
    \end{align*}
    where the final equality holds because they are both finite sums.
    The general statement is covered by these three cases.
  \end{proof}
\end{lemma}

\subsection{Properties of the sampled taper}\label{app:lemmas}

\begin{lemma}\label{res:sample:phaseproperties}
    The following properties hold
    \begin{align*}
        \forall\freq\in\RRd,\; w_{p}(\freq) &= \conj{w_{p}(-\freq)}, \\
        \forall\freq,\freq'\in\RRd,\; w_{p}(\freq+\freq') &= w_{p}(\freq)w_{p}(\freq'),\\
        \aliasfreq_1,\aliasfreq_2\in\Aliasfreq_{p} &\Rightarrow \aliasfreq_1\pm\aliasfreq_2\in\Aliasfreq_{p}.
    \end{align*}
    \begin{proof}
      This is immediate from the Definition.
    \end{proof}
\end{lemma}

\subsection{Properties of the aliased spectra}

\begin{lemma}\label{lemma:alias:grid:specific}
    Consider two different grids $\grid$ and $\grid'$, with sampling intervals $\Delta$, $\Delta' \in\QQ_{>0}^d$.
    Write $\Aliasfreq$ and $\Aliasfreq'$ to be the sets of aliasing frequencies for the first and second grid respectively, i.e.
    \begin{align*}
        \Aliasfreq = \ZZ^d \oslash \Delta, \quad \Aliasfreq' = \ZZ^d \oslash \Delta'
    \end{align*}
    then we can write
    \begin{align*}
        \Aliasfreq \cap \Aliasfreq' = a \circ \ZZ^d \oslash \Delta
    \end{align*}
    where $a=(a_{1},\ldots,a_{d})^\top$ is a vector such that $a_{j}/a_{j}'=\Delta_{j}/\Delta_{j}'$ and $a_{j},a_{j}'$ are coprime.
    
    \begin{proof}
        Firstly, note that if $\Aliasfreq_{j}=\ZZ/\Delta_j$, then $\Aliasfreq = \prod_{j=1}^d \Aliasfreq_{j}$, and similarly for $\Aliasfreq'$.
        Therefore, we may prove the result for each dimension separately.
        
        In particular, we wish to show that

        \begin{align*}
            \Aliasfreq_{j} \cap \Aliasfreq_{j}' = a_{j} \circ \ZZ/\Delta_{j}
        \end{align*}        

        We begin by showing that if $\aliasfreq\in\ZZ/\Delta_j$, then $\aliasfreq\in\Aliasfreq_j\cap\Aliasfreq'$.
        In particular, for any $z\in\ZZ$, set $\aliasfreq=a_{j} z/\Delta_{j}$ with $a_{j}\in\ZZ$ as defined in the Lemma.
        Then $a_{j} z\in\ZZ$, so $\aliasfreq\in\Aliasfreq_{j}$.
        Now, $\aliasfreq = a_{j} z/\Delta_{j} = a_{j}' z/\Delta_{j}'\in\Aliasfreq'$.
        
        Now for the other direction, let $\aliasfreq\in\Aliasfreq_{j}\cap\Aliasfreq_{j}'$, then there exists $z,z'\in\ZZ$ such that
        \begin{align*}
            z/\Delta_{j} = \aliasfreq = z'/\Delta_{j}'.
        \end{align*}
        This is always satisfied by $\aliasfreq = 0$. 
        If $\aliasfreq\neq0$, then
        \begin{align*}
            z/z' = \Delta_{j}/\Delta_{j}' = a_{j}/a_{j}'
        \end{align*}
        can only be satisfied if $\Delta_{j}/\Delta_{j}'\in\QQ$.
        Finally therefore $z' = z a_{j}'/a_{j}$, and so since the $a$s are coprime, $z=z''a_{j}$.
        Thus $\phi = z''a_{j}/\Delta_{j}$ for some $z''\in\ZZ$, as required.
    \end{proof}
\end{lemma}

From \cref{lemma:alias:grid:specific} we see that the joint aliasing effect also lies on a regular grid.

\begin{lemma}\label{lemma:alias:boundedcontinuous}
    Given \cref{assumption:spectra:exists}, the aliased spectra $\asdf{p,q}$ is bounded and continuous for all $p,q\in[P]$. Furthermore, we have
    \begin{align*}
        \sum_{\aliasfreq\in\Aliasfreq_{p}\cap\Aliasfreq_{q}} \abs{\sdf{p,q}(\freq+\aliasfreq)} < \infty, \qquad \forall \freq\in\RRd.
    \end{align*}

    \begin{proof}
        In the case where one of the two processes is not sampled on a grid, then $\asdf{p,q}=\sdf{p,q}$ which is bounded and continuous by the assumption that $\reducedcumulantmeasure{p,q}$ is totally finite \citep{daley2003introduction}.
        Let $\reducedcumulantmeasure{p,q}^{+}$ and $\reducedcumulantmeasure{p,q}^{-}$ be the positive and negative parts of the Jordan-Hahn decomposition for $\reducedcumulantmeasure{p,q}$ (see \cite{daley2003introduction} for example).
        Then from the assumption that $\reducedcumulantmeasure{p,q}$ is totally finite, $\reducedcumulantmeasure{p,q}^{+}(\RRd)+\reducedcumulantmeasure{p,q}^{-}(\RRd)<\infty$, and by construction $\reducedcumulantmeasure{p,q}=\reducedcumulantmeasure{p,q}^{+}-\reducedcumulantmeasure{p,q}^{-}$.
        In this case we have for any $\freq\in\RRd$,
        since $\abs{e^{-2\pi i \ip{\freq}{\lag}}} \leq 1$ for all $\lag\in\RRd$, we can use the dominated convergence theorem to establish continuity.
        Furthermore, we also have for all $\freq\in\RRd$
        \begin{align*}
            \abs{\sdf{p,q}(\freq)}
            &= \abs{\int_{\RRd} e^{-2\pi i \ip{\freq}{\lag}} \reducedcumulantmeasure{p,q}(\de\lag)} \\
            &= \abs{\int_{\RRd} e^{-2\pi i \ip{\freq}{\lag}} \reducedcumulantmeasure{p,q}^{+}(\de\lag) - \int_{\RRd} e^{-2\pi i \ip{\freq}{\lag}} \reducedcumulantmeasure{p,q}^{-}(\de\lag) } \\
            &\leq \reducedcumulantmeasure{p,q}^{+}(\RRd) + \reducedcumulantmeasure{p,q}^{-}(\RRd) \\
            &<\infty.
        \end{align*}
        The final statement then holds trivially, as $\Aliasfreq_{p}\cap\Aliasfreq_{q} = \set{0}$, and so
        \begin{align*}
            \sum_{\aliasfreq\in\Aliasfreq_{p}\cap\Aliasfreq_{q}} \abs{\sdf{p,q}(\freq+\aliasfreq)} = \abs{\sdf{p,q}(\freq)} < \infty.
        \end{align*}
        
        We need only check the cases where there is aliasing, in other words, when we have two random fields.
        In this case, we have
        \begin{align*}
            \asdf{p,q}(\freq)
            &= \sum_{\aliasfreq\in \Aliasfreq_{p} \cap \Aliasfreq_{q}} \sdf{p,q}(\freq+\aliasfreq)w_p(\aliasfreq)\conj{w_q(\aliasfreq)}\\
            &= \sum_{\aliasfreq\in \Aliasfreq_{p} \cap \Aliasfreq_{q}} \sdf{p,q}(\freq+\aliasfreq)e^{-2\pi i \ip{(\gridoffset_p-\gridoffset_q)}{\freq}}.
        \end{align*}
        Now by \cref{lemma:alias:grid:specific}, the intersection is a grid.
        In addition, by \cref{assumption:spectra:exists}
        \begin{align*}
            \abs{\sdf{p,q}(\freq)} < \frac{C}{(1+\norm{\freq})^{d+\delta}}
        \end{align*}
        for some $C>0$ and $\delta>0$.
        Therefore, we can write
        \begin{align*}
            \abs{\asdf{p,q}(\freq)} 
            &\leq \sum_{\aliasfreq\in \Aliasfreq_{p} \cap \Aliasfreq_{q}} \abs{\sdf{p,q}(\freq+\aliasfreq)}\\
            &\leq C \sum_{\aliasfreq\in \Aliasfreq_{p} \cap \Aliasfreq_{q}} \frac{1}{(1+\norm{\freq+\aliasfreq})^{d+\delta}} \\
            & < \infty.
        \end{align*}
        This established both boundedness and finiteness of the sum.
        Since $\sdf{p,q}(\freq)$ is continuous by the previous argument applied to $\reducedcumulantmeasure{p,q}$, this also establishes continuity of $\asdf{p,q}(\freq)$.
    \end{proof}
\end{lemma}

\begin{lemma}\label{lemma:posdef:alias}
    If the spectral density matrix $f(\freq)$ is positive definite for all $\freq\in\RRd$, then the aliased spectral density matrix $\tilde{f}(\freq)$ is positive definite for all $\freq\in\RRd$.
    \begin{proof}
        Begin by recalling that $\tilde{f}$ is the matrix with elements
        \begin{align*}
            \asdf{p,q}(\freq) &= \sum_{\aliasfreq\in\Aliasfreq_{p}\cap\Aliasfreq_{q}} \sdf{p,q}(\freq + \aliasfreq) w_{p}(\aliasfreq) \conj{w_{q}(\aliasfreq)}, \qquad \freq\in\RRd.
        \end{align*}
        So we could rewrite the matrix as
        \begin{align*}
            \tilde{f}(\freq) &= \sum_{\aliasfreq\in\Aliasfreq} \left\{w(\aliasfreq) w(\aliasfreq)^H\right\} \circ f(\freq+\aliasfreq)
        \end{align*}
        where $\Aliasfreq=\bigcup_{p=1}^P\Aliasfreq_{p}$ and
        \begin{align*}
            w(\aliasfreq) = [w_p(\aliasfreq)\indicator_{\Aliasfreq_{p}}(\aliasfreq)]_{p\in[P]}.
        \end{align*}
        
        We need to show that for any $x\in\CC^P\setminus\set{0}$,
        \begin{align*}
            x^H \tilde{f}(\freq) x > 0.
        \end{align*}
        Now by linearity
        \begin{align*}
            x^H \tilde{f}(\freq) x = \sum_{\aliasfreq\in\Aliasfreq} x^H \left\{w(\aliasfreq) w(\aliasfreq)^H\right\} \circ f(\freq+\aliasfreq) x.
        \end{align*}
        Firstly $0 \in \Aliasfreq$, and in this case $w(0)=1$.
        Therefore for one term, the summand is positive.
        All that remains is to show that all of the remaining summands are non-negative.
        
        Take some $\aliasfreq \in \Aliasfreq\setminus\set{0}$. Let $S = \left\{w(\aliasfreq) w(\aliasfreq)^H\right\} \circ f(\freq+\aliasfreq)$.
        By the Schur product theorem \citep{Schur1911Book}, the Hadamard product of two positive semi-definite matrices is positive semi-definite.
        Now $w(\aliasfreq)w(\aliasfreq)^H$ is positive semi-definite because for any $x\in\CC^P$, we have $x^H (w(\aliasfreq) w(\aliasfreq)^H) x = \abs{w(\aliasfreq)^H x}^2 \geq 0$.
        Therefore $S$ is positive semi-definite, and so the summands $x^H Sx$ are non-negative.
        As a result, $\tilde{f}(\freq)$ is positive definite as
        \begin{align*}
            x^H \tilde{f}(\freq) x \geq x^H f(\freq) x > 0.
        \end{align*}
        and by assumption $f(\freq)$ is positive definite for all $\freq\in\RRd$.
    \end{proof}
\end{lemma}

\section{The non-oracle case}\label{app:nonoracle}

Typically when developing the theory of spectral analysis, it is assumed that the process is either mean zero, or that the mean is easily estimated and removed, and so we may act as though the mean is known.
We also make this assumption in the theory developed prior to this point, however, it is important to justify this decision.
In our case, the mean is often not zero (a simple point process for instance), and so this should be explicitly addressed.
Doing no mean removal results in substantial bias, as was noted in the original paper introducing spectra of point processes \citep{bartlett1963spectral}.
Doing mean removal greatly reduces this bias, as shown by \cite{rajala2023what}.
However, for finite samples this bias is not completely removed by subtracting the sample mean, though it is greatly reduced in most scenarios.
The most extreme case of this is when no taper is applied, and the sample mean used. 
In this case the tapered Fourier transform is always zero at zero wavenumber.
As we will see from the results in this section, in general, the error introduced by not knowing the true intensity/mean can only be large for wavenumbers where $H(\freq)$ is large (in particular, when doing multitaper estimation, for wavenumbers whose norm is less than the bandwidth).

Say that the intensity (mean) of the process is estimated with a weighted mean, i.e.
\begin{align*}
    \intensityest{p} = \int_\RRd g_{p}(\lag) \randmeasure{p}(\lag)
\end{align*}
for some function $g_{p}$ supported on $\region$ satisfying the assumptions of Lemma~\ref{res:cov_process_freq}.
Studying the mean of the periodogram in this case, we see that the mean is the same as the oracle case, plus some extra error terms.

\begin{proposition}[Expectation of the periodogram with unknown mean]
    Say that the processes in question satisfy the assumptions of Proposition~\ref{res:bias:finite} (the expectation of the periodogram), but that the mean is estimated by a weighted mean as just described. Then
    \begin{align*}
        \EE{\pgram{p,q}{m}(\freq)} = & \int_\RRd H_{p;m}(\freq-\freq')\conj{H_{q;m}(\freq-\freq')} \sdf{p,q}(\freq')\de\freq' \\
        & + H_{p;m}(\freq)\conj{H_{q;m}(\freq)} (G_{p}(0)-1)(G_{q}(0)-1) \intensity{p}\intensity{q} \\
        & + H_{p;m}(\freq)\conj{H_{q;m}(\freq)} \int_\RRd G_{p}(-\freq')\conj{G_{q}(-\freq')} \sdf{p,q}(\freq')\de\freq' \\
        & - \conj{H_{q;m}(\freq)} \int_\RRd H_{p;m}(\freq-\freq') \conj{G_{q}(-\freq')} \sdf{p,q}(\freq')\de\freq' \\
        & - \conj{H_{q;m}(\freq)} \int_\RRd G_{p}(-\freq') \conj{H_{q;m}(\freq-\freq')}  \sdf{p,q}(\freq')\de\freq'
    \end{align*}
    for $\freq\in\RRd$.
    \begin{proof}
        Write $\randmeasure{p}^0(\de u) = \randmeasure{p}(\de u) - \intensity{p}\de u$.
        Begin by noting that
        \begin{align*}
            \intensityest{p} = \int_\RRd g_{p}(\lag) \randmeasure{p}^0(\lag) + \intensity{p}G_{p}(0).
        \end{align*}
        Therefore we can write
        \begin{align*}
            \dft{p}{m}(\freq) &= \int_\RRd H_{p;m}(\lag) e^{-2\pi i \ip{\lag}{\freq}} \randmeasure{p0}(\de\lag) + H_{p;m}(\freq)\intensity{p} \left\{1-G_{p}(0)\right\} - H_{p;m}(\freq)\int_\RRd g_{p}(\lag) \randmeasure{p}^0(\de\lag).
        \end{align*}
        Terms involving products of the middle term with one of the other in the periodogram will have zero expectation, and so
        \begin{align*}
            \EE{\pgram{p,q}{m}(\freq)} =&\; \EE{\int_\RRd H_{p;m}(\lag) e^{-2\pi i \ip{\lag}{\freq}} \randmeasure{0,p}(\de\lag) \conj{\int_\RRd H_{q;m}(\lag) e^{-2\pi i \ip{\lag}{\freq}} \randmeasure{q}^0(\de\lag)}} \\
            &+ H_{p;m}(\freq)\intensity{p} \left(1-G_{p}(0)\right) \conj{H_{q;m}(\freq)\intensity{q} \left(1-G_{q}(0)\right)} \\
            &+ H_{p;m}(\freq)\conj{H_{q;m}(\freq)}\EE{\int_\RRd g_{p}(\lag) \randmeasure{p}^0(\de\lag) \conj{\int_\RRd g_{q}(\lag) \randmeasure{q}^0(\de\lag)}} \\
            &- \conj{H_{q;m}(\freq)}\EE{\int_\RRd H_{p;m}(\lag) e^{-2\pi i \ip{\lag}{\freq}} \randmeasure{p}^0(\de\lag) \conj{\int_\RRd g_{q}(\lag) \randmeasure{q}^0(\de\lag)}} \\
            &- H_{p;m}(\freq)\EE{\int_\RRd g_{p}(\lag) \randmeasure{p}^0(\de\lag) \conj{\int_\RRd H_{q;m}(\lag) e^{-2\pi i \ip{\lag}{\freq}} \randmeasure{q}^0(\de\lag)}}
        \end{align*}
        with the remaining terms having zero expectation.
        All that remains is applying \cref{res:cov_process_freq}.
    \end{proof}
\end{proposition}

This is the usual bias, plus terms that scale like the size of the Fourier transform of the taper.
Outside of the bandwidth of the taper, this is negligible.
If an unbiased estimator of $\lambda$ is used (for either process), then the second term is zero.
The remaining terms are nowhere near the magnitude of bias observed if no mean correction is performed; however, for small samples they may still generate noticeable bias.
One approach to address this may be to down weight certain tapers in the multitaper estimator, to remove tapers whose Fourier transform is large at the wavenumber being investigated. Even if this wavenumber is in the bandwidth of the tapers, some of the Fourier transforms will still have little magnitude at this wavenumber (just not all of them).
In any case, this issue impacts few wavenumbers, often by negligible amounts in practice.

\section{Linear interpolation}\label{app:interp}
\subsection{Definitions and basic properties}

We use multilinear interpolation to interpolate functions defined on a grid $\grid\subset\RRd$ to a function defined on $\RRd$.
This form of interpolation is the common form of multivariate ``linear'' interpolation, generalising bilinear and trilinear interpolation to $d$ dimensions.
The interpolation works by applying linear interpolation in each dimension recursively, and so is defined by the following definition.

\begin{definition}
    Let $\mathcal{I}$ be the multilinear interpolation operator, that maps a function of a grid $f:\grid\rightarrow\RR$ to a function of $\RRd$ so that $\mathcal{I}[f]:\RRd\rightarrow\RR$ with
    \begin{align*}
        \mathcal{I}[f](u) &= \frac{1}{\elemprod{\Delta}} \sum_{z\in\grid} \indicator_{[0,\Delta)}(u-z) \sum_{v\in\set{0,1}^d} f(z+v\circ \Delta) \prod_{j=1}^d (u_j-z_j)^{v_j} (\Delta_j-u_j+z_j)^{1-v_j}
    \end{align*}
    for any $u\in\RRd$, where $[0,\Delta)=\prod_{j=1}^d [0,\Delta_j)$ and $z_j$, $v_j$ denote the $j$\Th elements of $z$ and $v$ respectively.
\end{definition}

\begin{definition}
    Let the interpolation weight for a given $z$ and $v$ be the function $\mathcal{I}_{v,z}:\RRd\rightarrow\RR$ defined by
    \begin{align*}
        \mathcal{I}_{v,z}(u) &= \prod_{j=1}^d (u_j-z_j)^{v_j} (\Delta_j-(u_j-z_j))^{1-v_j}\indicator_{[0,\Delta_j)}(u_j-z_j) .
    \end{align*}
\end{definition}

Consider two functions $f,g$ with domain $\RRd$ and codomain $\RR$.
Let
\begin{align*}
    \ipalt{f}{g}_2 = \int_{\RRd} f(u)g(u) \de u
\end{align*}
denote the $L^2$ inner product.

\begin{proposition}\label{supp:res:interp_vs_function}
    The inner product between a multilinear interpolation of $g:\grid\rightarrow\RRd$ and some other function $h:\RRd\rightarrow\RR$ is given by
    \begin{align*}
        \ipalt{\mathcal{I}[g]}{h}_2
        &= \frac{1}{\elemprod{\Delta}} \sum_{z\in\grid} \sum_{v\in\set{0,1}^d} g(z+v\circ \Delta) \ipalt{h}{\mathcal{I}_{v,z}}_2.
    \end{align*}
\begin{proof}
    Follows from rearranging the definition and linearity of inner products.
\end{proof}
\end{proposition}

Now, many quantities of interest can be derived by considering $\ipalt{h}{\mathcal{I}_{v,z}}_2$ for appropriate choices of $h$.
In particular, in the next two sections, we consider $h$ to be an interpolated function, and then the Fourier basis, so that we can explore the effect on orthogonality and wavenumber domain concentration.

\subsection{Linear interpolation with linear interpolation}\label{app:interp:withinterp}

\begin{lemma}\label{supp:res:Ivz:integral}
    If $v_j,v_j'\in\set{0,1}$ then
    \begin{align*}
        \int_{0}^{\Delta_j} x_j^{v_j+v_j'} (\Delta_j-x_j)^{2-v_j-v_j'} 
        &= \frac{\Delta_j^3}{6} \left[1+\delta_{v_j,v_j'} \right] .
    \end{align*}
    \begin{proof}
        We have the four cases $v_j,v_j'\in\set{0,1}$.
        When $v_j=v_j'=0$
        \begin{align*}
            \int_{0}^{\Delta_j} x_j^{v_j+v_j'} (\Delta_j-x_j)^{2-v_j-v_j'} \de x_j
            &= \int_{0}^{\Delta_j} (\Delta_j-x_j)^{2} \de x_j \\
            &= \int_{0}^{\Delta_j} y_j^2 \de y_j \\
            &= \frac{1}{3} \Delta_j^3.
        \end{align*}
        which is equivalent to the case $v_j=1,v_j'=1$. Finally, by symmetry we need only consider $v_j=0,v_j'=1$
        \begin{align*}
            \int_{0}^{\Delta_j} x_j (\Delta_j-x_j) \de x_j
            &= \Delta_j\int_{0}^{\Delta_j} x_j \de x_j - \int_{0}^{\Delta_j} x_j^2 \de x_j \\
            &= \frac{1}{2}\Delta_j^3 - \frac{1}{3}\Delta_j^3 \\
            &= \frac{1}{6}\Delta_j^3.
        \end{align*}
        Therefore, we have
        \begin{align*}
            \int_{0}^{\Delta_j} x_j^{v_j+v_j'} (\Delta_j-x_j)^{2-v_j-v_j'}  &= \begin{cases}
                \frac{1}{3} \Delta_j^3 & \text{if } v_j\neq v_j' \\
                \frac{1}{6} \Delta_j^3 & \text{if } v_j=v_j'.
            \end{cases}
        \end{align*}
    \end{proof}
\end{lemma}

\begin{lemma}\label{supp:res:Ivz:zeros}
    For any $z,z'\in\grid$ and $v,v'\in\set{0,1}^d$ we have
    \begin{align*}
        z\neq z' &\implies \forall u \in \RRd,\; \mathcal{I}_{v,z}(u)\mathcal{I}_{v',z'}(u) = 0, 
        \\
        z= z' &\implies \forall u \in \RRd \;\text{s.t.}\; u-z\notin[0,\Delta),\; \mathcal{I}_{v,z}(u)\mathcal{I}_{v',z'}(u) = 0.
    \end{align*}
    \begin{proof}
        This follows from the presence of the indicator functions in the definition of $\mathcal{I}_{v,z}$. In particular, because we have something multiplied by
        \begin{align*}
            \indicator_{[0,\Delta)}(u-z)\indicator_{[0,\Delta)}(u-z')
        \end{align*}
        which is zero in the cases indicated in this Lemma.
    \end{proof}
\end{lemma}

\begin{lemma}\label{supp:res:Ivz:ip}
    We have that for any $z,z'\in\grid$ and $v,v'\in\set{0,1}^d$
    \begin{align*}
        \ipalt{\mathcal{I}_{v',z'}}{\mathcal{I}_{v,z}}_2
        &= \delta_{z,z'} \prod_{j=1}^d \frac{\Delta_j^3}{6} \left[1+\delta_{v_j,v_j'} \right].
    \end{align*}
    \begin{proof}
        From  we see
        \begin{align*}
            \ipalt{\mathcal{I}_{v',z'}}{\mathcal{I}_{v,z}}_2 
            &= \int_\RRd \mathcal{I}_{v',z'}(u) \mathcal{I}_{v,z}(u) \de u \\
            &= \delta_{z,z'} \prod_{j=1}^d \int_{0}^{\Delta_j} x_j^{v_j+v_j'} (\Delta_j-x_j)^{2-v_j-v_j'} \de x_j & (\text{by \cref{supp:res:Ivz:zeros}}) \\
            &= \delta_{z,z'} \prod_{j=1}^d \frac{\Delta_j^3}{6} \left[1+\delta_{v_j,v_j'} \right]. & (\text{by \cref{supp:res:Ivz:integral}})
        \end{align*}
    \end{proof}
\end{lemma}

\begin{proposition}\label{supp:res:ip_interp}
    We have that
    \begin{align*}
        \ipalt{\mathcal{I}[g]}{\mathcal{I}[h]}_2 &= \elemprod{\Delta} \sum_{z\in\grid} \sum_{v,v'\in\set{0,1}^d} g(z+v\circ \Delta)h(z+v'\circ \Delta) \prod_{j=1}^d \frac{1}{6} \left[1+\delta_{v_j,v_j'} \right].
    \end{align*}
    \begin{proof}
        Using \cref{supp:res:Ivz:ip,supp:res:interp_vs_function}
        \begin{align*}
            \ipalt{\mathcal{I}[g]}{\mathcal{I}[h]}_2 &= 
            \frac{1}{\elemprod{\Delta}} \sum_{z\in\grid} \sum_{v\in\set{0,1}^d} g(z+v\circ \Delta) \ipalt{I[h]}{\mathcal{I}_{v,z}}_2 \\
            &= \frac{1}{\elemprod{\Delta}^2} \sum_{z\in\grid} \sum_{v,v'\in\set{0,1}^d} g(z+v\circ \Delta)h(z+v'\circ \Delta) \prod_{j=1}^d \frac{\Delta_j^3}{6} \left[1+\delta_{v_j,v_j'} \right] \\
        \end{align*}
    \end{proof}
\end{proposition}
Now we can rescale the interpolated tapers to have an $L^2$ norm of one, and use this to check the cross inner products are still close to zero.
If they are not, we can fix this by increasing the grid resolution.
The best way to see why is by considering the Fourier transform of the interpolated sequence.

\subsection{Linear interpolation with Fourier basis}\label{app:interpolation:fouriertransform}
The following proposition tells us how to calculate the Fourier transform of a multilinearly interpolated taper, which we will require for mean removal of the point patterns, and for examining the quality of our tapers.

\begin{proposition}\label{res:taper_ft}
    Let $\phi_\freq(\lag) = e^{-2\pi i\ip{\lag}{\freq}}$ and $g:\grid\rightarrow\RR$ then
    \begin{align*}
        \ipalt{\mathcal{I}[g]}{\phi_\freq}_2
        &= G^{(\grid)}(\freq) \prod_{j=1}^d \sinc^2(\pi \Delta_j \freq_j)
    \end{align*}
    where $\freq_j$ is the $j$\Th element of $\freq$.
    \begin{proof}
        Firstly, notice that
        \begin{align*}
            \ipalt{\phi_\freq}{\mathcal{I}_{v,z}}_2 e^{2\pi i \ip{(z+v\circ \Delta)}{\freq}} &= e^{2\pi i \ip{(z+v\circ \Delta)}{\freq}} \prod_{j=1}^d \int_{0}^{\Delta_j} x^{v_j}(\Delta_j-x_j)^{1-v_j} e^{-2\pi i \freq_j (x_j+z_j)} \de x_j \\
            &= \prod_{j=1}^d \int_{0}^{\Delta_j} x^{v_j}(\Delta_j-x_j)^{1-v_j} e^{-2\pi i \freq_j (x_j-v_j\Delta_j)} \de x_j,
        \end{align*}
        is invariant to $z$.
        Therefore
        \begin{align*}
            \ipalt{\mathcal{I}[g]}{\phi_\freq}_2
            &= \frac{1}{\elemprod{\Delta}} \sum_{z\in\grid} \sum_{v\in\set{0,1}^d} g(z+v\circ \Delta) \ipalt{\phi_\freq}{\mathcal{I}_{v,z}}_2 \\
            &= \frac{1}{\bar \Delta} \sum_{v\in\set{0,1}^d} \left(\sum_{z\in\grid}  g(z+v\circ \Delta) e^{-2\pi i \ip{(z+v\circ \Delta)}{\freq}} \right) \ipalt{\phi_\freq}{\mathcal{I}_{v,z}}_2 e^{2\pi i \ip{(z+v\circ \Delta)}{\freq}} \\
            &= \frac{1}{\bar \Delta^2}G^{(\grid)}(\freq) \sum_{v\in\set{0,1}^d} \ipalt{\phi_\freq}{\mathcal{I}_{v,z}}_2 e^{2\pi i \ip{(z+v\circ \Delta)}{\freq}}.
        \end{align*}
        Now we have
        \begin{align*}
            \sum_{v\in\set{0,1}^d} \ipalt{\phi_\freq}{\mathcal{I}_{v,z}}_2 e^{2\pi i \ip{(z+v\circ \Delta)}{\freq}} &= \sum_{v\in\set{0,1}^d} \prod_{j=1}^d \int_{0}^{\Delta_j} x^{v_j}(\Delta_j-x_j)^{1-v_j} e^{-2\pi i \freq_j (x_j-v_j\Delta_j)} \de x_j \\
            &= \prod_{j=1}^d \sum_{v=0}^1 \int_{0}^{\Delta_j} x^{v_j}(\Delta_j-x_j)^{1-v_j} e^{-2\pi i \freq_j (x_j-v_j\Delta_j)} \de x_j \\
            &= \prod_{j=1}^d \int_{0}^{\Delta_j} (\Delta_j-x_j) e^{-2\pi i \freq_j x_j} +x_j e^{-2\pi i \freq_j (x_j-\Delta_j)} \de x_j \\
            &= \prod_{j=1}^d \sinc^2(\pi \Delta_j \freq_j) \Delta_j^2,
        \end{align*}
        because if $k_j\neq 0$
        \begin{align*}
            & \hspace{-4em} \int_{0}^{\Delta_j} (\Delta_j-x_j) e^{-2\pi i \freq_j x_j} +x_j e^{-2\pi i \freq_j (x_j+\Delta_j)} \de x_j \\
            &= \int_{0}^{\Delta_j} (\Delta_j-x_j) e^{-2\pi i \freq_j x_j} \de x_j  + \int_{0}^{\Delta_j} (\Delta_j-x_j) e^{2\pi i \freq_j x_j} \de x_j \\
            &= 2\Re{\int_{0}^{\Delta_j} (\Delta_j-x_j) e^{2\pi i \freq_j x_j} \de x_j} \\
            &= 2\Re{\frac{2\pi i \Delta_j\freq_j-e^{2\pi i \Delta_j\freq_j}+1}{2\pi^2\freq_j^2}} \\
            &= \frac{\sin^2(\pi\Delta_j\freq_j)}{\pi^2\freq_j^2},
        \end{align*}
        otherwise if $k_j=0$,
        \begin{align*}
            \int_{0}^{\Delta_j} (\Delta_j-x_j) e^{-2\pi i \freq_j x_j} +x_j e^{-2\pi i \freq_j (x_j+\Delta_j)} \de x_j &= \int_0^{\Delta_j} \Delta_j \de x_j = \Delta_j^2.
        \end{align*}
        The stated result follows.
    \end{proof}
\end{proposition}
In practice, we scale $g$ such that $\norm{\mathcal{I}[g]}_2=1$.

\section{Bias in marked point processes}
\subsection{The expectation of the biased periodogram}\label{app:marked:expectation}

Periodograms have been proposed for marked point processes, both in the univariate \citep{renshaw2002two} and multivariate \citep{eckardt2019analysing} cases, however, these estimators are unfortunately systematically biased.
Without loss of generality, consider the univariate case, where $X$ is the point locations and $W_x$ is the mark at that location. 
We call this process the marked process, and the point process without the marks the ground process \citep{daley2003introduction}.
Breaking slightly from our previous notation, write $\lambda_g$ for the intensity of the ground process, $\lambda_m$ for the intensity of the marked process, and $\mu$ for the mean of the marks.
\citep{renshaw2002two} write an alternate discrete Fourier transform for the marked case as
\begin{align*}
    \mathfrak{F}(\freq) &= \frac{1}{\sqrt{\ell(\region)}} \sum_{x \in X \cap \region} (W_x-\mu) e^{-2\pi i x\freq}.
\end{align*}
\cite{renshaw2002two} only define this for rectangular regions but for convenience we study the general case here.
The proposed periodogram is then $\abs{\mathfrak{F}(\freq)}^2$.

Denote the discrete Fourier transform of the ground process (using only the points) and marked process by
\begin{align*}
    J_g(\freq) &= \frac{1}{\sqrt{\ell(\region)}} \sum_{x \in X \cap \region} e^{-2\pi i x\freq} - \lambda_g \int_{\region} e^{-2\pi i x\freq} \de x, \\
    J_m(\freq) &= \frac{1}{\sqrt{\ell(\region)}} \sum_{x \in X \cap \region} W_x e^{-2\pi i x\freq} - \lambda_m \int_{\region} e^{-2\pi i x\freq} \de x,
\end{align*}
respectively.
The mean correction we propose is different to that of \cite{renshaw2002two} in that we subtract $\lambda_m$ not $\mu$, and we do this everywhere, not just at the atoms of the mark sum measure.
In any case, $\lambda_m=\mu\lambda_g$ \citep[equation 5.1.19]{illian2008statistical}, and so
\begin{align*}
    \mathfrak{F}(\freq)
    &= \frac{1}{\sqrt{\ell(\region)}} \sum_{x \in X \cap \region} (W_x-\mu) e^{-2\pi i x\freq} \\
    &= \frac{1}{\sqrt{\ell(\region)}} \sum_{x \in X \cap \region} W_x e^{-2\pi i x\freq} - \lambda_m H(\freq) - \mu \left[\frac{1}{\sqrt{\ell(\region)}} \sum_{x \in X \cap \region} e^{-2\pi i x\freq} - \lambda_g H(\freq)\right]\\
    &= J_m(\freq) - \mu J_g(\freq),
\end{align*}
where for convenience we define
\begin{align*}
    H(\freq) &= \int_{\region} \frac{1}{\sqrt{\ell(\region)}} e^{-2\pi i x\freq} \de\freq.
\end{align*}

Therefore
\begin{align*}
    \abs{\mathfrak{F}(\freq)}^2 
    &= I_{mm}(\freq) + \mu^2 I_{gg}(\freq) - 2\mu \Re{I_{mg}(\freq)} \\ 
    &\rightarrow f_{mm}(\freq) + \mu^2 f_{gg}(\freq) - 2\mu \Re{f_{mg}(\freq)},
\end{align*}
as the region grows. 
This does not equal $f_{mm}(\freq)$, our desired estimand.

\subsection{The definition of spectra for marked processes}\label{app:marked:spectra}
In fact, the definition of the spectral density function given in \cite{renshaw2002two} and \cite{eckardt2019analysing} is not the same as the definition we use in this paper.
However, it is not equal to $f_{mm}(\freq) + \mu^2 f_{gg}(\freq) - 2\mu \Re{f_{mg}(\freq)}$.
In particular, \cite{renshaw2002two} define the spectral density function as the Fourier transform of the reduced factorial moment measure of the mark sum measure (though it is written differently in the paper, it is equivalent).
In contrast, we Fourier transform the reduced cumulant measure, matching the definition for point processes, random fields and general random measures \citep{daley2003introduction}.
In particular, define
    \begin{itemize}
        \item $\lambda_g$: the intensity of the ground process
        \item $\mu$: the mean of the mark (conditional on a point existing) 
        \item $\lambda_m = \lambda_g \mu$: the mark intensity
        \item $\mu_2$: the second moment of the mark (conditional on a point)
        \item $\rho_{gg}, \rho_{mm}$: the second-order product density of the ground process and mark sum measure respectively (factorial moment density)
        \item $\Breve{C}_m$: the covariance measure of the mark sum measure
        \item $k_{mm}(x) = {\rho_{mm}(x)}/{\rho_{gg}(x) \mu^2}$: the mark correlation, note this definition is from \cite{eckardt2019partial} and \cite{illian2008statistical}, who divide by $\mu^2$. \cite{eckardt2019analysing} only divide by $\mu$, though this doesn't impact the point of what follows.
    \end{itemize}

    Define the mean product of marks as
    \begin{align*}
        U(x) 
        &= \rho_{gg}(x) k_{mm}(x) \\
        &= \rho_{mm}(x) \frac{1}{\mu^2}
    \end{align*}
    Then \cite{renshaw2002two} define the spectra to be
   \begin{align*}
        S(\freq) 
        &= \int_\RRd U(x) e^{2\pi i \freq\cdot x}\de x.
    \end{align*}

    In contrast, the spectral density function of the mark sum measure is given by
    \begin{align*}
        f_m(\freq) &= \int_{\RRd} e^{2\pi i \freq\cdot x} \Breve{C}_m(\de x).
    \end{align*}
    Now we have for $B \in \borel{\RRd}$
    \begin{align*}
        \Breve{C}_m(B) &= \int_B \rho_{mm}(x) - \lambda_m^2\de x + a_m \delta(B)
    \end{align*}
    where the value of the atom $a_m$ can be found from Proposition 8.1.IV of \cite{daley2003introduction}.
    One can therefore see that the difference between the two definitions is that \cite{renshaw2002two} use the reduced factorial moment measure (up to some scaling), which is not the same as the reduced covariance measure.
    Typically it is not totally finite, and omitting the atom at zero means the spectral density function can be negative, which removes the interpretation as a variance of a wavenumber domain process.
    
    More specifically, we have
    \begin{align*}
        f_m(\freq) 
        &= \int_{\RRd} e^{2\pi i \freq\cdot x} \Breve{C}_m(\de x) \\
        &= \int_{\RRd} e^{2\pi i \freq\cdot x} \left(\rho_{mm}(x) - \lambda_m^2\right)\de x + a_m \\
        &= \int_{\RRd} e^{2\pi i \freq\cdot x} \mu^2 U(x) \de x - \lambda_m^2 \delta(\freq) + a_m
    \end{align*}
    and therefore
    \begin{align*}
        S(\freq) &= \frac{1}{\mu^2} \left(f_m(\freq) +  \lambda_m^2\delta(\freq) - a_m \right).
    \end{align*}
    Notice that one of the two spectra will not be a function in the usual sense.
    Consider the case when the ground point process has a spectral density $f_g$, and the marks are independent of the ground process and are IID, then the ground and mark spectra are related by
    \begin{align*}
        f_m(\freq) &= \mu_1^2 f_g(\freq) + \lambda_g (\mu_2-\mu^2)
    \end{align*}
    \citep{daley2003introduction}.
    So in this case, our definition of spectral density function is a density, whilst the Renshaw spectra is not.
    Furthermore, if we set the marks to have a point mass at 1, we obtain $\mu_2=\mu=1$, and so
    \begin{align*}
        f_m(\freq) &= f_g(\freq)
    \end{align*}
    as one would expect. Again this is not true of Renshaw spectra.

    Whilst this is a choice, it is sensible to choose the cumulant measure to Fourier transform, a choice which is made in the point process case, even in \cite{renshaw2002two}.
    Consider a basic null model where we have multiple independent Poisson processes with IID marks.
    Under the definition of \cite{renshaw2002two} and \cite{eckardt2019analysing}, the spectral density matrix function would be zero (except for a point mass at zero).
    This means that the spectral density matrix is not invertible at most frequencies, and thus partial coherence would not be well define even in the most simple case, in contrast to our definition.
    Furthermore, our definition is an exact generalisation of the unmarked case, as setting the marks to a point mass at one recovers the unmarked case, both in the spectral density function and our proposed estimators.
\subsection{Small simulation study}
\begin{figure}

\includegraphics{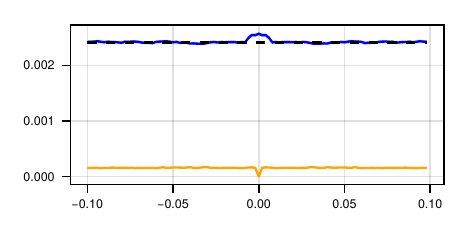}
    \caption{The true spectral density function of the IID marked Poisson process (dashed), the average of the multitaper estimates (blue) and the average of the alternate periodogram (orange).
    Taken at the slice $\freq_2=0$.
    }
    \label{fig:supp:bias}
\end{figure}

To verify the bias in the alternate periodogram, we performed a small simulation study.
We simulate Poisson processes with intensity $\lambda=0.001$, with marks drawn from an IID Normal(1.5,0.4) distribution.
We generate 1000 replications and at each estimate the spectra with multitapering and the alternate periodogram.
We show a comparison of the true spectra to two estimates in Fig.~\ref{fig:supp:bias}.
To ease comparison, we take a slice at $k_2=0$.
We can see that indeed the existing methodology is significantly biased, but the multitaper estimator is not.

\section{Irregular sampling}\label{app:irregular_sampling}
Recall that we assume we have a mean-zero random field $Y$ and point process $X$, the locations where we will sample $Y$. 
Then we can regard this sampled process as a marked point process, say  $\randmeasurebase$, with marks $Y(x)$ for $x\in X$.
Then, if the marks are independent of the locations, from \cite{daley2003introduction}, page 338, we have
\begin{align*}
    f_{\randmeasurebase\randmeasurebase}(\freq)
    &= \int_{\RRd}f_{YY}(\freq-\freq') f_{XX}(\freq')\de\freq' + \intensity{X} f_{YY}(\freq), \qquad \freq \in \RRd
\end{align*}
where $f_{\randmeasurebase\randmeasurebase}$ is the spectral density function of the marked process, $f_{YY}$ is the spectral density function of the random field, $f_{XX}$ is the spectral density function of the point process, and $\intensity{X}$ is the intensity of the point process.
If the point process used for sampling is Poisson, then
\begin{align*}
    f_{\randmeasurebase\randmeasurebase}(\freq)
    &= \intensity{X}\var{Y(0)} + \intensity{X}^2 f_{YY}(\freq), \qquad \freq \in \RRd.
\end{align*}

In fact, this corresponds to the approach proposed in \cite{matsuda2009fourier}. 
To see this, note that \cite{matsuda2009fourier} also assume that the random field is mean zero.
They assume that you sample the process with some random points with pdf given by $g$ supported on some subset of $[0,1]^d$ say $\tilde\region$.
They then assume that we observe the process at random locations $x_i = A\circ u_i$ where $u_i$ is a realisation of a random variable with pdf $g$, and the non-negative vector $A$ defines the region scaling.
They then scale the periodogram (which corresponds to our periodogram in the marked case with mean-zero marks) by a factor.
In particular, let $G=\norm{g}_2^2$ and $S = \prod_{j=1}^d [0, A_j]$, the box the region is rescaled too.
Then write the rescaled region as $\region$.
For simplicity we consider the untapered Fourier transform here.
The untapered Fourier transform from \cite{matsuda2009fourier} is
\begin{align*}
    \tilde{J}(\freq) &= G^{-1/2} \ell(S)^{1/2} n^{-1} \sum_{j=1}^n Y(x_j) e^{-2\pi i \ip{\freq}{x_j}} \\
    &=G^{-1/2} \ell(S)^{1/2} n^{-1} \ell(\region)^{1/2} \frac{1}{\ell(\region)^{1/2}} \sum_{j=1}^n Y(x_j) e^{-2\pi i \ip{\freq}{x_j}}\\
    &= G^{-1/2} \ell(S)^{1/2} n^{-1} \ell(\region)^{1/2} J(\freq) \\
    &= C J(\freq).
\end{align*}
where $J(\freq)$ is taken to be our marked tapered Fourier transform with taper $\ell(\region)^{-1/2}\indicator_{\region}$ (if we were to view the collection of points as a point process), and $C = G^{-1/2} \ell(S)^{1/2} n^{-1} \ell(\region)^{1/2}$. 
Of course, this will not satisfy all of the assumptions of our method (we would always use tapers) but we do not need the results that require the $L^1$ assumption here and this simplifies the point (it is about scaling).

In particular, assume that the sampling is uniform over the region, then
\begin{align*}
    G = \int_{\tilde{\region}} \frac{1}{\ell(\tilde{\region})^2} \de x = \frac{1}{\ell(\tilde{\region})}.
\end{align*}
Now we see that
\begin{align*}
    C 
    &= G^{-1/2} \ell(S)^{1/2} n^{-1} \ell(\region)^{1/2} \\
    &= \ell(S)^{1/2} n^{-1} \ell(\region)^{1/2} \ell(\tilde\region)^{1/2} \\
    &= \frac{\ell(\region)}{n}.
\end{align*}
Now \cite{matsuda2009fourier} point out that there will be an additional bias term which should be removed, which is
\begin{align*}
    B
    &= \frac{\ell(S)}{Gn}\var{Y} \\
    &= \frac{\ell(S)\ell(\tilde{\region})}{n} \var{Y} \\
    &= \frac{\ell(\region)}{n}\var{Y}.
\end{align*}
Putting all this together then, the periodogram version of the estimator proposed by \cite{matsuda2009fourier} is
\begin{align*}
    \tilde{I}(\freq) &= \frac{I(\freq)}{\hat\lambda_X^2} - \frac{\var{Y}}{\hat\lambda_X}
\end{align*}
where $\hat\lambda_X = n/\ell(\region)$.
Contrasting this with the equation from \cite{daley2003introduction}
\begin{align*}
    f_{\randmeasurebase\randmeasurebase}(\freq) &= \intensity{X}\var{Y} + \intensity{X}^2 f_{YY}(\freq) \\
    \Rightarrow f_{YY}(\freq) &= \frac{f_{\randmeasurebase\randmeasurebase}(\freq) - \intensity{X}\var{Y}}{\intensity{X}^2} \\
    &= \frac{f_{\randmeasurebase\randmeasurebase}(\freq)}{\intensity{X}^2} - \frac{\var{Y}}{\intensity{X}}.
\end{align*}
Therefore, the estimator one would construct from our marked process methodology using the relation given by \cite{daley2003introduction} coincides precisely with that proposed by \cite{matsuda2009fourier} in this case (we are just saying that $n$ is random).

Of course, this does not tell us about the case when the sampling is not uniform (which in the framework of \cite{matsuda2009fourier} would correspond to inhomogeneity which we could not deal with).
However, the approach of \cite{matsuda2009fourier} does not resolve the other, probably larger, issue of preferential sampling.
By this we mean, when the point locations at which we sample the field are not independent of the field.
The methodology proposed by \cite{matsuda2009fourier} also does not address dependence within the sampling locations, independent of the field.

\section{Additional details from the main simulation study}\label{app:simulation:details}

Say that the $p$\Th process is a log-Gaussian random field, and the $q$\Th and $r$\Th processes are point processes which are conditionally inhomogeneous Poisson processes with rate function $Y_p(\loc)$, but which are conditionally independent of each other.
Let $c$ be the covariance function of the Gaussian random field, from \cite{moller1998log} we have 
\begin{align*}
    \intensity{q} &= e^{\intensity{p}+c(0)/2} \\
    \reducedcumulantdens{q,q}(\lag) &= \left(\intensity{q}\right)^2[e^{c(\lag)}-1]+\intensity{q}\delta(\lag).
\end{align*}
By similar arguments one also has
\begin{align*}
    \reducedcumulantdens{q,p}(\lag) &= \left(\intensity{q}\right)^2[e^{c(\lag)}-1] \\
    \reducedcumulantdens{q,r}(\lag) &= \left(\intensity{q}\right)^2[e^{c(\lag)}-1] \\
\end{align*}

In the wavenumber domain, the spectral density functions are therefore
\begin{align*}
    \sdf{q,q}(\freq) &= \left(\intensity{q}\right)^2 \int_\RRd \left( e^{c(\lag)}-1\right)e^{-2\pi i\ip{\lag}{\freq}} \de\lag+\intensity{q}, \\
    \sdf{p,q}(\freq) &= \left(\intensity{q}\right)^2 \int_\RRd \left( e^{c(\lag)}-1\right)e^{-2\pi i\ip{\lag}{\freq}} \de\lag, \\
    \sdf{p,r}(\freq) &= \left(\intensity{q}\right)^2 \int_\RRd \left( e^{c(\lag)}-1\right)e^{-2\pi i\ip{\lag}{\freq}} \de\lag.
\end{align*}

For the final model, we approximate the true spectral density function through simulations.
In particular, we simulate 1000 realizations on the full 1000m by 500m square, and then estimate the spectral density matrix function using a small bandwidth.
We then average the results to get a good approximation of the true spectral density matrix function.
From this, we then compute coherence and partial coherence for the comparisons.

\begin{figure}
    \centering

\includegraphics{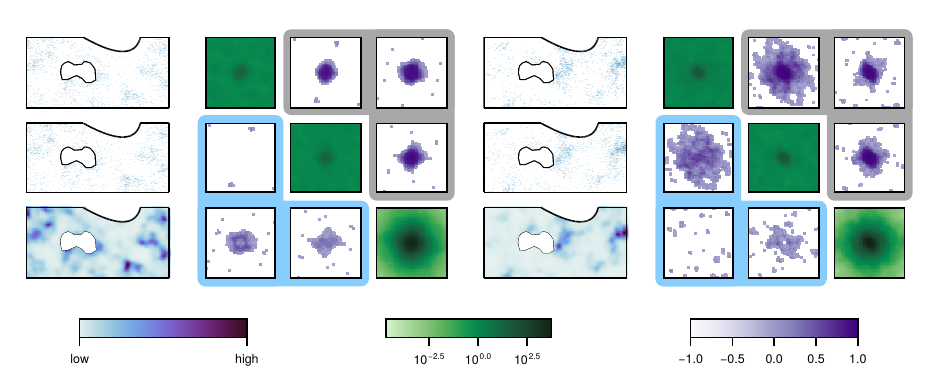}
    \caption{
        A realization of models 2 (left) and 3 (right). The spatial data has axes 0 to 1000 and 0 to 500, and the wavenumber domain plots show both $\freq_1$ and $\freq_2$ ranging from -0.05 to 0.05.
        Colorbars indicate from left to right, the value of the field from low to high, the value of the log standardized marginal spectra and the value of the signed coherence and signed partial coherence.
        In both cases, the swamp is excluded.
    }
    \label{fig:supp:simulation_examples}
\end{figure}

Example realizations and corresponding spectral estimates for models 2 and 3 are shown in Fig.~\ref{fig:supp:simulation_examples}.
We can see that the estimated spectra do a good job of reflecting the underlying structure of the model from which the data was simulated.

\section{Additional simulation models and results}
\subsection{Colocation model}\label{supp:sim:coloc}
Consider a multivariate random field model, in particular, a colocation model, see \cite{waagepetersen2016analysis} for a description.
In our case, we consider one shared latent field, and two individual fields, so that the output fields are given by
\begin{align*}
    Y_{1}(\loc) &= U_{1}(\loc) + \alpha_{1} X(\loc), \\
    Y_{2}(\loc) &= U_{2}(\loc) + \alpha_{2} X(\loc),
\end{align*}
for some $\alpha_{1},\alpha_{2}\in\RR$.
In our case, the random fields $U_{1},U_{2},X$ are all assumed to be independent of each other.
In this case, we have that for $1\leq p,q \leq 2$,
\begin{align*}
    f_{p,q}^{(Y)}(\freq) &= f_{p,q}^{(U)}(\freq)\delta_{p,q} + \alpha_{p}\alpha_{q} f^{(X)}(\freq),
\end{align*}
where subscripts indicate the spectrum corresponds to that kind of process.

Furthermore, we assume the processes $U_{1},U_{2},X$ are Gaussian with the same marginal covariance structure, but we choose different grids for the sampling two processes $Y_{1}$ and $Y_{2}$.
Recall the grid for the $j$\Th process  is specified by its spatial offset $\gridoffset_j$, and sampling interval $\Delta_{j}$, which in this case are both in $\RR^2$. In particular,

The grids are defined by
\begin{align*}
    \gridoffset_{1} = (0,0), &\qquad \Delta_{1} = (5,5),\\
    \gridoffset_{2} = (0,3), &\qquad \Delta_{2} = (10,15).
\end{align*}
And the covariance function is Mat\'ern, with length scale $\ell=20$ and smoothness parameter $\nu=3$ (and variance $\sigma=1.0)$. We take $\alpha_{1}=0.8$ and $\alpha_{2}=0.5$.

\begin{figure}[h]

\includegraphics{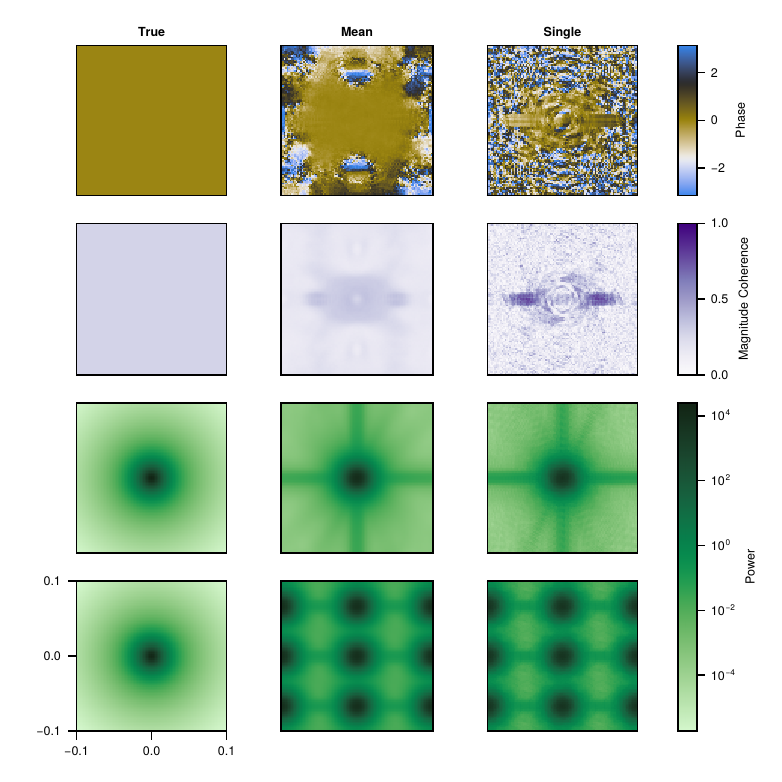}
    \caption{The log marginal spectra, coherence and phase for the colocation model.
    In each case, we show and the ground truth (left), the average over 1000 replications (middle) and the estimate from a single realization (right). In the case of phase, the average estimate uses a circular average.}
    \label{fig:supp:sim:coloc}
\end{figure}

We run the simulation 1000 times. We show the average of the log spectrum, and the coherence and phase in Fig.~\ref{fig:supp:sim:coloc}.
The average estimate is close to the true value, and the estimate from a single realization is also close to the true value.
However, as expected we can see the effect of aliasing in the second process, as that grid is courser.

\subsection{Shifted point pattern}
Say that we have some initial point pattern $\randmeasure{p}$, whose covariance has no atoms other than at zero. 
Construct $\randmeasure{q}$ by shifting every point in the original pattern by some fixed $\tau\in\RRd\setminus \set{0}$, i.e. so that $\randmeasure{q}(B)=\randmeasure{p}(B-\tau)$.
Then there is an atom present in the cross-covariance (though it is not at zero), in particular
\begin{align*}
    \reducedcumulantdens{p,q}(\lag)
    &= \intensity{p}\delta(\lag+\tau) + \reducedcumulantdens{[p,p]}(\lag+\tau) = \reducedcumulantdens{p,p}(\lag+\tau), \qquad \lag\in\RRd,
\end{align*}
where $\reducedcumulantdens{[p,p]}$ is the factorial covariance density of the point process $\randmeasure{p}$.
Thus the cross-spectral density function is given by
\begin{align*}
    \sdf{p,q}(\freq) &= \int_\RRd \reducedcumulantdens{p,p}(\lag+\tau) e^{-2\pi i \ip{\lag}{\freq}} \de\lag \\
    &= \sdf{p,p}(\freq) e^{2\pi i \ip{\tau}{\freq}},
\end{align*}
for $\freq\in\RRd$.

Though not a realistic process, this shifted example is still important as these simple cases are helpful for developing intuition about the cross-spectral density function and related quantities.
In particular, the two processes in question have perfect coherence, but with some fixed phase.
In the time series case, taking a process and its deterministic shift, we also have perfect coherence and some phase determined by the shift.
As in the time series case, such examples allow us to build intuition about phase in terms of spatial shifts.

\begin{figure}[h]

\includegraphics{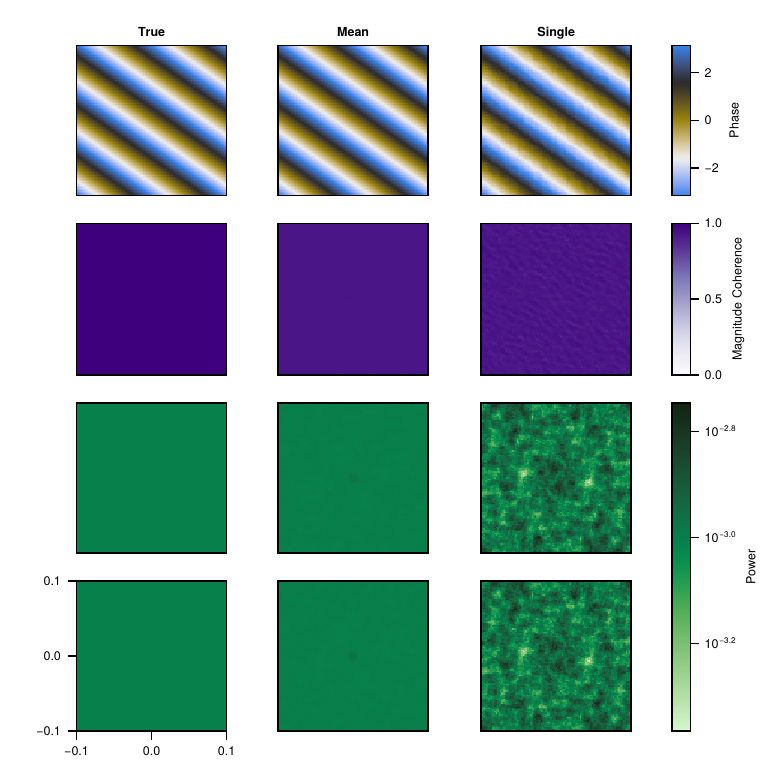}
    \caption{
    The log marginal spectra, coherence and phase for the shifted model.
    In each case, we show and the ground truth (left), the average over 1000 replications (middle) and the estimate from a single realization (right). In the case of phase, the average estimate uses a circular average.
    }
    \label{fig:sim:shift}
\end{figure}

Consider a Poisson process with intensity $\lambda=0.001$, and then make a second by shifting the first by $(10.5,15)$, again performing 1000 replications.
The average vs true log marginal spectra, coherence and phase are shown in Fig.~\ref{fig:sim:shift}.
We see that as in the theory, the estimate obtains a constant coherence with a phase which is linear, though it is plotted modulus $2\pi$ on the interval $(-\pi,\pi]$.
\subsection{Partial correlation of normal random variables}
\begin{figure}[htp]
    \centering

\includegraphics{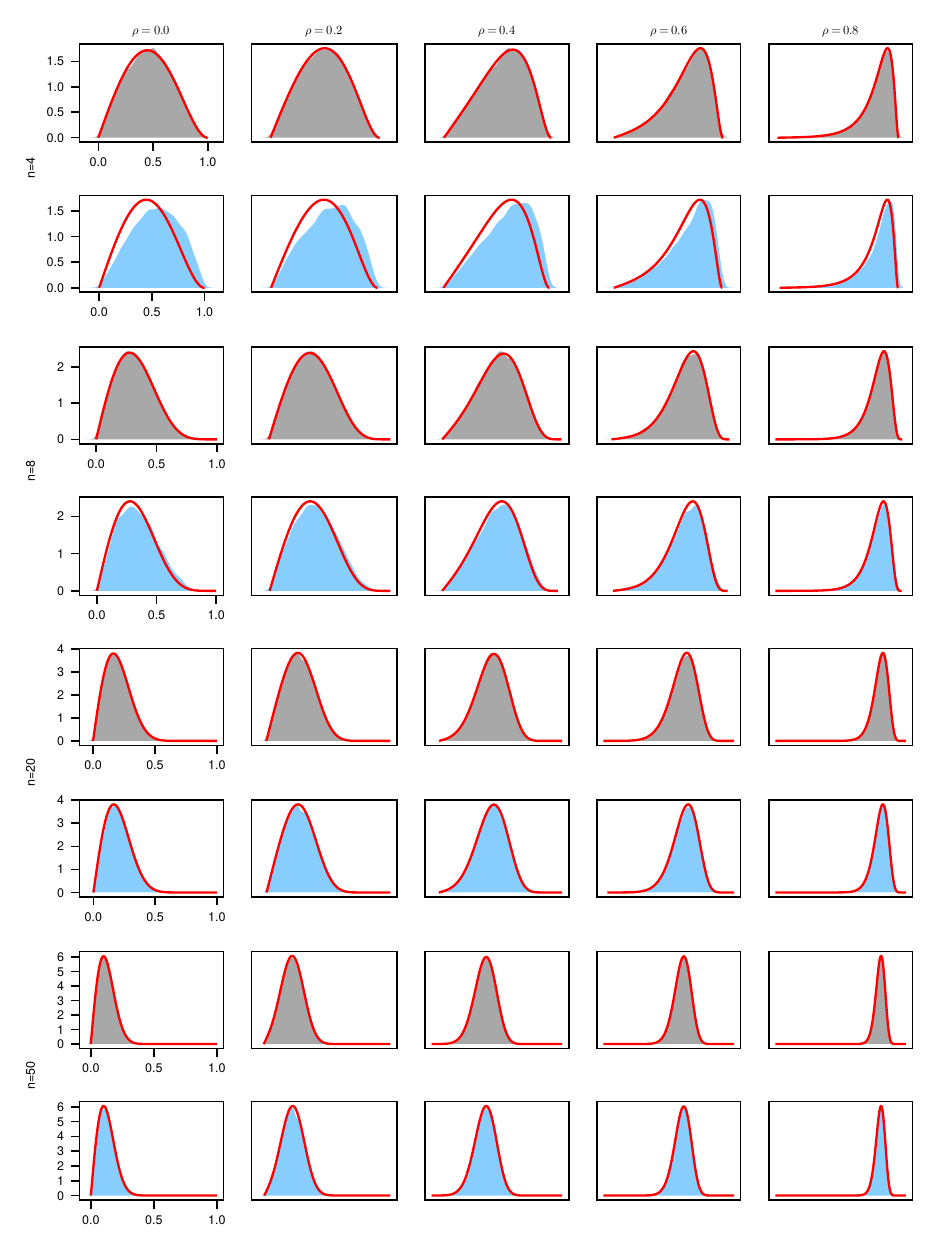}
    \caption{The empirical distributions of the magnitude correlation and partial magnitude correlation for normal random variables vs the truth (shown by the red line).
    Rows are organized so they alternate correlation (gray) and partial correlation (blue).}
    \label{fig:supp:partial_correlation}
\end{figure}

We use the distribution of magnitude correlation for magnitude partial correlation of normal random variables in the paper.
To examine this, we consider a simple simulation experiment. In particular, define

\begin{align*}
    z &\sim \mathcal{CN}(0,1,0) \\
    \epsilon &\sim \mathcal{CN}(0,1,0) \\
    \nu &\sim \mathcal{CN}(0,1,0) \\
    \eta &= \rho \epsilon + (1-\rho^2)^{1/2}  \nu \\
    x &= z + \epsilon \sigma \\
    y &= z + \eta \sigma
\end{align*}
and consider the correlation of $x$ and $y$, and the partial correlation of $x$ and $y$ given $z$.
Assume that we obtain $n$ replications of $x,y,z$.
We compute the sample magnitude correlation and partial magnitude correlation for a range of $\rho$ in simulations (assuming the mean is known to be zero and fixing $\sigma=1$).
We perform this experiment 10000 times for each value of $\rho$ and multiple values of $n$.
We show the results in Fig.~\ref{fig:supp:partial_correlation}.
Note that except for very small $n$, the distributional approximation is very good for the partial magnitude correlation (of course it is exact for magnitude correlation).

\section{Additional application figures}\label{app:bci:extrafigures}

\begin{figure}

\includegraphics{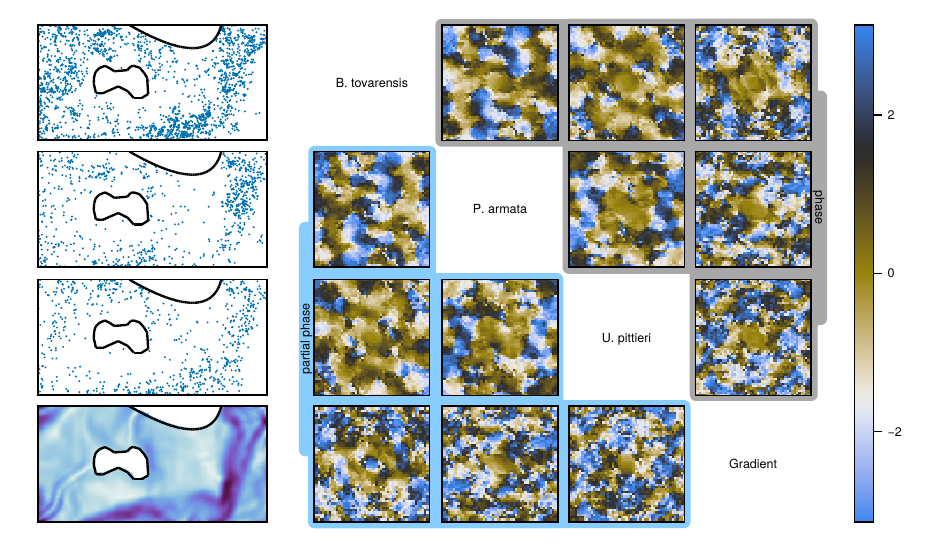}
    \caption{
        The phase (upper triangle of the plot matrix) and partial phase (lower triangle) of \textit{B. tovarensis}, \textit{P. armata}, \textit{U. pittieri} and the gradient of the terrain.
        The spatial data has axes 0 to 1000 and 0 to 500, and the wavenumber domain plots ranging from -0.05 to 0.05 in both x and y axes.
    }
    \label{fig:supp:bci:phase:gradient}
\end{figure}

The phase for the Barro Colorado Island data for the first analysis (\textit{B. tovarensis}, \textit{P. armata}, \textit{U. pittieri} and gradient) is shown in Fig.~\ref{fig:supp:bci:phase:gradient}.
We see that for large values of the marginal spectra, they are close to zero as expected.

\clearpage
\bibliographystyle{apalike}
\bibliography{paper-ref}

\end{document}